\documentclass[12pt]{article}
\usepackage{epsfig}
\usepackage{here}
\usepackage{a4}
\usepackage{amssymb}

\begin{document}

\def\Journal#1#2#3#4{{#1} {\bf #2}, #3 (#4)}

\def\etal{{\it et\ al.}}
\def\NCA{\em Nuovo Cim.}
\def\NIM{\em Nucl. Instrum. Methods}
\def\NIMA{{\em Nucl. Instrum. Methods} A}
\def\NPB{{\em Nucl. Phys.} B}
\def\PLB{{\em Phys. Lett.}  B}
\def\PRL{\em Phys. Rev. Lett.}
\def\PRC{{\em Phys. Rev.} C}
\def\PRD{{\em Phys. Rev.} D}
\def\ZPC{{\em Z. Phys.} C}
\def\ASP{{\em Astrop. Phys.}}
\def\JETP{{\em JETP Lett.\ }}

\def\numunue{\nu_\mu\rightarrow\nu_e}
\def\numunutau{\nu_\mu\rightarrow\nu_\tau}
\def\nuebar{\bar\nu_e}
\def\nue{\nu_e}
\def\nutau{\nu_\tau}
\def\numubar{\bar\nu_\mu}
\def\numu{\nu_\mu}
\def\ra{\rightarrow}
\def\numubarnuebar{\bar\nu_\mu\rightarrow\bar\nu_e}
\def\nuebarnumubar{\bar\nu_e\rightarrow\bar\nu_\mu}
\def\osc{\rightsquigarrow}

\thispagestyle{empty}
\begin{flushright}
{\tt ICARUS/TM-2001/10}\\ 
December 21, 2001
\end{flushright}
\vspace*{0.5cm}
\begin{center}
{\Large{\bf On the energy and baseline optimization
to study effects related to the $\delta$-phase (CP-/T-violation)
in neutrino oscillations at a Neutrino Factory} }\\
\vspace{.5cm}
A. Bueno\footnote{Antonio.Bueno@cern.ch},
M. Campanelli\footnote{Mario.Campanelli@cern.ch},
S. Navas-Concha\footnote{Sergio.Navas.Concha@cern.ch}
and A. Rubbia\footnote{Andre.Rubbia@cern.ch}

\vspace*{0.3cm}
Institut f\"{u}r Teilchenphysik, ETHZ, CH-8093 Z\"{u}rich,
Switzerland

\end{center}
\vspace{0.3cm}
\begin{abstract}
\noindent
   In this paper we discuss the detection of $CP$ and $T$-violation
effects in the framework of a neutrino factory.
   We introduce three quantities, which are good discriminants
for a non vanishing complex phase ($\delta$) in the $3\times 3$ neutrino
mixing matrix:
$\Delta_{\delta}$, $\Delta_{CP}$ and $\Delta_T$.
   We find that these three discriminants (in vacuum) 
all scale with $L/E_{\nu}$, where $L$ is the baseline and $E_\nu$ the neutrino energy.
Matter effects modify the scaling, but these effects are large enough
to spoil the sensitivity only for
baselines larger than 5000~km.
So, in the hypothesis of constant neutrino factory power (i.e. number of muons
inversely proportional to muon energy), the sensitivity on the
$\delta$-phase is independent of the baseline chosen.
   Specially interesting is the
direct measurement of $T$-violation from the ``wrong-sign'' electron channel
(i.e. the $\Delta_T$ discriminant),
which involves a comparison of the $\nue\ra\numu$ and $\numu\ra\nue$
oscillation rates. However, the $\numu\ra\nue$ measurement requires
magnetic discrimination of the electron charge, experimentally very challenging in a
neutrino detector. 
Since the direction of the electron curvature has to be estimated before
the start of the electromagnetic shower, low-energy neutrino beams
and hence short baselines, are preferred.
   In this paper we show, as an example, the exclusion regions in the
$\Delta m^2_{12} - \delta$ plane using the $\Delta_{CP}$ and $\Delta_{T}$
discriminants for two concrete cases keeping
the same $L/E_{\nu}$ ratio (730 km/ 7.5 GeV and 2900 km/30 GeV).
We obtain a similar excluded region provided that the
electron detection efficiency is $\sim$20\% and the charge confusion 0.1\%.
The $\Delta m^2_{12}$ compatible with the LMA solar data can be tested with
a flux of 5$\times 10^{21}$ muons.
   We compare these results with the fit of the visible energy distributions.

\end{abstract}

\newpage
\pagestyle{plain} 
\setcounter{page}{1}
\setcounter{footnote}{0}

\section{Introduction}
The firmly established disappearance of muon neutrinos of cosmic ray 
origin~\cite{atmevid} strongly points 
toward the existence of neutrino
oscillations~\cite{pontecorvo}. 

The first generation long baseline (LBL) experiments ---
K2K~\cite{k2k}, MINOS~\cite{minos}, OPERA~\cite{opera} and
ICARUS~\cite{icarus,Arneodo:2001tx,icanoe} --- will search for
a conclusive and unambiguous signature of the oscillation
mechanism using artificial neutrino beams produced by 
the ``traditional'' meson-decay method.
They will provide the first precise
measurements of the parameters governing the main
muon disappearance mechanism. 

In contrast, a neutrino factory\cite{geers,nufacwww} is understood 
as a machine where
low energy muons of a given charge are accelerated in a storage
ring.
Neutrino factories raised the interest of
the physics community, since they appear natural follow-ups to the current 
experimental LBL program and could
open the way to future muon colliders. 

The real physics potential of a neutrino factory comes
from its ability to test in a very clean and high statistics
environment all possible flavor oscillation transitions.
As many studies have
shown~\cite{nufacfis}, the
physics potential of such facilities are very vast.
An entry-level neutrino factory could
test the LSND signal in a background free environment\cite{Bueno:1998xy}.
A neutrino factory source would be of
sufficiently high intensity to perform very long baseline
(transcontinental) experiments.

The ability to measure all possible neutrino oscillations will provide very
stringent information on all the elements of the neutrino mixing matrix and on
the mass pattern of the neutrinos. In a $3\times 3$-mixing scenario, the mixing
matrix, which should be unitary, is determined by three angles and a complex
phase. As studies have shown~\cite{nufacfis}, precise determination of two
angles and of the largest mass difference will be obtained. In addition, a test
of the unitarity of the matrix could be performed.

Apart from being able to measure very precisely all the magnitude of the
elements of the mixing matrix, the more challenging and most interesting
goal of the neutrino factory will be the search for effects related to the
complex phase of the mixing matrix. The complex phase will in general alter
the neutrino flavor oscillation probabilities, and will most strikingly
introduce a difference of transition probabilities between neutrinos and
antineutrinos (so called $CP$-violation effects), and between 
time-reversed transitions (so called $T$-violation
effects)\cite{Arafune:1997hd}. 
Neutrino factories should provide the intense 
and well controlled beams needed to perform these studies.

Two fundamental experimental parameters for these studies are (1) the energy
$E_\mu$ of the stored muon beam which determines the neutrino energy $E_\nu$
spectrum and (2) the baseline $L$ between the source and the detector. In
earlier works, one has advocated ŒŒhigh energy neutrinos¹¹, with a consensus
around $E_\mu\simeq 30-50\rm\ GeV$ coupled to long baselines $L\simeq 3000\ \rm
km$. In Ref.\cite{sato}, a much lower muon energy in the range $E_\mu\simeq 1\
\rm GeV$ coupled to short baselines $L\simeq 100\ \rm km$ has been
proposed. This apparent contradiction was recently addressed by
Lipari\cite{Lipari:2001ds}.

In this paper, we expand our early work\cite{Bueno:2000fg} on the detection
of $CP$-violation at a neutrino factory. In particular, we concentrate on
the general strategies to detect effects related to the complex phase of
the neutrino mixing matrix. 
The complete and comprehensive detection of the effect is very difficult 
for terrestrial experiments, as it would require $L/E_\nu$ values simultaneously in the
range of solar and atmospheric neutrinos. We concentrate on medium $L/E_\nu$ and
demonstrate  that the relevant effects scales as $L/E_\nu$ and that there is an
optimal choice for the ratio $L/E_\nu$, given by the neutrino mass
pattern. Hence this yields an priori large freedom for the choice of $L$ and
$E_\nu$. We show however that for too large $L$, matter effects destroy the
sensitivity to the complex phase, so baselines smaller than $L\simeq 5000\ \rm
km$ are preferred. In addition, owing to the linear rise of the neutrino
cross-section with neutrino energy at high energy, the statistical significance
of the effect scales with $E_\nu$, and hence grows linearly with $L$, for
$L/E_\nu$ constant.

The choice of baseline $L$ is particularly critical, in the sense that at
the time that a neutrino factory would start running, there will be already
existing experiments at baselines from 730~km from FNAL and CERN: (1) at
Soudan, MINOS with its 5.4~kton fiducial mass will have been fully
operational, and similarly, (2) at GranSasso,
the OPERA experiment and a multi-kton ICARUS detector. In 
view of the existence of such massive devices, it is worth considering if these
detectors at their current baselines could be reused or improved in the context of
the neutrino factory. If new sites have to be found in order
to satisfy the requirements of longer baselines, major new ``investments''
will be required.

As far as the neutrino energy is concerned, it is clear that the average
neutrino energy $E_\nu$ scales linearly with the muon beam energy
$E_\mu$. A non-negligible aspect of the neutrino factory is the need to
accelerate quickly the muons to the desired energy, and so, it is expected
that higher energies will be more demanding that lower ones. Eventually,
cost arguments could determine the muon energy. It could therefore be
that lower energy, more intense neutrino factories could be more advantageous
than higher energy, less intense ones. 

\section{The effects related to $\delta$ in vacuum}
In a three-family neutrino oscillation scenario,
the flavor eigenstates
$\nu_\alpha(\alpha= e,\mu,\tau)$ are related to the mass eigenstates
$\nu'_i(i=1,2,3)$ by the mixing matrix U
\begin{equation}
\nu_\alpha=U_{\alpha i}\nu'_i
\end{equation}
and it is customary to parameterize it as:
\begin{equation}
U(\theta_{12},\theta_{13},\theta_{23},\delta)=\left(
\begin{tabular}{ccc}
$c_{12}c_{13}$      & $s_{12}c_{13}$   &  $s_{13}e^{-i\delta}$ \\
$-s_{12}c_{23}-c_{12}s_{13}s_{23}e^{i\delta}$ &
$c_{12}c_{23}-s_{12}s_{13}s_{23}e^{i\delta}$ & $c_{13}s_{23}$ \\
$s_{12}s_{23}-c_{12}s_{13}c_{23}e^{i\delta}$ &
$-c_{12}s_{23}-s_{12}s_{13}c_{23}e^{i\delta}$ & $c_{13}c_{23}$ 
\end{tabular}\right)
\end{equation}
with $s_{ij}=\sin\theta_{ij}$ and $c_{ij}=\cos\theta_{ij}$. 

We adopt the usual neutrino mass assignment in which the smallest $\Delta
m^2_{21}\equiv m^2_2-m^2_1$ 
is assigned to the solar neutrino deficit and the largest mass difference, 
$\Delta m^2_{31}\approx\Delta m^2_{32}$ is describing the atmospheric
neutrino observations. 
We neglect accordingly the important results from LSND, awaiting further 
clarification from the MiniBOONE experiment.  We therefore consider the following
indicative set of oscillation parameters\footnote{Note that in studies of
effects related to $\delta$, the signs of
the $\Delta m^2$'s are important.}, compatible with current
experimental observations (neglecting the LSND claim):

\begin{equation}
\Delta m_{32}^2 = 3 \times 10^{-3}\rm eV^2,\ \ \ \ \sin^2 \theta_{23} = 0.5
\label{eq:atm}
\end{equation}

The value of the angle $\theta_{13}$ is currently not known, 
and the best experimental bound comes from CHOOZ\cite{chooz} which limits 
$\sin^2 2\theta_{13} \lesssim 0.1$. We take:

\begin{equation}
\sin^2 2\theta_{13} = 0.05 
\label{eq:chooz}
\end{equation}

For the solar parameters, we assume values compatible with the LMA-solution, 
which is known to yield optimal conditions for $CP$-violation studies (see
Ref.~\cite{golden}):
\begin{equation}
\Delta m^2_{21}=1\times 10^{-4}\rm eV^2,\ \ \ \  \sin^2 \theta_{12}=0.5
\label{eq:sol}
\end{equation}

With this mass assignment and with the parameterization of the mixing matrix
described above, a neutrino factory can provide precise information on the
largest squared mass difference $\Delta m^2_{13}\approx \Delta m^2_{23}$ and on both
angles $\theta_{23}$ and $\theta_{13}$.

In the case where neutrinos (or antineutrinos) propagate in vacuum, the
probability for flavor transition has a simple behavior. It can be written as
\begin{eqnarray}\label{eq:oscillcp}
P(\nu_\alpha\osc\nu_\beta;E_\nu,L)  & = & \sum_{jk} J_{\alpha\beta j k}
e^{-i\Delta m^2_{jk}L/2E}
\end{eqnarray}
where $E_\nu$ is the neutrino energy and $L$ the baseline between the source
and the detector.
The Jarlskog factor is $J_{\alpha\beta j k} = U_{\beta j} U^*_{\beta k} U^*_{\alpha j}
U_{\alpha k}$. For antineutrinos, we must replace $U\ra U^*$, i.e. $J_{\alpha\beta j k}
\ra J^*_{\alpha\beta j k}=J_{\alpha\beta k j}=J_{\beta\alpha j k}$.

Since in general we have $J_{\alpha\beta j k}\neq J^*_{\alpha\beta j k}$, the probability for neutrinos
$P(\nu_\alpha\ra\nu_\beta;E_\nu,L)$ can be different from that for
anti-neutrinos $P(\bar\nu_\alpha\ra\bar\nu_\beta;E_\nu,L)$, 
leading to the CP-violating effects. Since 
$J_{\alpha\beta j k}=J^*_{\beta\alpha j k}$, there are also T-violating
effects. 
From the definition of $J$ also follows that (CPT-invariance)
\begin{equation}
P(\nu_\alpha\osc\nu_\beta;E_\nu,L) = P(\bar\nu_\beta\osc\bar\nu_\alpha;E_\nu,L)
\end{equation}

The oscillation probability can be made more explicit by
separating real and imaginary contributions of $J_{\alpha\beta j k}$:
\begin{eqnarray}\label{eq:oscill}
 P(\nu_\alpha\osc\nu_\beta;E_\nu,L)  & = & \sum_{j}J_{\alpha\beta j j}+
\sum_{j>k} \left( J_{\alpha\beta j k}e^{-i\Delta m^2_{jk}t/2E}+
J^*_{\alpha\beta j k}e^{+i\Delta m^2_{jk}t/2E} \right) \\
& = & \delta_{\alpha\beta}-4\sum_{j>k}  \Re e J_{\alpha\beta j k}\sin^2\left(\Delta_{jk}\right) \\ \nonumber
&   & + 4\sum_{j>k} \Im m J_{\alpha\beta j k}\sin\left(\Delta_{jk}\right)\cos\left(\Delta_{jk}\right)
\end{eqnarray}
Here 
\begin{equation}
\Delta_{jk}\equiv\Delta m^2_{jk}\frac{L}{4E_\nu}
\end{equation}
and $\Delta_{12}+\Delta_{23}+\Delta_{31}=0$.
For anti-neutrinos, we must
replace $\Im m J_{\alpha\beta j k}\ra -\Im m J_{\alpha\beta j k}$. Hence,
the first
two terms of $P$ form a CP-even term, and the last one is a CP-odd term.

In the rest of the paper, we concentrate on transitions between electron
and muon neutrinos. 
One can explicitly calculate the probabilities, inserting the elements of
the chosen 
mixing matrix parameterization:
\begin{eqnarray}
 P(\nu_e\ra\nu_\mu)   = P(\bar\nu_\mu\ra\bar\nu_e)   = \nonumber \\ 
4c^2_{13}
\Bigl[  \sin^2\Delta_{23} s^2_{12} s^2_{13}
      s^2_{23} + c^2_{12}
      \left( \sin^2\Delta_{13}s^2_{13}s^2_{23} + 
        \sin^2\Delta_{12}s^2_{12}
         \left( 1 - \left( 1 + s^2_{13} \right) s^2_{23} \right)  \right)  \Bigr]  
\nonumber \\ 
  -  \frac{1}{2}c^2_{13}\sin (2\theta_{12})s_{13}\sin (2\theta_{23})
 \ \underline{\cos\delta}
     \left[ \cos 2\Delta_{13} - \cos 2\Delta_{23} - 
       2\cos(2\theta_{12})\sin^2\Delta_{12}\right]  \nonumber \\ 
  + \frac{1}{2}c^2_{13}\ \underline{\sin\delta}
\sin(2\theta_{12})s_{13}\sin (2\theta_{23})
\left[ \sin2\Delta_{12} - \sin2\Delta_{13} + 
\sin 2\Delta_{23} \right]
\label{eq:probnenm}
\end{eqnarray}
This expression has been split in a first part independent from the phase
$\delta$, and in the two parts proportional respectively to $\cos\delta$ and $\sin\delta$.
To obtain the probabilities for $\numu\ra\nue$ and $\bar\nue\ra\bar\numu$,
we must replace $\delta\longrightarrow -\delta$, with the effect of
changing $\sin\delta\longrightarrow -\sin\delta$ and 
$\cos\delta\longrightarrow \cos\delta$. The term proportional to
$\sin\delta$ is the CP- or T-violating term, while 
the $\cos\delta$ term equally modifies the probability for 
both $CP$-conjugate states.

The behavior in the range where the energy is too large or the baseline too short to be
sensitive to the smallest mass difference, i.e. $|\Delta m^2_{12}|\ll E_\nu/L$ or
$|\Delta_{12}|\ll 1$, is found by $\Delta_{12}\ra 0$ and
$\Delta_{13}\ra\Delta_{23}$.
Both terms depending on $\cos\delta$ and $\sin\delta$ vanish, and we recover the well known probability
\begin{equation}
\label{eq:wellknownonemass}
P(\nu_e \to \nu_\mu) \simeq\sin^2(2\theta_{13})s^2_{23} 
\sin^2\left(\Delta_{23}\right)
\end{equation}

Going back to the general case,
the $\nue\ra\numu$ oscillation probability
can be simplified by introducing values for
$\theta_{12}$ and $\theta_{23}$ consistent with current experimental
knowledge. Indeed, for $\theta_{12}=\theta_{23}=\pi/4$, we find
\begin{eqnarray}
\label{eq:nuenumudel}
 P(\nu_e\ra\nu_\mu)   & = & \frac{1}{2}c^2_{13}
 \Bigl\{
c^2_{13}\sin^2\Delta_{12}+
s_{13}\Bigl[ 2\sin\Delta_{12}
\sin(\Delta_{13}+\Delta_{23}-\delta)+\\ \nonumber
& & \sin\delta\sin 2\Delta_{12}+ 
 2s_{13}\left(\sin^2 \Delta_{13}+\sin^2\Delta_{23} \right)\Bigr]
\Bigr\}
\end{eqnarray}

From this dependence, we see that a precise measurement of
the $\nue\ra\numu$ oscillation probability can yield information
of the $\delta$-phase provided that the other oscillation
parameters in the expression are known sufficiently accurately.
We also note that the phase $\delta$ changes the neutrino energy position
of the maximum of the oscillation, because of the term 
$\sin(\Delta_{13}+\Delta_{23}-\delta)$.

Clearly, the dependence of the parameter $\delta$ is a priori most
``visible'' in the energy-baseline range such that 
$|\Delta_{12}|=|\Delta m^2_{21}|L/4E_\nu\simeq 1$ and
$|\Delta_{23}|=|\Delta m^2_{23}|L/4E_\nu\simeq 1$. This is the region
where all terms in Eq.~(\ref{eq:nuenumudel}) contribute.

When $|\Delta_{12}|\ll 1$ and $|\Delta_{23}|\simeq 1$, we obtain
\begin{eqnarray}
 P(\nu_e\ra\nu_\mu) & \simeq & \frac{1}{2}c^2_{13} \Bigl\{ c^2_{13}\Delta^2_{12}+
 2s^2_{13}\left(\sin^2 \Delta_{13}+\sin^2\Delta_{23} \right) \\ \nonumber
& &  +2\Delta_{12}s_{13}\Bigl[ 
\sin(\Delta_{13}+\Delta_{23})\cos\delta+\left(1-\cos(\Delta_{13}+\Delta_{23})\right)\sin\delta 
\Bigr]
\Bigr\}
\end{eqnarray}

At even higher $E_\nu$ or smaller $L$, we further have, when 
both $|\Delta_{12}|\ll 1$ and $|\Delta_{13}|,|\Delta_{23}|\ll 1$:
\begin{eqnarray}
 P(\nu_e\ra\nu_\mu) & \simeq & \frac{1}{2}c^2_{13} \Bigl\{ c^2_{13}\Delta^2_{12}+
 2s^2_{13}\left(\Delta^2_{13}+\Delta^2_{23} \right) 
 +2\Delta_{12}s_{13}
(\Delta_{13}+\Delta_{23})\cos\delta
\Bigr\}
\label{eq:probhighe}
\end{eqnarray}
and the dependence on the phase is only through $\cos\delta$. From this
follows a degeneracy under the change of sign of $\delta$, as was pointed
out in Ref.~\cite{Lipari:2001ds}. So, in this $L$ and $E_\nu$ range, a precise
determination of the oscillation probability can no longer determine the
sign of $\delta$.

\begin{figure}
\centering
\epsfig{file=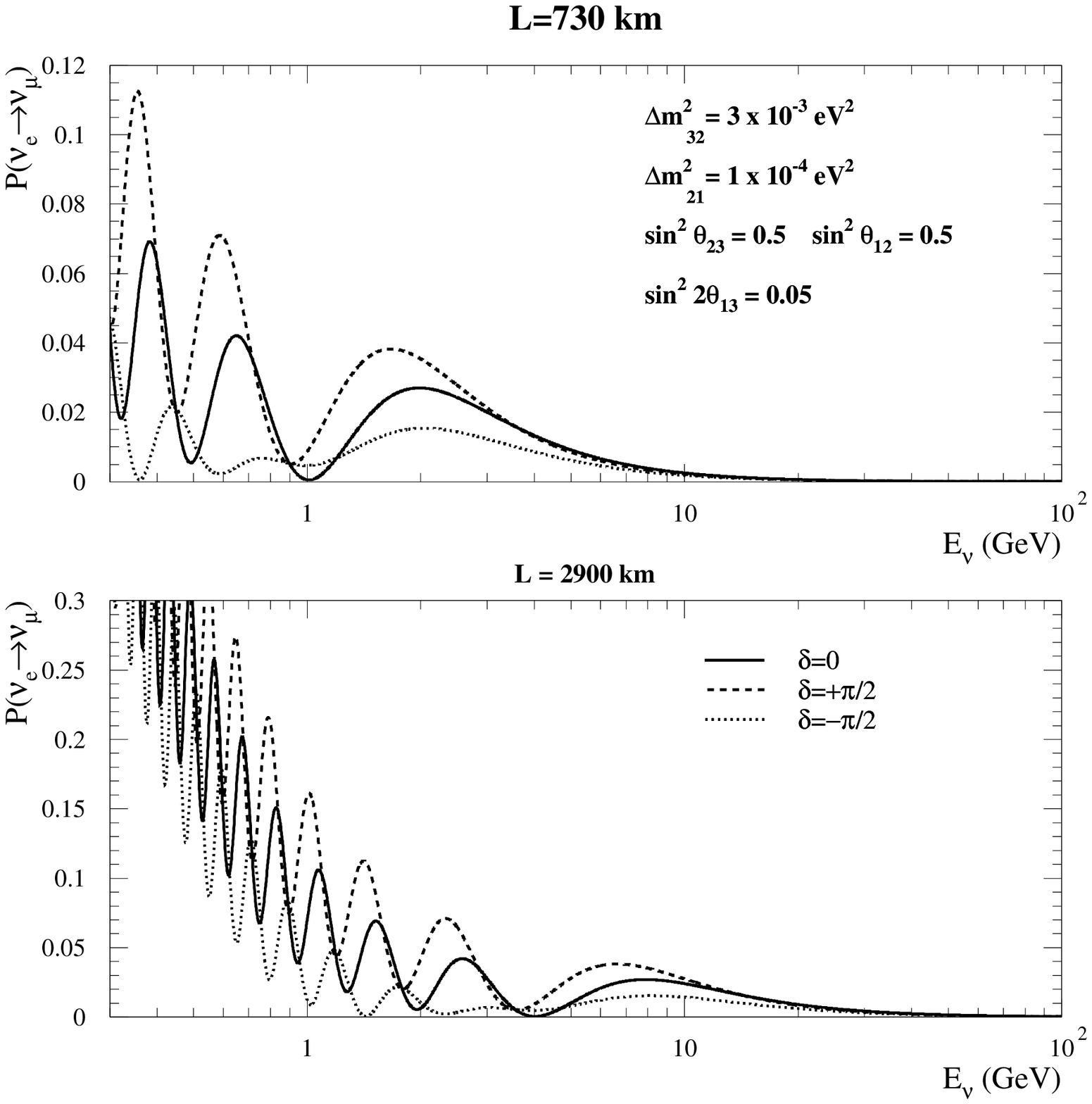,width=15cm}
\caption{Probability for $\nue\ra\numu$ oscillations in vacuum for two
baselines $L=730\ \rm km$ and 2900~km as a function of neutrino energy
$E_\nu$.
The probabilities are computed for three values of the $\delta$-phase:
$\delta=0$ (line), $\delta=+\pi/2$ (dashed), $\delta=-\pi/2$ (dotted).
The other oscillation parameters are
$\Delta m^2_{32}=3\times 10^{-3}\ \rm eV^2$,
$\Delta m^2_{21}=1\times 10^{-4}\ \rm eV^2$,
$\sin^2 \theta_{23} = 0.5$, $\sin^2 \theta_{12} = 0.5$,
and $\sin^2 2\theta_{13} = 0.05$.}
\label{fig:pnenmvac}
\end{figure}

The behavior for the two baselines $L=730\ \rm km$ and 2900~km
as a function of neutrino energy $E_\nu$
is explicitly shown in Figure~\ref{fig:pnenmvac}.
The probabilities are computed for three values of the $\delta$-phase:
$\delta=0$ (line), $\delta=+\pi/2$ (dashed), $\delta=-\pi/2$ (dotted).
The other oscillation parameters are those described in Eqs.~\ref{eq:atm}
and \ref{eq:sol}: $\Delta m^2_{32}=3\times 10^{-3}\ \rm
eV^2$, $\Delta m^2_{21}=1\times 10^{-4}\ \rm
eV^2$, $\sin^2 \theta_{23} = 0.5$, $\sin^2 \theta_{12} = 0.5$, 
and $\sin^2 2\theta_{13} = 0.05$.

A region corresponding to the oscillation of the ``first maximum'' is clearly
visible on the curves. We define the energy  of the ``first maximum'' as
follows
\begin{eqnarray}
\label{eq:oscmax}
\Delta_{32}=\frac{\pi}{2} & \longrightarrow &
E^{max}_\nu\equiv \frac{\Delta m^2_{32}}{2\pi}L \nonumber \\
& \longrightarrow & E^{max}_\nu(\rm GeV)\simeq\Delta m^2_{32}(\rm
eV^2)\left(\frac{2\times 1.27}{\pi}\right){L(\rm km)}
\end{eqnarray}
which yields $E^{max}_{\nu}\simeq 2\rm\ GeV$ at 730~km,
$E^{max}_{\nu}\simeq 8\rm\ GeV$ at 2900~km and
$E^{max}_{\nu}\simeq 20\rm\ GeV$ at 7400~km
for $\Delta m^2_{32}=3\times 10^{-3}\rm\ eV^2$. This energy corresponds to
the point of maximum oscillation induced by $\Delta m^2_{32}$ and coincides
with the maximum when $\delta=0$. It will be useful when we discuss the
point of maximum sensitivity to the $\delta$-phase.

A similar relation can also be defined in terms of $L/E_\nu$ which is
\begin{eqnarray}
\label{eq:leoscmax}
\left(\frac{L}{E_\nu}\right)^{max}\equiv \frac{2\pi}{\Delta m^2_{32}} 
 \longrightarrow 
\left(\frac{L(\rm km)}{E_\nu(\rm GeV)}\right)^{max}
\simeq\left(\frac{\pi}{2\times 1.27 \Delta m^2_{32}(\rm eV^2)}\right)
\end{eqnarray}
which is approximately equal $(L/E_\nu)^{max}\simeq 400\rm\ km/GeV$
for $\Delta m^2_{32}=3\times 10^{-3}\rm\ eV^2$.

The dependence at high energy is not easily understood from a plot
of the probabilities as a function of energy. In addition, while
the effect is expected to be most ``visible'' at low energies, 
experiments at low energy are a priori more
difficult to perform than at high energy, because of the rising cross-section, easiness
of detection and neutrino flux considerations. In fact, what is more
important is not the probability behavior, but rather how will the number
of events, observed in a given experiment at a given baseline $L$ within a certain neutrino energy
range, depend on the phase $\delta$, relative to the total number of
observed events in absence of $\delta$ phase effects. Hence, in the following, we
consider for definiteness the neutrino flux of a neutrino factory and will
in order to ease the comparison between various energies and baselines
introduce a ``rescaled'' probability.

\section{The neutrino factory (NF)}
\label{sec:nf}
The decay of the muon is a very well known process. Within the Standard
Model and to leading order, 
the neutrino energy distribution in the muon rest frame in $\mu^-$ decays
is given by
\begin{eqnarray}
\frac{d^2N_{\nu_\mu}}{dxd\Omega}\propto\frac{2x^2}{4\pi}[(3-2x)+(1-2x)P_\mu
\cos\theta]\\
\frac{d^2N_{\bar{\nu_e}}}{dxd\Omega}\propto\frac{12x^2}{4\pi}[(1-x)+(1-x)P_\mu
\cos\theta]
\end{eqnarray}
where $P_\mu$ is the average polarization of the muon beam, $x\equiv
2E_\nu/m_\mu$,
and $\theta$ is the angle between the momentum vector of the neutrino 
and the mean angle of the muon polarization. 
The beam polarization can be carefully measured via the spectrum of the electrons
from muon decay; thus the spectra of the two components of the neutrino
beam can be known with good accuracy, and so the ratio of
the fluxes reaching the far detector.

Assuming perfect focusing, an unpolarized muon beam and for very
relativistic muons, 
the energy spectrum of neutrinos detected at small angles with respect to
the muon 
flight direction as in the case of long baseline experiment, is given by
\begin{equation}
\phi_{\nu_\mu}(z)\propto 2z^2(3-2z);\ \ \ \ \phi_{\bar{\nu}_e}(z)\propto 12z^2(1-z)
\end{equation}
where $z={E_\nu}/{E_\mu}$. A key point is that angular opening of the neutrino
beam will shrink due to the Lorentz boost as $\gamma^{-2}=(E_\mu/m_\mu)^{-2}$
and hence the total neutrino fluency at a far detector will increase as
$E_\mu^2$. This means that the flux can be expressed as
\begin{equation}
\Phi_{\nu_\mu}(E_\mu,L,z)=\frac{E_\mu}{L^2}\phi_{\nu_\mu}(z); \ \ \ \ \Phi_{\bar{\nu}_e}(E_\mu,L,z)= \frac{E_\mu}{L^2}\phi_{\bar{\nu}_e}(z)
\end{equation}

The scaling of the neutrino event rate can be expressed as
\begin{eqnarray}
N_{\nu}&=&\int^{E_\mu}_0 \Phi_{\nu_\mu}\left(E_\mu,L,E_\nu\right)\sigma_\nu(E_\nu)dE_\nu\\
&=&\frac{E_\mu^2}{L^2}\int^{1}_0 \phi_{\nu}(z) \sigma_\nu(zE_\mu) dz
\end{eqnarray}
since $dE_\nu = E_\mu dz$. By approximating the neutrino cross-section as a
linear function of the neutrino energy, i.e. $\sigma_\nu(zE_\mu)\simeq
\sigma^0_\nu \times E_\nu$, where $\sigma^0_\nu$ is a constant, we obtain
\begin{eqnarray}
N_{\nu}&\simeq &\frac{E^3_\mu}{L^2}\sigma^0_\nu \int^{1}_0 \phi_{z} z dz\propto
\frac{E^3_\mu}{L^2}
\end{eqnarray}
The integral over $z$ is independent of $E_\mu$ and we find the known rapid
growth of neutrino events with $E_\mu^3$, characteristic of the neutrino
factory.

\section{The rescaled probabilities}
\label{rescaledprob}
In order to compare effects at different energies and various baselines, we
define a ``rescaled probability'' parameter that allows a direct comparison 
of effects.

Since to a good approximation (when $z\lesssim 0.6$ for
electron-neutrinos and 
always for muon-neutrinos) the neutrino 
energy distribution behaves like $E_\nu^2$ (see section~\ref{sec:nf}) and
in addition, the neutrino flux scales like $L^{-2}$ due to the beam divergence,
we define the ``rescaled probability'' parameter $p(\nu_\alpha\osc\nu_\beta;E_\nu,L)$ as
\begin{equation}
p(\nu_\alpha\osc\nu_\beta;E_\nu,L)\equiv P(\nu_\alpha\osc\nu_\beta;E_\nu,L)\times \frac{E^2_\nu}{L^2}
\end{equation}
We note that
\begin{enumerate}
\item it approximately correctly ``weighs'' the probability by the neutrino energy
spectrum $E_\nu^2$ of the neutrino factory spectrum;
\item it can be directly compared at different baselines, since it contains
the $L^{-2}$ attenuation of the neutrino flux with distance $L$;
\item $p$ tends to a constant for $E_\nu\rightarrow \infty$, hence
the high energy behavior can be easily studied.
\end{enumerate}

We start by illustrating the behavior of the rescaled probability
for $\nue\ra\numu$ oscillations (see Eq.~\ref{eq:probhighe}) in the case where
both $|\Delta_{12}|\ll 1$ and $|\Delta_{13}|,|\Delta_{23}|\ll 1$. We note
that the expression depends on the second power of the
$\Delta_{jk}$'s. The resulting $(L/E_\nu)^2$ dependence is canceled
in the rescaled probability:
\begin{eqnarray}
\label{eq:respemucanc}
 p(\nu_e\ra\nu_\mu) & \simeq & \frac{1}{2}c^2_{13} \Bigl\{ c^2_{13}\Delta^2_{12}+
 2s^2_{13}\left(\Delta^2_{13}+\Delta^2_{23} \right) 
 +2\Delta_{12}s_{13}
(\Delta_{13}+\Delta_{23})\cos\delta
\Bigr\}\times \frac{E^2_\nu}{L^2} \nonumber \\ 
& \simeq & 
\frac{1}{32}c^2_{13} \Bigl\{ c^2_{13}(\Delta m^2_{12})^2+ 
 2s^2_{13}\Bigl((\Delta m^2_{13})^2+
 (\Delta m^2_{23})^2 \Bigr) \nonumber \\
& &  +8\Delta m^2_{12}s_{13}
(\Delta m^2_{13}+\Delta m^2_{23})\cos\delta
\Bigr\}
\end{eqnarray}
As expected, the result is independent of the energy $E_\nu$ and 
the baseline $L$, and the rescaled probability ``constant'' is modified
with a $\cos\delta$ dependent term. The sign of this term depends as
expected on the sign of the $\Delta m^2_{jk}$'s.

\begin{figure}
\centering
\epsfig{file=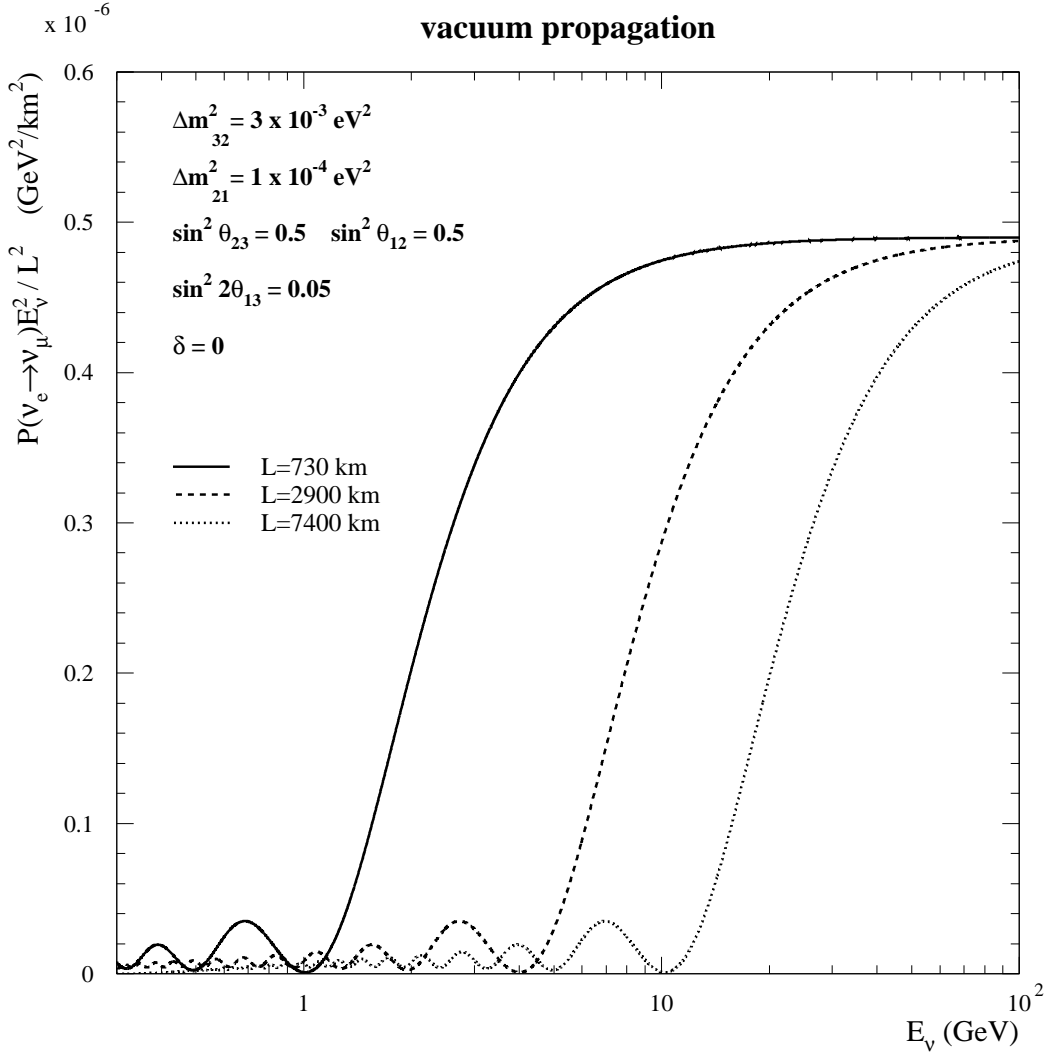,width=15cm}
\caption{Rescaled probability (see text) for $\nue\ra\numu$ oscillations in vacuum for three
baselines $L=730\ \rm km$ (line), 2900~km (dashed) and 7400~km (dotted) as a function of neutrino energy
$E_\nu$.
The oscillation parameters are $\Delta m^2_{32}=3\times 10^{-3}\ \rm
eV^2$, $\Delta m^2_{21}=1\times 10^{-4}\ \rm
eV^2$, $\sin^2 \theta_{23} = 0.5$, $\sin^2 \theta_{12} = 0.5$, 
$\sin^2 2\theta_{13} = 0.05$ and $\delta=0$.}
\label{fig:respnenmvac}
\end{figure}

Figure~\ref{fig:respnenmvac} shows the rescaled probability for $\nue\ra\numu$ oscillations in vacuum for three
baselines $L=730\ \rm km$ (line), 2900~km (dashed) and 7400~km (dotted) as a function of neutrino energy
$E_\nu$. In this case, we chose $\delta=0$ in order to illustrate the
effect of the rescaling. The interesting oscillations at low energy driven
by the interference of the $\Delta_{12}$, $\Delta_{13}$ and $\Delta_{23}$
terms are largely
``damped'' by the rescaling, and the situation has dramatically changed
compared to Figure~\ref{fig:pnenmvac}. 

We also note that the asymptotic
value reached at high energy is independent of the distance $L$, reflecting the
well-known result that the oscillated number of events is independent of
distance $L$ at high energy, since the $L^2$ growing of the oscillation cancels the
$L^{-2}$ attenuation of the flux with distance.

\begin{figure}
\centering
\epsfig{file=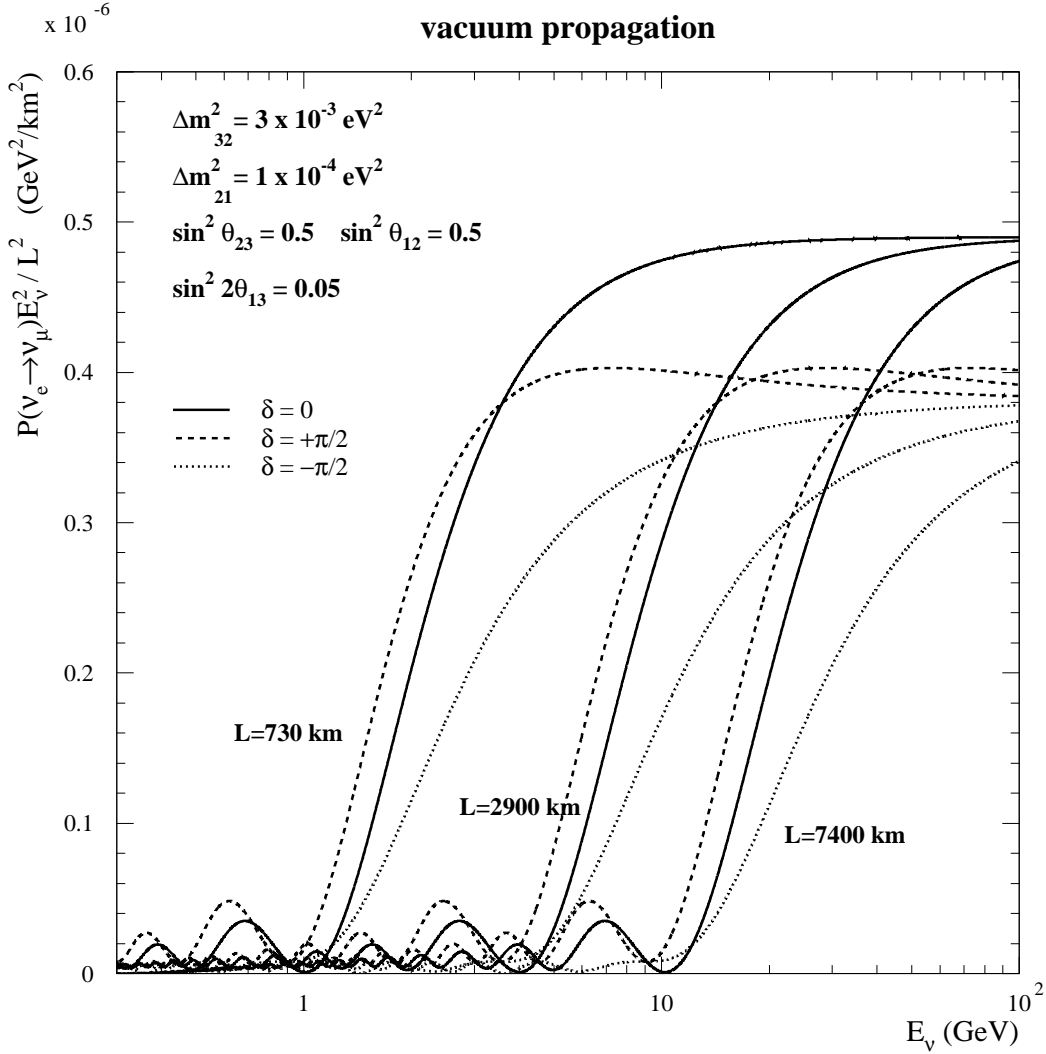,width=15cm}
\caption{Rescaled probability (see text) for $\nue\ra\numu$ oscillations in vacuum for three
baselines $L=730\ \rm km$ (line), 2900~km (dashed) and 7400~km (dotted) as a function of neutrino energy
$E_\nu$.
For each baseline, the probabilities are computed for three values of the $\delta$-phase:
$\delta=0$ (upper), $\delta=+\pi/2$ (middle), $\delta=-\pi/2$ (lower).
The oscillation parameters are $\Delta m^2_{32}=3\times 10^{-3}\ \rm
eV^2$, $\Delta m^2_{21}=1\times 10^{-4}\ \rm
eV^2$, $\sin^2 \theta_{23} = 0.5$, $\sin^2 \theta_{12} = 0.5$, 
$\sin^2 2\theta_{13} = 0.05$.}
\label{fig:respnenmvacd}
\end{figure}

The dependence on the phase $\delta$ is illustrated in
Figure~\ref{fig:respnenmvacd}. For $\delta=\pm\pi/2$, the $\cos\delta$ term
vanishes and the probability is reduced at high energy compared to the
$\delta=0$ case. The asymptotic value reached in the high energy limit is
obviously independent of the baseline and is
different than that reached for $\delta=0$.
For $\delta=\pm\pi/2$, it converges to the same values when the
condition for $|\Delta_{12}|\ll 1$ and $|\Delta_{13}|,|\Delta_{23}|\ll 1$
is true. This depends on the baseline $L$ considered. For $L=730(2900)\rm\ km$,
the probabilities for $\delta=\pm\pi/2$ are indistinguishable to within 10\%
for $E_\nu\gtrsim 20(80)\rm\ GeV$. 

\section{Propagation in matter}
Since neutrino factories will be associated to long baseline, it is not
possible to avoid 
including effects associated to the neutrino propagation through the Earth matter.
The simplest way to take into account these effects is to maintain the
formalism developed for propagation in vacuum and to replace the mixing angles and the neutrino
mass differences by ``effective'' values. 

The general exact and precise treatment of
the effects induced by the propagation through the Earth can be quite
complicated, due to the a priori complexity of the matter profile in the
Earth. We will hence assume that the effects due to matter can be
parameterized by one parameter, i.e. the ``average'' density $\rho$, to reproduce
approximately the effect of traversing the Earth. This assumption was
recently revisited in Ref.~\cite{Ota:2000hf}, where it was concluded that for
a baseline of 3000~km or less, the effect of the matter profile is not important.
For very long baselines like $L\approx 7400\ \rm km$, they conclude that
the Earth profile is important. This however does not pose us a problem,
since, as we will show below, we will not consider that baseline in the
context of the searches for $\delta$-phase-induced effects. 
We hence all along use the approximation of constant density.

In the case of two-neutrino oscillation, in which the mixing is driven by a
single mixing angle $\theta$, and propagating in matter of 
constant density $\rho$, it is rather easy to understand
the behavior of the oscillation probability for neutrinos or
antineutrinos. 
An important quantity for
matter effects is $D$, defined as
\begin{equation}
D(E_\nu,\rho)\equiv 2\sqrt{2}G_F n_e E_\nu=7.56\times 10^{-5}eV^2(\frac{\rho}{g cm^{-3}})
(\frac{E_\nu}{GeV}) \equiv -D(E_{\bar{\nu}},\rho)
\end{equation}
where $n_e$ is the electron density and $\rho$ the matter density. For 
antineutrinos, $D$ is replaced by $-D$. 

In the case of three neutrino mixing, the situation is mathematically more
complex. Approximate oscillation probabilities have been considered by
perturbative expansion in the oscillation 
parameters\cite{Koike:2000hf,Arafune:1997hd,Cervera:2000kp}.
Exact formulas have been derived\cite{zaglauer,pakvasa}
and we make use of this formalism as
implemented in Ref.~\cite{Bueno:2000fg}.
For completeness, we reproduce here the expressions. 
The mass eigenvalues in matter $M_1$, $M_2$ and $M_3$ are: 
\begin{eqnarray}
M_1^2 & = & m_1^2 +\frac{A}{3} - \frac{1}{3}\sqrt{A^2-3B}S -
\frac{\sqrt{3}}{3}\sqrt{A^2-3B}\sqrt{1-S^2}\\
M_2^2 & = & m_1^2 +\frac{A}{3} - \frac{1}{3}\sqrt{A^2-3B}S +
\frac{\sqrt{3}}{3}\sqrt{A^2-3B}\sqrt{1-S^2}\\
M_3^2 & = & m_1^2 +\frac{A}{3} + \frac{2}{3}\sqrt{A^2-3B}S
\end{eqnarray}
where $A$, $B$ and $S$ are given in the Appendix of Ref.~\cite{Bueno:2000fg}.
For the mixing angles in matter the analytical expressions read: 
\begin{eqnarray}
\sin^2\theta^m_{12} & = & \frac{-(M_2^4-\alpha M_2^2+\beta)\Delta M^2_{31}}
{\Delta M^2_{32}(M_1^4 -\alpha M_1^2+\beta)-\Delta
M^2_{31}(M^4_2-\alpha M_2^2 + \beta)} \\
\sin^2\theta^m_{13} & = & \frac{M_3^4-\alpha M_3^2+\beta}{\Delta
M^2_{31}\Delta M^2_{32}} \\
\sin^2\theta^m_{23} & = &
\frac{G^2s^2_{23}+F^2c^2_{23}+2GFc_{23}s_{23}c_\delta}{G^2+F^2} 
\end{eqnarray}
where $\alpha$, $\beta$, $G$ and $F$ are found in the Appendix
of Ref.~\cite{Bueno:2000fg}, and also an expression for the $\delta$-phase
in matter, i.e. $\delta^m$.

It can be shown that in the relevant situation in which $|\Delta
m^2_{21}|\ll |\Delta m^2_{31}|\approx |\Delta m^2_{32}|$, matter effects
in three-family mixing decouple to two independent
two-family mixing scenarios. This is essentially because matter effects
become important when the parameter $D(E_\nu,\rho)$ is similar to one of the $|\Delta
m^2|$'s and hence the matter effects driven by $|\Delta
m^2_{21}|$ will occur for fixed density at a very different energy $E_\nu$
than for $|\Delta m^2_{31}|\approx |\Delta m^2_{32}|$. The term
$|\Delta m^2_{21}|$ (resp. $|\Delta m^2_{31}|$) will produce a MSW resonance in 
the $\theta_{12}$(resp. $\theta_{13}$) angle.

Given the energies and baselines considered, we illustrate analytically our arguments
for a two-family mixing matter effect in $\nue\ra\numu$ oscillations, in
which we identify the 2-mixing
angle $\theta$ with the 3-mixing $\theta_{13}$ and the mass difference
squared as $|\Delta m^2_{31}|\approx |\Delta m^2_{32}|$. 
In the case $|\Delta_{12}|\ll|\Delta_{13}|\approx|\Delta_{23}|\approx 1$
in matter, the effective mixing angles $\theta_{23}^m$, $\theta_{12}^m$,
$\theta_{13}^m$ and $\delta^m$ will be
approximately equal to (see e.g. Figure~1 of Ref.~\cite{Bueno:2000fg})
\footnote{In reality, the angles $\sin^2\theta^m_{12}$ and
$\sin^2\theta^m_{23}$ tend to rise slightly for $D>\Delta m^2_{32}$,
because the non-vanishing $\Delta m^2_{21}$ splitting removes the
degeneracy between muon and tau flavors (see Ref.~\cite{Bueno:2000fg}), but
we neglect this effect.}
\begin{eqnarray}
\label{eq:effang}
\sin^2\theta^m_{23}(D) & \simeq & \sin^2\theta_{23} \nonumber \\
\sin^22\theta^m_{12}(D) & \rightarrow & \approx 0 \nonumber \\
\sin^2 2\theta^m_{13}(D)&=&\frac{\sin^2 2\theta_{13}}{\sin^2 2\theta_{13}+ (
\frac{D}{\Delta m^2_{32}}-\cos 2\theta_{13})^2} \nonumber \\
&\simeq&\frac{\sin^2 2\theta_{13}}{\sin^2 2\theta_{13}+ {\cal M}^2}
\nonumber\\
\delta^m\approx \delta
\end{eqnarray}
where ${\cal M} \equiv D/\Delta m^2_{32}-1$ and,
for the second line, we assume $\theta_{13}$ small. This term has the
well-known MSW resonance behavior, which predicts that the effective
angle $\sin^2 2\theta^m_{13}(D)\ra 1$ for $D\ra \Delta m^2_{32}$
(i.e. ${\cal M}\ra 0$), no matter
how small $\sin^2 2\theta_{13}$ is.
The effective mass
squared difference becomes
\begin{eqnarray}
\Delta M^2_{32}\simeq\Delta m^2_{32}\sqrt{\sin^2 2\theta_{13}+ 
{\cal M}^2}
\end{eqnarray}
which implies that the effective oscillation wavelength becomes very large
(resp. small)
close to (resp. far from)  the resonance energy.

The oscillation probability at small $L/E_\nu$ (see
Eq.~\ref{eq:wellknownonemass}) is in matter
\begin{eqnarray}
\label{eq:matterprob}
P^m(\nu_e \to \nu_\mu) & \simeq & \sin^2(2\theta^m_{13})s^2_{23} 
\sin^2\left(\Delta^m_{23}\right) \nonumber\\
&   \simeq & \frac{\sin^2 2\theta_{13}}{\sin^2 2\theta_{13}+ 
{\cal M}^2}s^2_{23} 
\sin^2\left(\Delta m^2_{32}\sqrt{\sin^2 2\theta_{13}+ 
{\cal M}^2}\frac{L}{4E_\nu}\right) 
\end{eqnarray}

In the case where the neutrino energy is
at the MSW resonance (i.e. ${\cal M}=0$ and 
$E_\nu={\cal E}^{res}\cos2\theta_{13}\Delta m^2_{32}$), the probability is
approximately
\begin{eqnarray}
\label{eq:pmatatres}
P^m(\nu_e \to \nu_\mu) & \simeq & 
s^2_{23} 
\sin^2\left(\Delta m^2_{32}\sqrt{\sin^2 2\theta_{13}}
\frac{L}{4{\cal E}^{res}\cos2\theta_{13}\Delta m^2_{32}}
\right) \nonumber \\
& \simeq & 
s^2_{23} \sin^2\left(
\frac{\tan 2\theta_{13}}{4{\cal E}^{res}}L\right) \nonumber \\
&  \simeq & 
s^2_{23} \left(
\frac{\tan 2\theta_{13}}{4{\cal E}^{res}}\right)^2 L^2
\end{eqnarray}
where in the last line we assumed $L<\tan2\theta_{13}/4{\cal
E}^{res}\approx
14000\rm\ km$. In this case,
the probability grows with $L^2$, giving rise to the enhanced
oscillation probability at large distances through matter. 

For neutrino energies $E_\nu$ above the resonance energy $E^{res}_\nu$,
where one can approximate ${\cal M}^2>1\gg \sin^2 2\theta_{13}$, the
probability is
\begin{eqnarray}
\label{eq:pmataboveres}
P^m(\nu_e \to \nu_\mu) & \simeq & 
\frac{\sin^2 2\theta_{13}}{{\cal {M}}^2}s^2_{23} 
\sin^2\left(\Delta m^2_{32}
\sqrt{{\cal M}^2}\frac{L}{4E_\nu}\right)
\end{eqnarray}
and this expression will vanish for growing ${\cal M}$ since the sine
function will not cancel the ${\cal M}^2$ at the denominator. These behaviors
will be rediscussed later in the context of the $\delta$-phase dependent terms.

To summarize, there will be two specific
neutrino energies of interest when neutrinos propagate through matter:
\begin{enumerate}
\item for $D\approx \Delta m^2_{32}$, we reach for neutrinos the MSW
resonance, in which the effective mixing angle
$\sin^2(2\theta^m_{13})\approx 1$. In terms of neutrino energy, this
implies
\begin{eqnarray}
E^{res}_\nu& = &\frac{\cos2\theta_{13}\Delta m^2_{32}}{2\sqrt{2}G_F
n_e} = {\cal E}^{res}\cos2\theta_{13}\Delta m^2_{32} \nonumber \\
& \simeq &\frac{1.32\times 10^4\cos2\theta_{13}\Delta
m^2_{32}(\rm eV^2)}{\rho(g/cm^3)}\ \rm in\ GeV
\end{eqnarray}
where ${\cal E}^{res}=(2\sqrt{2}G_Fn_e)^{-1}$. For density parameters
$\rho$
equal to $2.7$, $3.2$ and $3.7\rm\ g/cm^3$ one finds
$E^{res}_\nu \simeq 14.1$, $12.3$ and $10.7\rm\ GeV$ for $\Delta
m^2_{32}=3\times 10^{-3}\rm\ eV^2$.
\item for $D> 2 \Delta m^2_{32}$, the effective mixing angle for neutrinos
is always smaller than that in vacuum, i.e.
$\sin^2(2\theta^m_{13})< \sin^2(2\theta_{13})$.
In terms of neutrino energy, this
is equivalent to
\begin{equation}
E_\nu > 2E^{res}_\nu
\end{equation}
\item these arguments are independent of the baseline $L$ and depend only
on the matter density $\rho$.
\end{enumerate}

In the rest of the paper, we will solve the equations numerically in order
to properly introduce exact matters effects into our oscillation
probabilities. 

\begin{figure}
\centering
\epsfig{file=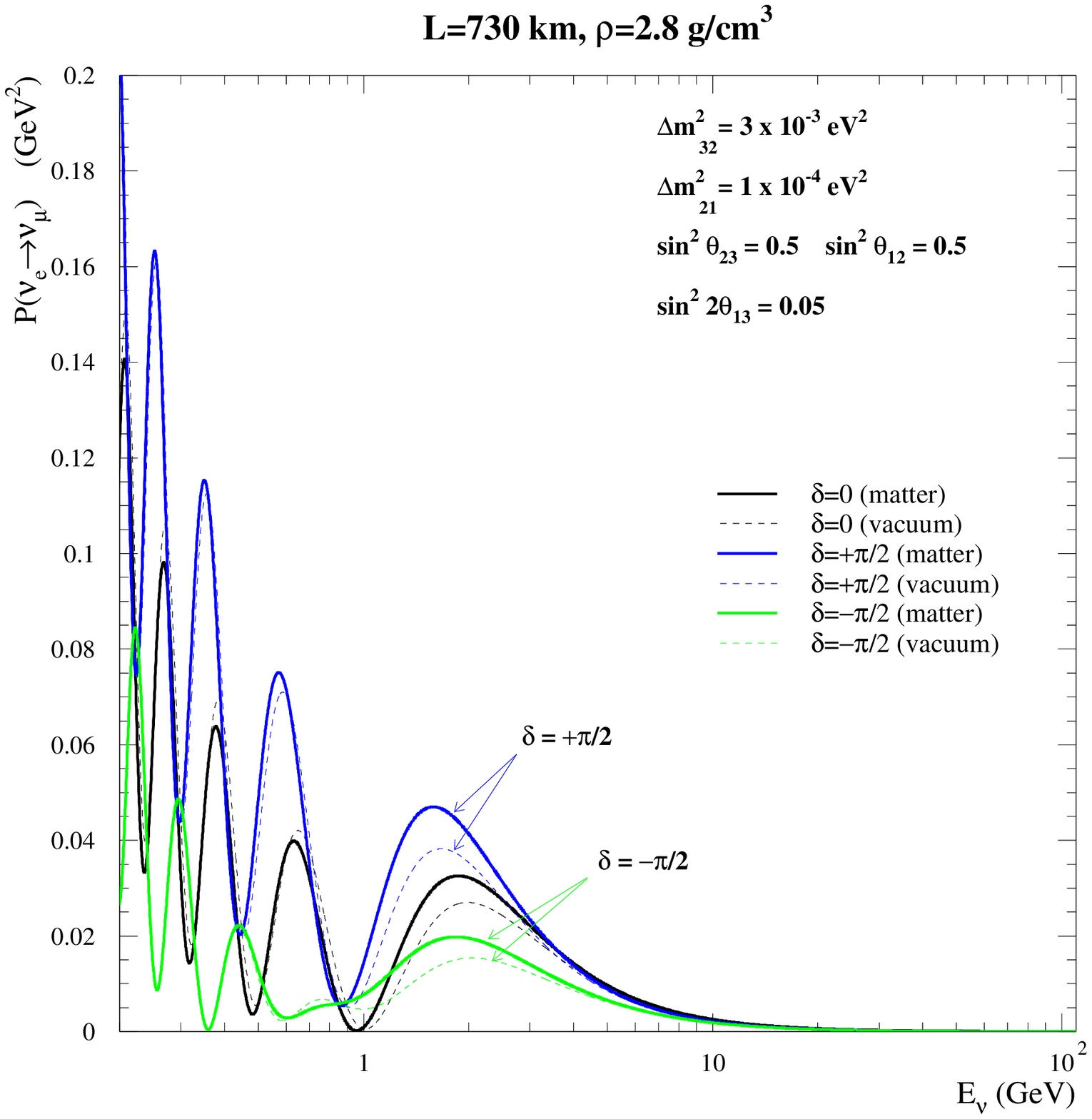,width=15cm}
\caption{Oscillation probability for $\nue\ra\numu$ oscillations for a 
baseline $L=730\ \rm km$ as a function of neutrino energy
$E_\nu$.
The probabilities are computed for neutrinos in matter (full line) and 
in vacuum (dotted line) and
for three values of the $\delta$-phase: $\delta=0$, $\delta=+\pi/2$ and
$\delta=-\pi/2$.
The oscillation parameters are $\Delta m^2_{32}=3\times 10^{-3}\ \rm
eV^2$, $\Delta m^2_{21}=1\times 10^{-4}\ \rm
eV^2$, $\sin^2 \theta_{23} = 0.5$, $\sin^2 \theta_{12} = 0.5$, 
$\sin^2 2\theta_{13} = 0.05$.}
\label{fig:probmatcp1}
\end{figure}

\begin{figure}
\centering
\epsfig{file=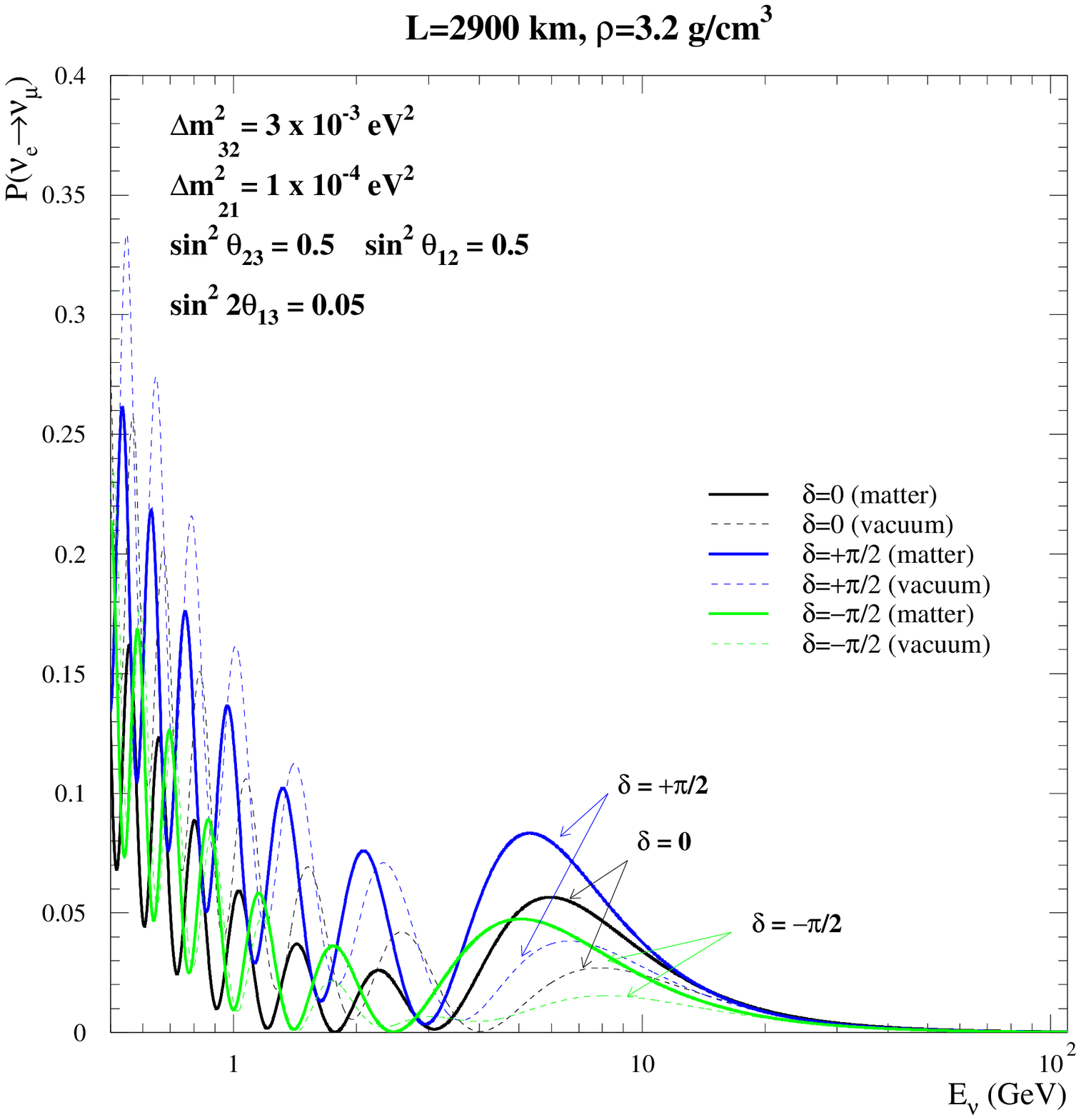,width=15cm}
\caption{Same as Figure~\ref{fig:probmatcp1} but for a baseline $L=2900\rm\ km$.}
\label{fig:probmatcp2}
\end{figure}

The distortion of the oscillation probabilities introduced
by the propagation through matter are shown in
Figures~\ref{fig:probmatcp1} and \ref{fig:probmatcp2}, where
the probability for $\nue\ra\numu$ oscillations for two
baselines $L=730\ \rm km$ and 2900~km are shown as 
a function of neutrino energy
$E_\nu$.
For each baseline, the probabilities are computed for three values of the $\delta$-phase:
$\delta=0$, $\delta=+\pi/2$, $\delta=-\pi/2$.

\begin{figure}
\centering
\epsfig{file=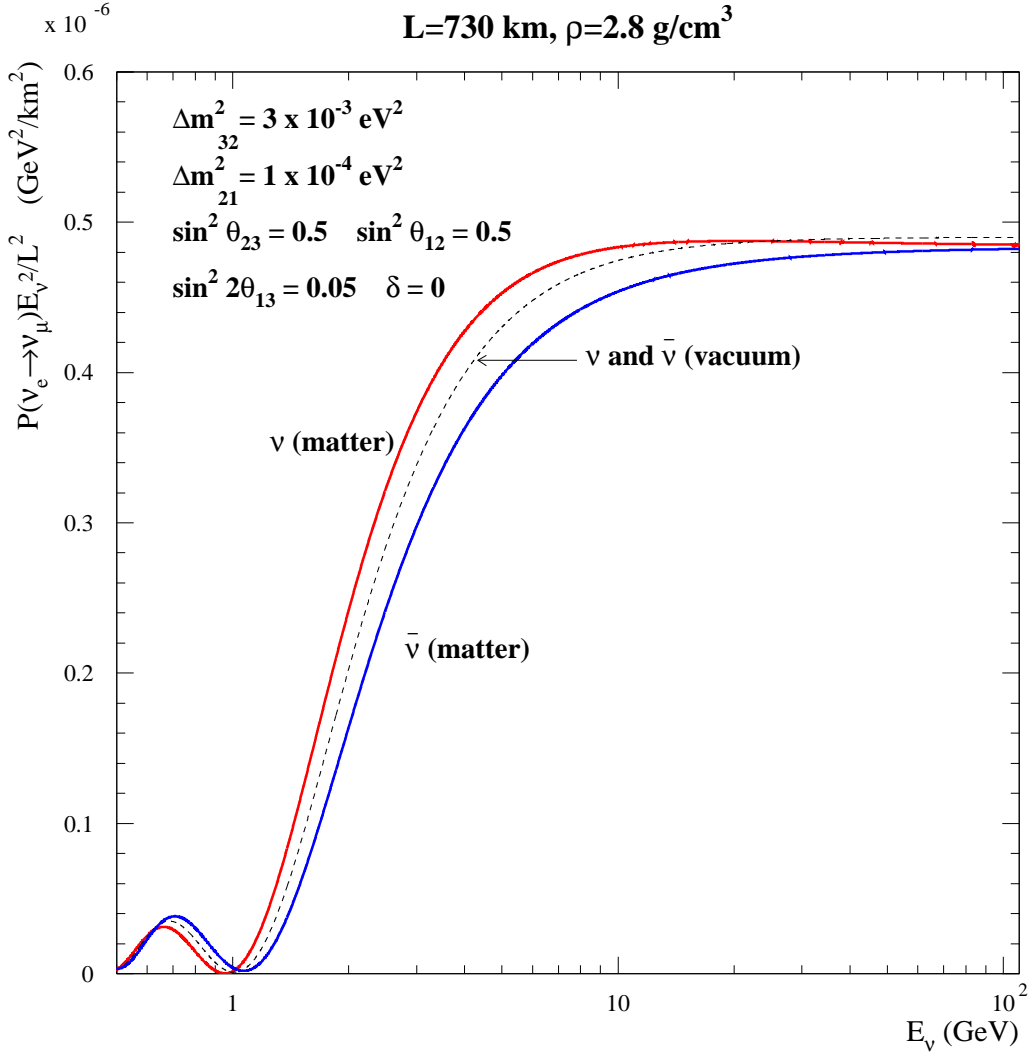,width=15cm}
\caption{Rescaled probability (see text) for $\nue\ra\numu$ oscillations for a
baseline $L=730\ \rm km$ as a function of neutrino energy
$E_\nu$.
The probabilities are computed for neutrinos in matter,
in vacuum (dotted line, same for neutrinos and antineutrinos) and for antineutrinos in matter.
The oscillation parameters are $\Delta m^2_{32}=3\times 10^{-3}\ \rm
eV^2$, $\Delta m^2_{21}=1\times 10^{-4}\ \rm
eV^2$, $\sin^2 \theta_{23} = 0.5$, $\sin^2 \theta_{12} = 0.5$, 
$\sin^2 2\theta_{13} = 0.05$ and $\delta=0$.}
\label{fig:resprobmat}
\end{figure}

\begin{figure}
\centering
\epsfig{file=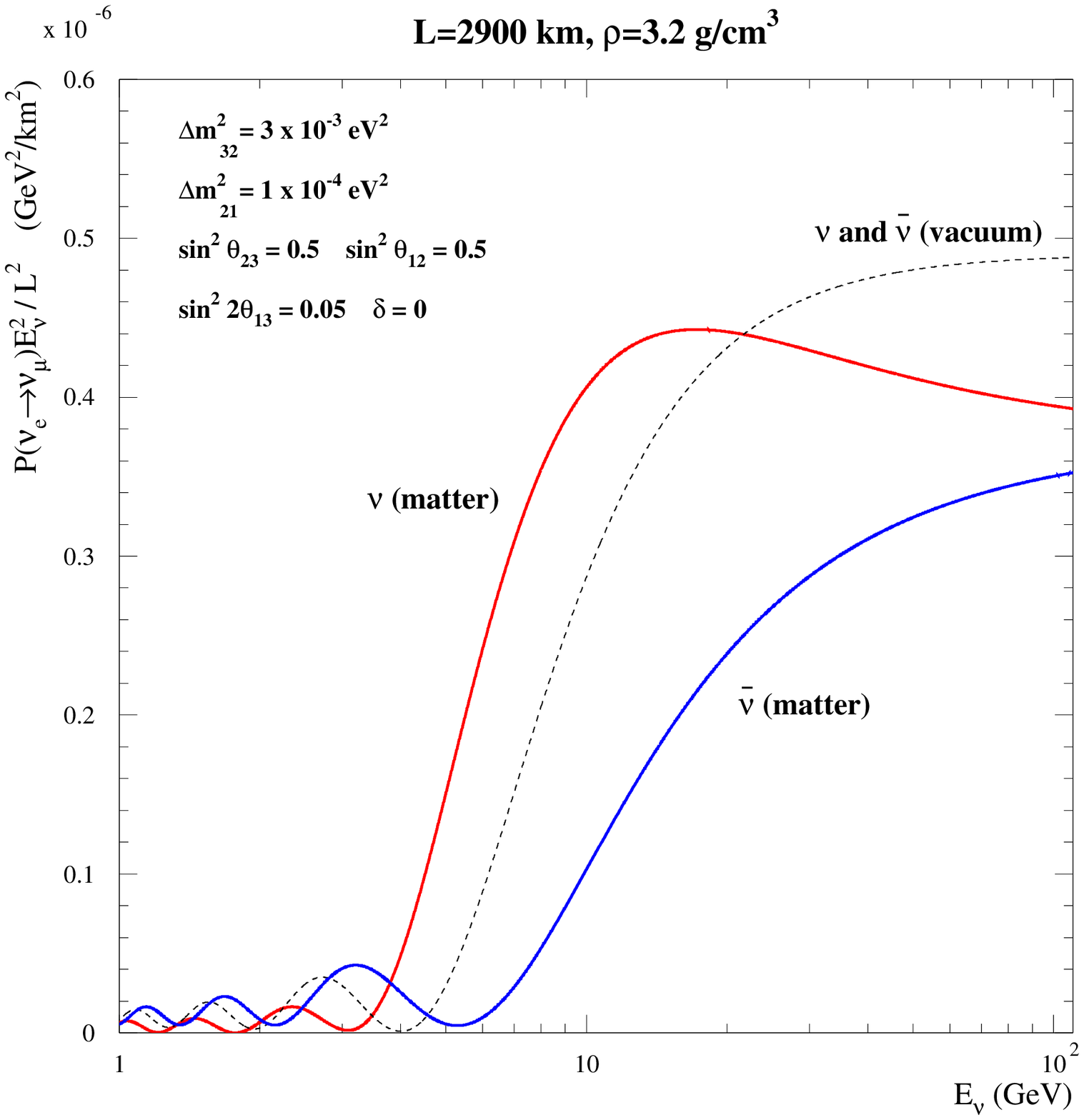,width=15cm}
\caption{Same as Figure~\ref{fig:resprobmat} but for a baseline $L=2900\rm\ km$.}
\label{fig:resprobmat2}
\end{figure}

\begin{figure}
\centering
\epsfig{file=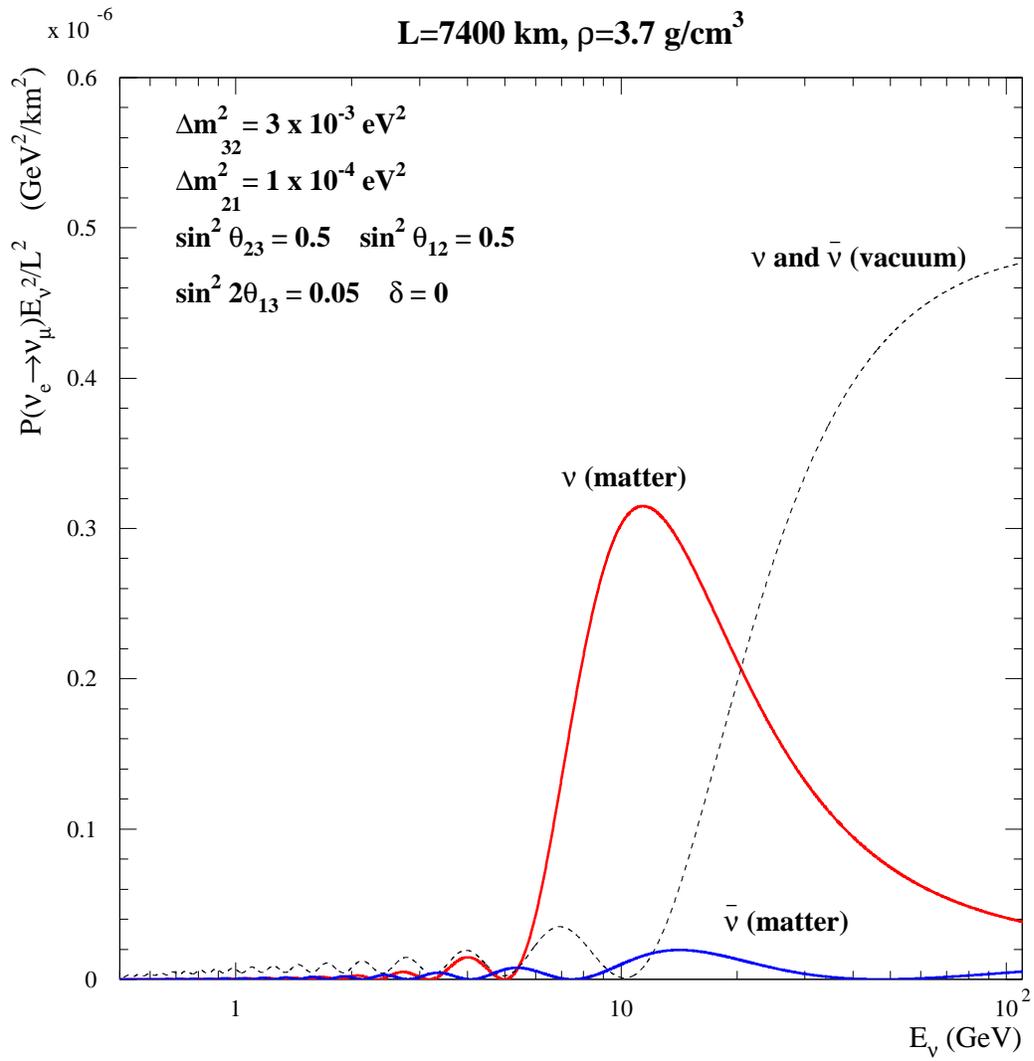,width=15cm}
\caption{Same as Figure~\ref{fig:resprobmat} but for a baseline $L=7400\rm\ km$.}
\label{fig:resprobmat3}
\end{figure}

We further illustrate the behavior in matter in
Figures~\ref{fig:resprobmat}, \ref{fig:resprobmat2} and \ref{fig:resprobmat3}, where
the rescaled probability for $\nue\ra\numu$ oscillations for three
baselines $L=730\ \rm km$, 2900~km and 7400~km are shown as 
a function of neutrino energy
$E_\nu$.
For each baseline, the probabilities are computed for neutrinos in matter,
in vacuum (dotted line, same for neutrinos and antineutrinos) and for antineutrinos in matter.
A phase $\delta=0$ has been assumed.

The characteristic behavior in matter can be readily seen. At the shortest
baseline, $L=730\ \rm km$, the effect of matter is small, i.e. the
probabilities are very similar to that in vacuum and the probability is
simply slightly enhanced (suppressed) for neutrinos (antineutrinos).
Between approximately $1<E_\nu<10\rm\ GeV$, the rescaled probability rises; this
corresponds to the oscillation at the ``first maximum'' (see equation~\ref{eq:oscmax}),
which yields $E^{max}_{\nu}\simeq 2\rm\ GeV$ for $3\times 10^{-3}\rm\ eV^2$.
Above $E_\nu\simeq 10\ \rm GeV$, the rescaled probability reaches a
constant. 

At the middle baseline $L=2900\ \rm km$, the effect of propagation through matter is quite
visible. The resonance for neutrinos can be seen at 
$E^{res}_\nu(GeV) \simeq 12\ \rm GeV$, which is quite close from the
``first maximum'', which is now given the baseline 
at $E^{max}_\nu\simeq 8\ \rm GeV$ (see equation~\ref{eq:oscmax}).
As expected, the probability in vacuum reaches at higher energies the same constant as for
the shortest baseline $L=730\ \rm km$. However, in matter
for $E_\nu \gtrsim 2E^{res}$, the probability becomes smaller than that in
vacuum. For antineutrinos, the probability tends to be suppressed by
matter, but tends to the same value as that of neutrinos at high energy.

For the longest baseline $L=7400\ \rm km$, the effect of matter is quite
strong. The ``first maximum'' is now at 
$E^{max}_\nu\simeq 20\ \rm GeV$ (see equation~\ref{eq:oscmax}), but this
corresponds to an energy very close to the condition 
for $D> 2 \Delta m^2_{32}$, in which the effective mixing angle for neutrinos
is smaller than that in vacuum. The probability at such large distances is
therefore quite ``suppressed'' by matter as opposed to what it would be in vacuum.

From these discussions, it is already clear that if one wants to study
oscillations in the region of the ``first maximum'', one should not choose
a too large baseline $L$, otherwise, matter effects will suppress the
oscillation probability. This will be further discussed in the next section.

\section{Detecting the $\delta$ phase at the NF and the $L/E_\nu$ scaling}
\label{discriminators}
We are now ready to investigate methods to optimally look for effects
related to the phase $\delta$ in neutrino oscillations. We consider that
propagation of neutrinos will always occur through matter and hence use the
exact numerical calculations for three family mixing, without any
approximation in the oscillation probabilities, nor in the treatment of
matter effects.

In order to further 
study the dependence of the $\delta$-phase, we consider the
following three quantities which are good discriminators for a
non-vanishing 
phase $\delta$:
\begin{enumerate}
\item $\Delta_\delta \equiv  P(\nu_e\ra\nu_\mu,\delta=+\pi/2)-
P(\nu_e\ra\nu_\mu,\delta=0)$\\
The discriminant $\Delta_\delta$ can be used in an experiment where
one is comparing the measured $\nue\ra\numu$ oscillation probability as a
function of the neutrino energy $E_\nu$ compared to a ``Monte-Carlo
prediction'' of the spectrum in absence of $\delta$-phase.
\item $\Delta_{CP}(\delta) \equiv  P(\nu_e\ra\nu_\mu,\delta)-P(\bar\nu_e\ra\bar\nu_\mu,\delta)$ \\
The discriminant $\Delta_{CP}$ can be used in an experiment by
comparing the appearance of $\nu_\mu$ (resp. $\bar\nu_\mu$) in a beam of
stored $\mu^+$ (resp. $\mu^-$) decays as a function of the
neutrino energy $E_\nu$.
\item $\Delta_{T}(\delta) \equiv
P(\nu_e\ra\nu_\mu,\delta)-P(\nu_\mu\ra\nu_e,\delta)$ or
$\bar\Delta_{T}(\delta) \equiv  P(\bar\nu_e\ra\bar\nu_\mu,\delta)-
P(\bar\nu_\mu\ra\bar\nu_e,\delta)$ \\
The discriminant $\Delta_{T}$ can be used in an experiment by
comparing the appearance of $\nu_\mu$ (resp. $\bar\nu_\mu$) {\bf and}
$\bar\nu_e$ (resp. $\nu_e$) and in a beam of
stored $\mu^+$ (resp. $\mu^-$) decays as a function of the
neutrino energy $E_\nu$.
\end{enumerate}
Each of these discriminants have their advantages and disadvantages.

The $\Delta \delta$-method consists in searching for distortions in the
visible energy spectrum of events produced by the $\delta$-phase. While
this method can in principle provide excellent determination of the phase limited
only by the statistics of accumulated events, in practice, systematic
effects will have to be carefully kept under control in order to look for a
small effect in a seen-data versus Monte-Carlo-expected comparison.
In addition, the precise knowledge of the other oscillation parameters will
be important, and as will be discussed below, there is a risk of degeneracy
between solutions and a possible strong correlation with the $\theta_{13}$
angle at high energy.

The $\Delta_{CP}$ is quite straight-forward, since it involves comparing
the appearance of so-called wrong-sign muons for two polarities of the
stored muon beam. 
It really takes advantage from the fact that
experimentally energetic muons are rather easy to detect and identify due
to their penetrating nature, and with the help of a magnetic field, their
charge can be easily measured, in order to suppress the non-oscillated
background from the beam. 
A non-vanishing $\Delta_{CP}$ should in principle be a direct proof
for a non-vanishing $\delta$-phase.
This method suffers, however, from the inability
to perform long-baseline experiment through vacuum. Indeed, matter effects
will largely ``spoil'' $\Delta_{CP}$ since it involves both neutrinos and
antineutrinos, which will oscillate very differently
through matter. Hence, the $\Delta_{CP}$ requires a good understanding of
the effects related to matter. In addition, it involves measuring neutrinos
and antineutrinos. The matter suppression of the antineutrinos will in
practice determine the statistical accuracy with which the discriminant can
be measured.

Finally, the $\Delta_{T}$ is the theoretically cleanest method, since it
does not suffer from the problems of $\Delta \delta$ and
$\Delta_{CP}$. Indeed, a difference in oscillation probabilities
between $\nue\ra\numu$ and
$\numu\ra\nue$ would be a direct proof
for a non-vanishing $\delta$-phase. In addition, matter affects both
probabilities in a same way, since it involves only neutrinos.
Unfortunately, it is experimentally very challenging to discriminate the
electron charge produced in the events, needed in order to suppress the
background from the beam. However, one can decide to measure only
neutrinos, which are enhanced by matter effects, as opposed to
antineutrinos in the $\Delta_{CP}$ which were matter suppressed, and hence
the statistical accuracy of the measurement will be determined by the
efficiency to recognize the electron charge, rather than by matter suppression.

\subsection{The $L/E_\nu$ scaling}
\label{le_scaling}
Regardless of their advantages and disadvantages, there is one thing in
common between the three discriminants $\Delta_\delta$, 
$\Delta_{CP}$ and $\Delta_{T}$: their behavior with respect to the
neutrino energy $E_\nu$ and the baseline $L$.
By explicit calculation, we find
\begin{eqnarray}
\label{eq:deldeldef}
\Delta_\delta&=&-\frac{1}{2}c^2_{13}\sin 2\theta_{12}s_{13}\sin
2\theta_{23}\times
\\ \nonumber
& & \Bigl[
\cos 2\Delta_{13}-\cos2\Delta_{23}-2\cos2\theta_{12}\sin^2\Delta_{12}
+\sin 2\Delta_{12}-\sin 2\Delta_{13}+\sin 2\Delta_{23}\Bigr] \\ \nonumber
&=& -\frac{1}{2}c^2_{13}s_{13}
\Bigl[
\cos 2\Delta_{13}-\cos2\Delta_{23}
+\sin 2\Delta_{12}-\sin 2\Delta_{13}+\sin 2\Delta_{23}\Bigr] 
\end{eqnarray}
where for the second line
we assumed for simplicity $\theta_{12}=\theta_{23}=\pi/4$, and similarly,
\begin{eqnarray}
\label{eq:delcptdef}
\Delta_{CP}=\Delta_{T} & = &
c^2_{13}s_{13}\sin2\theta_{12}\sin 2\theta_{23}\sin\delta
\Bigl[ \sin 2\Delta_{12}- \sin 2\Delta_{13} + \sin 2\Delta_{23}\Bigr]
\nonumber \\
& = &
-c^2_{13}s_{13}\sin2\theta_{12}\sin 2\theta_{23}\sin\delta
\Bigl[ \sin \Delta_{12}\sin \Delta_{13}\sin\Delta_{23}\Bigr]
\end{eqnarray}
As expected, both expressions vanish in the limit 
$\Delta m^2_{12}\ra 0$ where $\Delta m^2_{13}\ra \Delta m^2_{23}$.
Also, as one reaches the higher energies,
the terms $\Delta_{CP}=\Delta_{T}$ vanish as 
\begin{eqnarray}
\label{eq:delcptdefvac}
|\Delta_{CP}|=|\Delta_{T}| 
& \simeq & c^2_{13}s_{13}\sin2\theta_{12}\sin 2\theta_{23}\sin\delta
 \Delta m^2_{12}\left(\frac{L}{4E_\nu}\right)
\sin^2\left(\Delta m^2_{23}\frac{L}{4E_\nu}\right) \nonumber \\
& \simeq & c^2_{13}s_{13}\sin2\theta_{12}\sin 2\theta_{23}\sin\delta
\Delta m^2_{12}(\Delta m^2_{23})^2
\left(\frac{L}{4E_\nu}\right)^3
\end{eqnarray}
hence, in the very high energy limit at fixed baseline, the effects
decrease as $E_{\nu}^{-3}$. That the effects disappear at high energy is
expected, since in this regime, the ``oscillations'' of the various
$\Delta_{jk}$'s wash out.

At high energy, we can also approximate $\cos 2\Delta_{13}
\approx \cos2\Delta_{23}\approx 1$, and find
\begin{equation}
|\Delta_\delta| \simeq \frac{1}{2}\Delta_{CP}(\delta=\pi/2)= \frac{1}{2}\Delta_{T}(\delta=\pi/2)
\end{equation}
so the CP- or T-conjugation is equivalent to a change of phase
$\delta\rightarrow-\delta$, i.e. $\delta=+\pi/2\rightarrow\delta=-\pi/2$.

The important point is that all expressions depend upon some factors which
contain the various mixing angles, multiplied by oscillatory terms which
always vary like sine or cosine of $\Delta_{jk}$-terms (the terms in
squared brackets in the expressions above). 

Hence, we expect the various discriminants to scale like
$\Delta_{jk}\propto L/E_\nu$. 
The sensitivity to the $\delta$-phase will therefore follow the
behavior of the oscillation probability, and we therefore argue that the
maximum of the effect will occur around the ``first maximum'' of the
oscillations, i.e. for
$E^{max}_\nu\equiv \Delta m^2_{32}L/2\pi$ (see Eq.~(\ref{eq:oscmax})).

Strictly speaking, the maximum of the $\delta$-phase
sensitivity does not lie exactly at the ``first maximum'' as defined in 
Eq.~(\ref{eq:oscmax}). From Eq.~(\ref{eq:delcptdef}), we expect the
maximum to be ``shifted'' to higher values of $L/E_\nu$, since it
corresponds to the maximum of the term
\begin{equation}
 \sin \Delta_{12}\sin \Delta_{13}\sin\Delta_{23}\simeq 
 \Delta m^2_{12}\frac{L}{4E_\nu}\sin^2\left(\Delta m^2_{23}\frac{L}{4E_\nu}\right)
\end{equation}
which has the functional form $x\sin^2 x$ and, therefore, has its maximum
shifted to higher values of $x$ compared to $\sin^2 x$. This small shift is
smaller than the oscillation wavelength itself, and does not cause a
major problem, since experimentally we will always have sufficient energy
range to cover the full oscillation.

These considerations are strictly true only for propagation in vacuum. When
neutrinos propagate through matter, matter effects will alter these
conclusions. We will however show that as long as the baseline is smaller
than some distance such that the corresponding ``first maximum''
$E^{max}_\nu$ lies below
the MSW resonance neutrino energy $E^{res}_{\nu}$, the considerations
related to the $L/E_\nu$ scaling are still largely valid. 

\begin{figure}
\centering
\epsfig{file=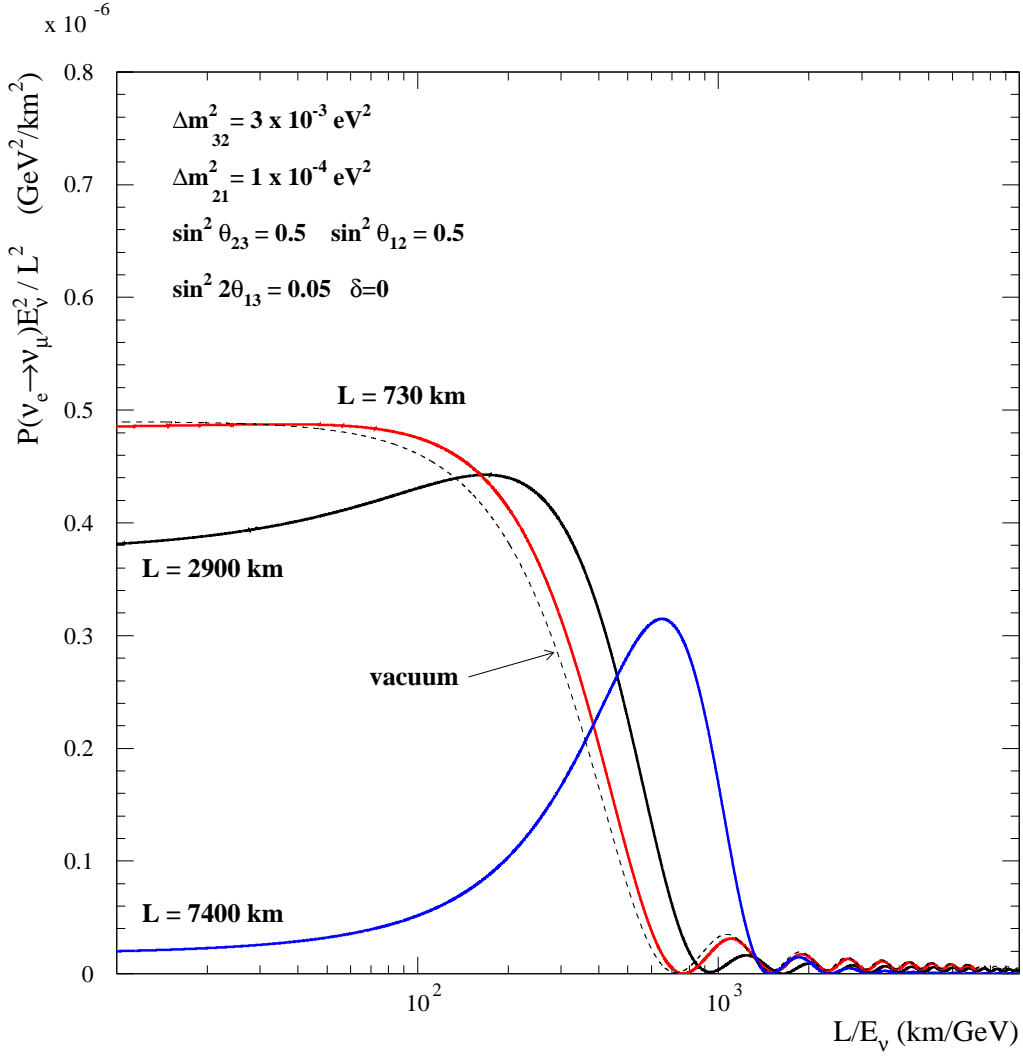,width=15cm}
\caption{Same as plots in Figure~\ref{fig:resprobmat} but only for
neutrinos
and as a function of 
$L/E_\nu$ for three
baselines $L=730\ \rm km$, 2900~km, 7400~km and in vacuum (dashed line,
independent of baseline).
The oscillation parameters are $\Delta m^2_{32}=3\times 10^{-3}\ \rm
eV^2$, $\Delta m^2_{21}=1\times 10^{-4}\ \rm
eV^2$, $\sin^2 \theta_{23} = 0.5$, $\sin^2 \theta_{12} = 0.5$, 
$\sin^2 2\theta_{13} = 0.05$ and $\delta=0$.}
\label{fig:reslovere}
\end{figure}

To illustrate this, we show in Figure~\ref{fig:reslovere}
the rescaled probability for $\nue\ra\numu$ oscillations
involving neutrinos and as a function of $L/E_\nu$, for three
baselines $L=730\ \rm km$, 2900~km, 7400~km and in vacuum 
(independent of baseline). We assume $\delta=0$. We readily observe that
for the shortest baseline $L=730\ \rm km$ the curve is very similar to the
one in vacuum. For the middle baseline $L=2900\ \rm km$, the curve is
modified with respect to the vacuum case, in the sense that the probability
around $(L/E_\nu)^{max}\simeq 400$ (see Eq.\ref{eq:leoscmax}) is enhanced,
but for $(L/E_\nu)\lesssim 130$ the probability is smaller than that in vacuum.
For the longest baseline $L=7400\rm\ km$, the curve is highly distorted by
matter and already for $(L/E_\nu)\lesssim 400$ is the probability smaller
than that in vacuum.

In what way does the matter effect alter the ability to look for effects
related to the $\delta$-phase? It is incorrect to believe that only the
$\Delta_{CP}$ discriminant will be affected by propagation through matter,
since it is the only one to a priori mix neutrinos and antineutrinos. In
reality, the ``dangerous'' effect of matter is to reduce the dependence of
the probability on the $\delta$-phase, and this for any kind of discriminant.

In matter, we would for example write the $\Delta_T$ discriminant
as
\begin{eqnarray}
|\Delta^m_{T}| & = &
(c^{m}_{13})^2s^m_{13}\sin2\theta^m_{12}\sin 2\theta^m_{23}\sin\delta^m
\Bigl[ \sin \Delta^m_{12}\sin \Delta^m_{13}\sin\Delta^m_{23}\Bigr]
\nonumber \\
& \simeq &
(c^{m}_{13})^2s^m_{13}\sin2\theta^m_{12}\sin 2\theta_{23}\sin\delta
\Bigl[ \sin \Delta^m_{12}\sin \Delta^m_{13}\sin\Delta^m_{23}\Bigr]
\end{eqnarray}
where because of our choice of $\Delta m^2_{jk}$'s,
we have $\theta^m_{23}\approx \theta_{23}$ and
$\delta^m\approx \delta$ (see Eq.~(\ref{eq:effang})). 
This implies that the $\delta$-phase discriminants have a different
structure that the terms that define the probability of the oscillation
(i.e. the non $\delta$-phase dependent terms).
The discriminants are the products of sines and cosines of {\it all} mixing
angles and of the $\Delta_{jk}$'s (see Eqs.~(\ref{eq:deldeldef}) and
(\ref{eq:delcptdef})). Because of this structure, their property in matter
is different.

The Jarlskog's determinant can be shown to be ``invariant'' under
matter effects\cite{Harrison:2000df} in the sense that the product
\begin{eqnarray}
J\Bigl( \Delta m^2_{12}\Delta m^2_{13}\Delta m^2_{23}\Bigr)\equiv
c^2_{13}s_{13}\sin2\theta_{12}\sin 2\theta_{23}\sin\delta
\Bigl( \Delta m^2_{12}\Delta m^2_{13}\Delta m^2_{23}\Bigr)
\end{eqnarray}
is {\it independent} from matter effects. This point was also discussed in
Ref.\cite{Lipari:2001ds}, where it was pointed out that this implies that the
$\delta$-phase terms will be independent from the matter effects as long as
we can approximate
\begin{eqnarray}
 \sin \Delta_{12}\sin \Delta_{13}\sin\Delta_{23}\approx 
\Delta_{12}\Delta_{13}\Delta_{23}.
\end{eqnarray}
This fact is not very revelant in the current context, 
since the best
sensitivity to the $\delta$-phase is expected just when this approximation
is {\it not} valid (i.e. $|\Delta_{13}|\approx |\Delta_{23}|\approx 1$), 
otherwise the effects wash out. As a consequence, we expect that the
CP-odd terms {\it do} depend on matter effects in the
relevant baseline and energy region. 

The behavior for neutrino energies above the MSW
resonance $E^{res}_\nu$ is determined by the fact that in this energy
regime, $\theta^m_{13}(E>E_\nu^{res})\rightarrow \pi/2$ and therefore
$c^{m}_{13}\ra 0$ and $s^m_{13}\ra 1$. 
More explicitly, one can show that
\begin{eqnarray}
(c^{m}_{13})^2s^m_{13} & \propto & E_\nu^{-2} \nonumber \\
\sin 2\theta^m_{23} & \simeq & \sin 2\theta_{23} = const. \nonumber \\
\sin 2\theta^m_{12} & \ra & const. \nonumber \\
\Delta M^2_{31} & \approx & \Delta M^2_{32} \propto E_\nu \nonumber \\
\Delta M^2_{21} & \approx & \Delta m^2_{32} = const
\end{eqnarray}
Therefore,
\begin{eqnarray}
|\Delta^m_{T}| & \propto & 
E_\nu^{-2} \sin\delta 
\Bigl[ \sin\left( \Delta m^2_{32}\frac{L}{4E_\nu}\right)
\sin^2 \left( \Delta M^2_{13}\frac{L}{4E_\nu}\right)\Bigr]
\propto 
E_\nu^{-3}
\end{eqnarray}
and one recovers a neutrino energy dependence identical to that in vacuum
(see Eq.~(\ref{eq:delcptdefvac})). Note also that the argument of the sine
function $\Delta M^2_{13}L/4E_\nu$ is {\it not} small (i.e. the
approximation $\sin x\simeq x$ is not valid). For our choice of oscillation
parameters, the mass difference is approximately equal to 
$\Delta M^2_{13}(\rm eV^2)\simeq 3\times 10^{-4}\times E_\nu(\rm GeV)$,
and hence the dependence on the baseline is 
\begin{eqnarray}
|\Delta^m_{T}| & \propto & 
E_\nu^{-2} \sin\delta 
\Bigl[ \sin\left( 1.27\Delta m^2_{32}\frac{L(\rm km)}{E_\nu(\rm GeV)}\right)
\sin^2 \left( 3.8\times 10^{-4} L(\rm km)\right)\Bigr]
\end{eqnarray}
Hence, the discriminant will first be enhanced and then be suppressed by
matter effects. The maximum is found when the sine squared function reaches
a maximum, or at approximately 4000~km under the assumption of high energy
neutrinos. We will verify numerically that this is really the case in 
Section~\ref{sec:direct}.

As anticipated, these discussions say that if one wants to study
oscillations in the region of the ``first maximum'', one should not choose
a too large baseline $L$, otherwise, matter effects will suppress the
oscillation probability. This is even more so true, as it will be recalled
below, that the magnitude of the effects related to the $\delta$-phase are
suppressed more rapidly than the oscillation.

The simplest way to express the condition on the matter is to
require that the energy of the ``first maximum'' be smaller than the MSW
resonance energy:
\begin{eqnarray}
2\sqrt{2}G_Fn_eE^{max}_\nu\lesssim\Delta m^2_{32}\cos 2\theta_{13}
\end{eqnarray}
and, by inserting the definition of $E^{max}_\nu\equiv\Delta m^2_{32}L/2\pi$ we get
\begin{eqnarray}
\label{eq:lmaxmatter}
L_{max}& \lesssim& \frac{\pi\cos 2\theta_{13}}{\sqrt{2}G_Fn_e} \approx
\frac{\pi\cos 2\theta_{13}}{2\times 1.27\times 7.56\times 10^{-5}(\rm
eV^2)\rho(\rm g/cm^3)} \nonumber \\ & \approx &\frac{1.5\times 10^4(\rm km)}{\rho(\rm
g/cm^3)}
\approx 5000\rm\ km
\end{eqnarray}

To summarize, we find that {\it the discriminants of the $\delta$-phase all
scale with $L/E_\nu$. The most favorable choice of neutrino energy $E_\nu$ and
baseline $L$ is in the region of the ``first maximum'' given by 
$(L/E_\nu)^{max}\simeq 400$ for $|\Delta m^2_{32}|=3\times 10^{-3}\rm\
eV^2$}.
This leaves a great flexibility in the choice of the actual neutrino
energy and the baseline, since only their ratio $L/E_\nu$ is
determinant. Because of the rising neutrino cross-section with energy, we
will see below that it will more favorable to go to higher energies if the
neutrino fluency is constant.{\it  Keeping the $L/E_\nu$ ratio constant, 
this implies an optimization at longer baselines $L$. One will hence gain
with the baseline $L$ until we reach $L_{max}\approx 5000\rm\ km$ beyond
which matter effects will spoil our sensitivity}.

\section{Detection of $\Delta_\delta$ at the NF}

We begin the discussion with the detection of the $\delta$-phase with the
method of the $\Delta_\delta$ discriminant. We recall that this method
implies the comparison of 
the measured $\nue\ra\numu$ oscillation probability as a
function of the neutrino energy $E_\nu$ compared to a ``Monte-Carlo
prediction'' of the spectrum in absence of $\delta$-phase.

\begin{figure}
\centering
\epsfig{file=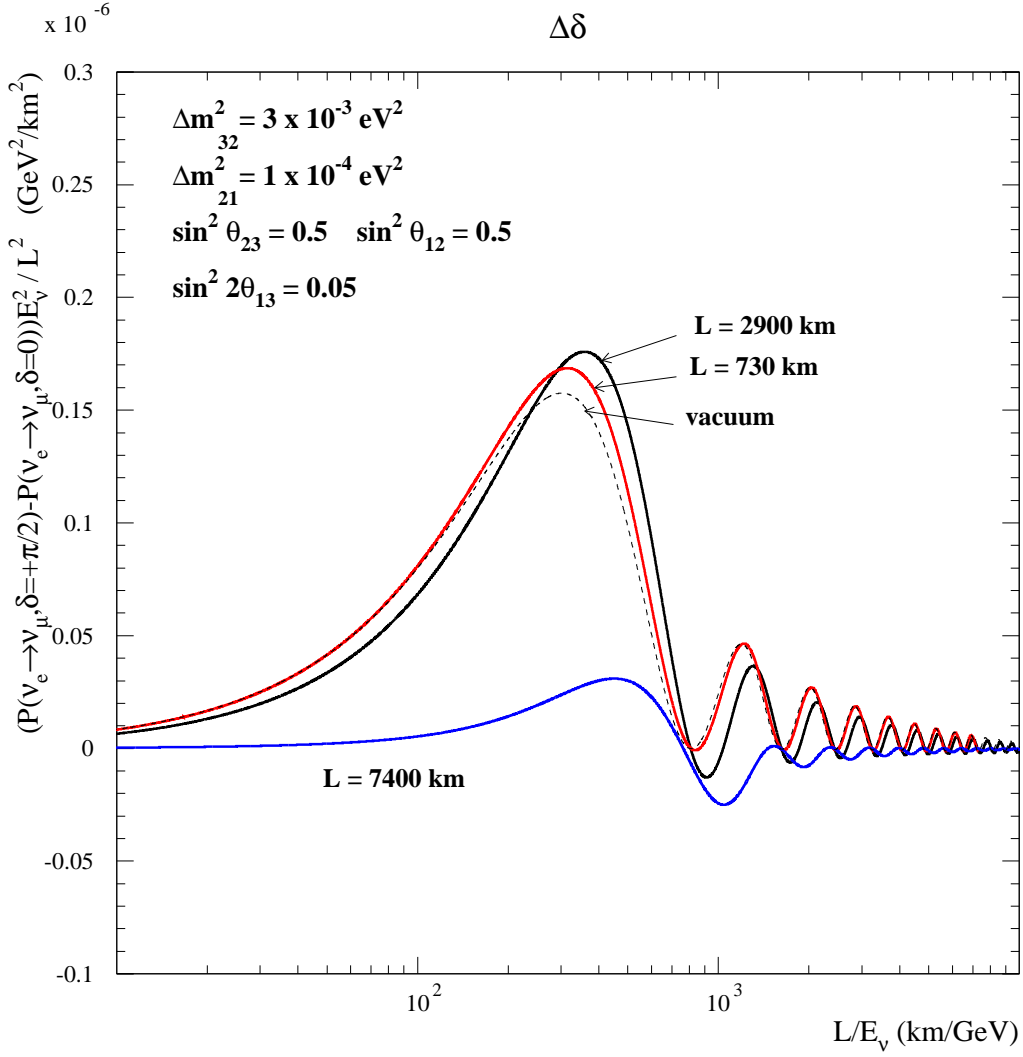,width=15cm}
\caption{The rescaled $\Delta_\delta$ discriminant (see text for
definition) as a function of the $L/E_\nu$ ratio,
computed for neutrinos propagating in matter
at three different baselines $L=730\rm\ km$, 2900~km and 7400~km (full lines),
and for neutrinos propagating in vacuum (dashed line).}
\label{fig:deldellovere}
\end{figure}

We first illustrate that the discriminant really scales like $L/E_\nu$
in Figure~\ref{fig:deldellovere}, where
the rescaled $\Delta_\delta$ discriminant is shown as
a function of the $L/E_\nu$ ratio.
The plot is computed for neutrinos propagating in matter
at three different baselines $L=730\rm\ km$, 2900~km and 7400~km, and for
neutrinos propagating in vacuum. Indeed, for the shortest two baselines
$L=730\rm\ km$ and 2900~km, the value of the discriminant is very similar
to that obtained in vacuum. For the longest baseline $L=7400\rm\ km$, well
above $L_{max}$ (see Eq.~(\ref{eq:lmaxmatter})), the effect is highly
suppressed by matter effect, since the ``first maximum'' corresponds for
the given baseline $L$ to an energy $E_\nu$ above the MSW resonance energy $E^{res}_\nu$.

From the figure, we clearly also see that the effect is largest 
around the ``first maximum'', in the region given by $(L/E_\nu)^{max}\simeq 400$.

Given the freedom in the energy $E_\nu$ and baseline $L$, provided the
ratio $L/E_\nu$ is appropriately chosen, and given the matter effects at
distances beyond $\approx 5000\rm\ km$, we restrict our choices to the two
baselines $L=730\rm\ km$ and 2900~km.

\subsection{The correlation with $\theta_{13}$}
When searching for effects related to the $\delta$-phase by comparing
the measured $\nue\ra\numu$ oscillation probability as a
function of the neutrino energy $E_\nu$ to a ``Monte-Carlo
prediction'' of the spectrum in absence of $\delta$-phase, requires
necessarily a precise knowledge of the other oscillation parameters
entering in the oscillation probability expression.

In particular, the knowledge of the angle $\theta_{13}$ could be quite
important. Indeed, the $\nue\ra\numu$ oscillation is primarily driven by
the $\theta_{13}$ angle and only thanks to a different energy dependence of
the terms proportional to $\delta$ than to those independent of
$\delta$ can one hope to determine $\theta_{13}$ and $\delta$ at the same
time!

This is however not true at high energy, when 
both $|\Delta_{12}|\ll 1$ and $|\Delta_{13}|,|\Delta_{23}|\ll 1$.
This can be explicitly shown for example
for simplicity in the limit of small $\theta_{13}$. The rescaled
probability is in this case (see Eq.~(\ref{eq:respemucanc})) a constant:
\begin{eqnarray}
 p(\nu_e\ra\nu_\mu) & \simeq & 
\frac{(\Delta m^2_{12})^2}{32} \Bigl\{ 1+ 
 2Ms^2_{13} +8Ns_{13}\cos\delta
\Bigr\}
\end{eqnarray}
where $M=((\Delta m^2_{13})^2+(\Delta m^2_{23})^2)/(\Delta m^2_{12})^2$ and
$N=(\Delta m^2_{13}+\Delta m^2_{23})/(\Delta m^2_{12})$. The absence of
``oscillations'' at high energy implies that a change of $\theta_{13}$ can
mimic a change of $\delta$.

\begin{figure}
\centering
\epsfig{file=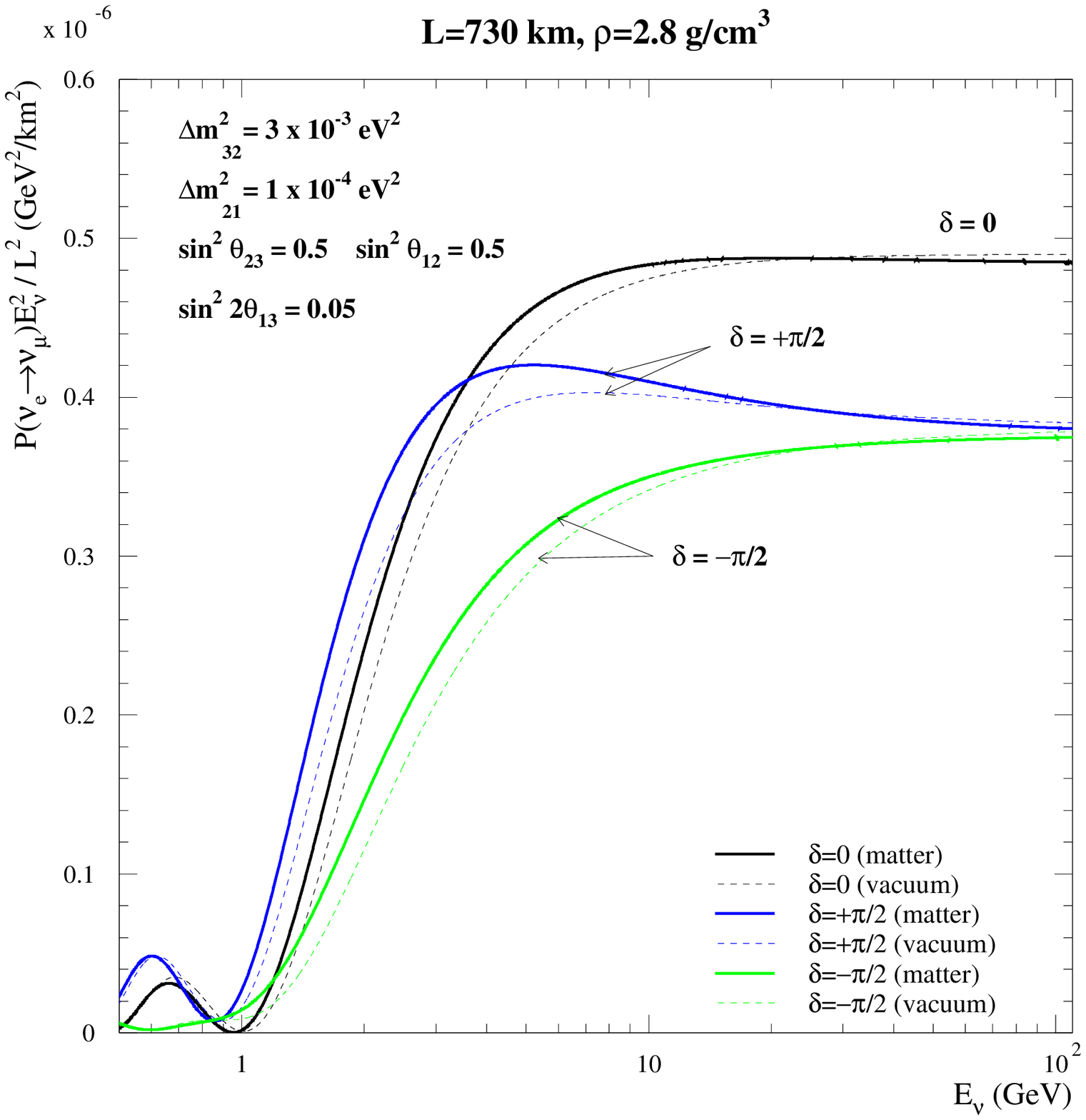,width=15cm}
\caption{Rescaled probability (see text) for $\nue\ra\numu$ oscillations for a
baseline $L=730\ \rm km$ as a function of neutrino energy
$E_\nu$.
The probabilities are computed for neutrinos in matter (full line)
and in vacuum (dotted line), and for three values of the 
$\delta$-phase: $\delta=0$, $\delta=+\pi/2$, $\delta=-\pi/2$.
The other oscillation parameters are $\Delta m^2_{32}=3\times 10^{-3}\ \rm
eV^2$, $\Delta m^2_{21}=1\times 10^{-4}\ \rm
eV^2$, $\sin^2 \theta_{23} = 0.5$, $\sin^2 \theta_{12} = 0.5$, 
$\sin^2 2\theta_{13} = 0.05$.}
\label{fig:respcp1}
\end{figure}

\begin{figure}
\centering
\epsfig{file=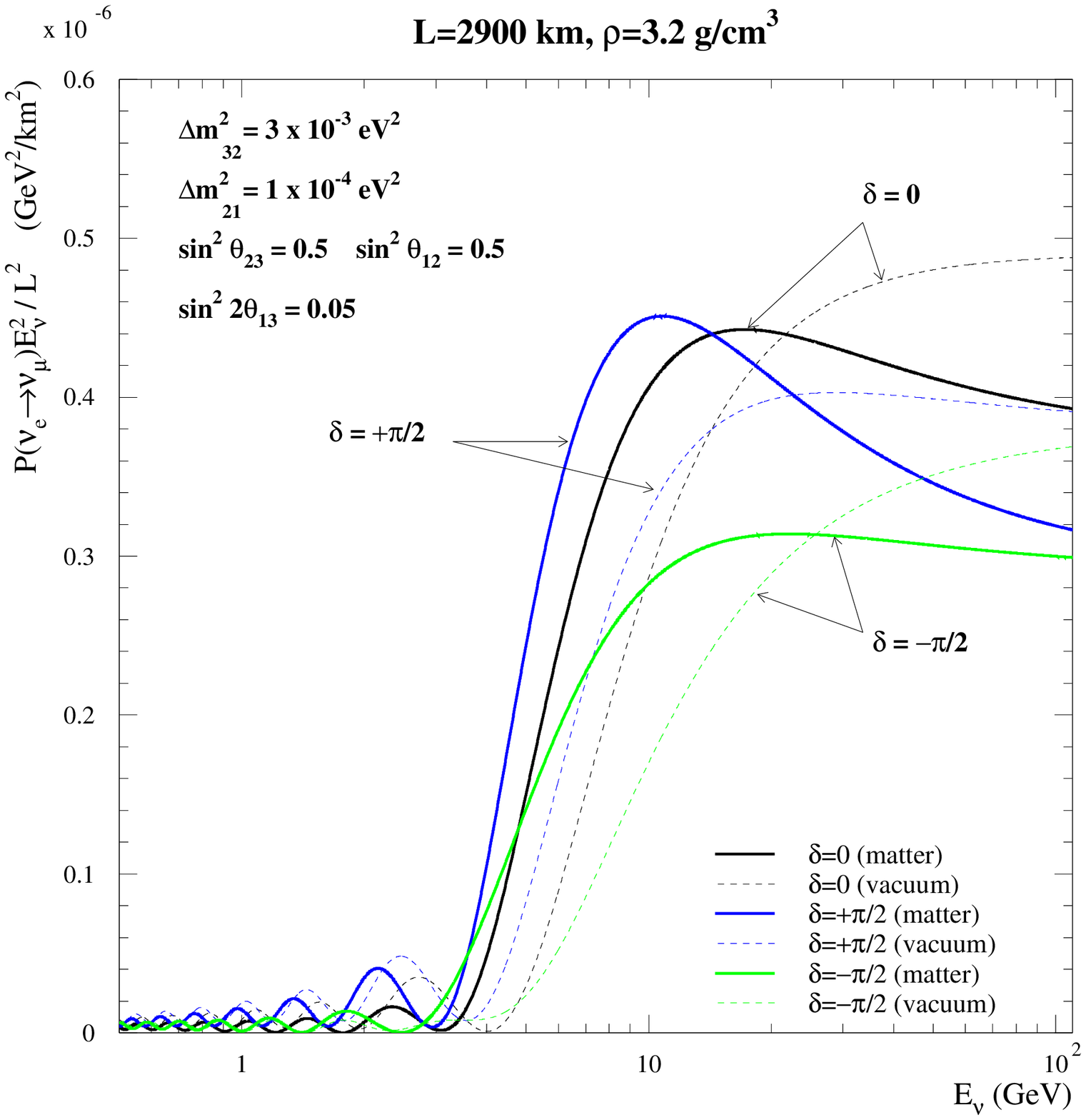,width=15cm}
\caption{Same as Figure~\ref{fig:respcp1} but for a baseline $L=2900\rm\ km$.}
\label{fig:respcp2}
\end{figure}

To illustrate this effect in a concrete example, we show in Figures~\ref{fig:respcp1}
and \ref{fig:respcp2} 
the rescaled probabilities for the two baselines $L=730\rm\ km$
and 2900~km as a function of neutrino energy
$E_\nu$.
The probabilities are computed for neutrinos in matter (full line)
and in vacuum (dotted line), and for three values of the 
$\delta$-phase: $\delta=0$, $\delta=+\pi/2$, $\delta=-\pi/2$.

At the shortest baseline $L=730\rm\ km$, this behavior can be clearly seen
for $E_\nu\gtrsim 20\rm\ GeV$. The two curves for $\delta=+\pi/2$ and
$\delta=-\pi/2$ tend to a constant, which is clearly different than the
constant for $\delta=0$, but that cannot be distinguished from a $\delta=0$
with a different $\theta_{13}$. In addition, we recall that the reason that
both $\delta=+\pi/2$ and $\delta=-\pi/2$ tend to the same constant, is due
to the fact that the probability at high energy depends only on
$\cos\delta$ (see Eq.~(\ref{eq:respemucanc})).

At the longer baseline $L=2900\rm\ km$, the situation is improved since
up to $\approx 100\rm\ GeV$, the rescaled probability is clearly not a
constant, but rather still falling with energy. We also point out that one
can visually see that most of the sensitivity at $L=2900\rm\ km$ is for
neutrino energies around $10\rm\ GeV$ or so. This is an important point to
take into account when considering the muon detection threshold of a given 
experimental setup.

\section{Direct detection of $T$-violation at the NF}
\label{sec:direct}
We now turn to the most challenging search for effects related to $\delta$
and consider the $\Delta_{T}$ discriminant. 
We recall here that the discriminant $\Delta_{T}$ implies the
comparison between the appearance of $\nu_\mu$ (resp. $\bar\nu_\mu$)
{\bf and} $\bar\nu_e$ (resp. $\nu_e$) in a beam of
stored $\mu^+$ (resp. $\mu^-$) decays as a function of the
neutrino energy $E_\nu$.

The $\Delta_{T}$ method takes in advantage that deals
with differences between neutrinos (or anti-neutrinos) only,
so matter effects give only a scale factor and no ``CP-fake'' violation
effects are expected for $\delta$=0. So, the comparison of
$\nu_\mu \rightarrow \nu_e$ and $\nu_e \rightarrow \nu_\mu$
oscillation probabilities offers a direct way to highlight a
complex component in the mixing matrix, independent of matter
and other oscillation parameters.

\begin{figure}[p]
\centering
\epsfig{file=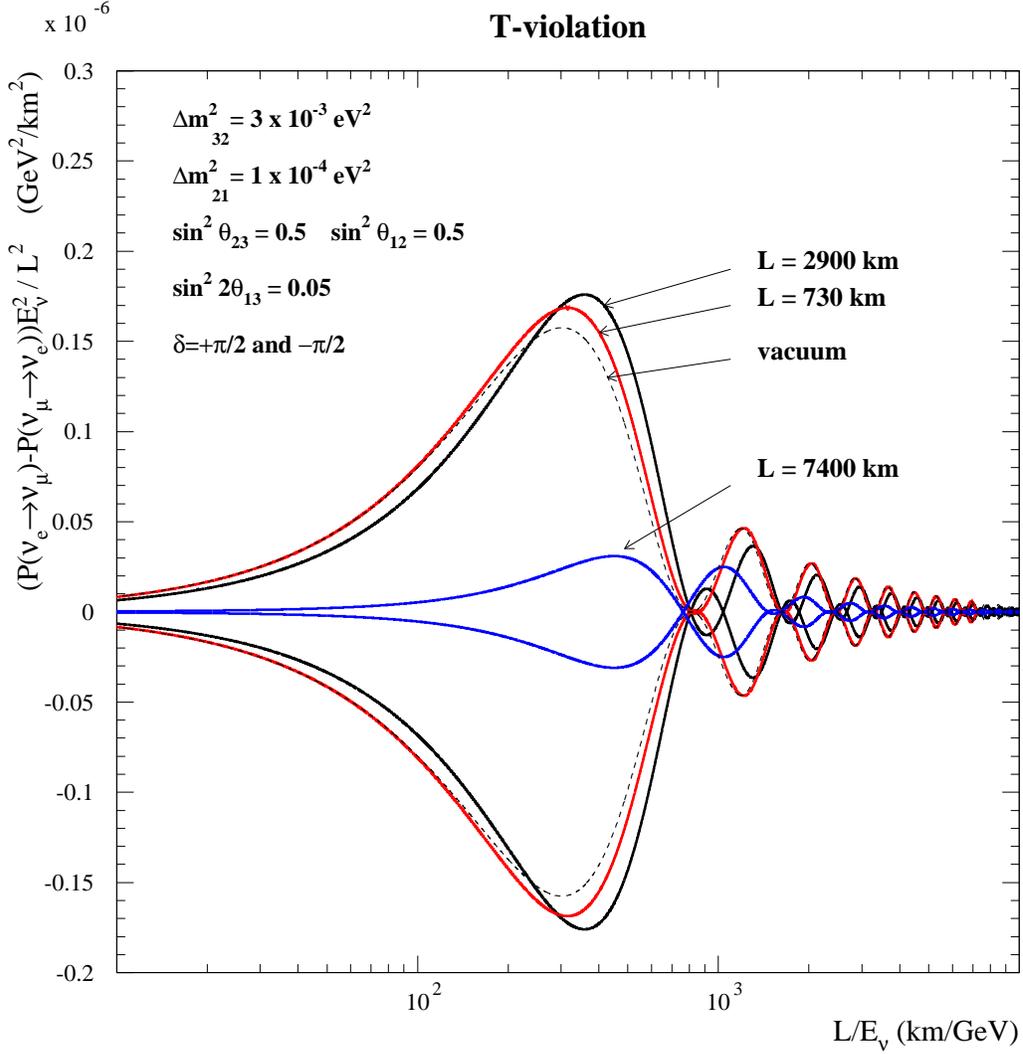,width=15cm}
\caption{The rescaled $\Delta_{T}$ discriminant (see text for
definition) as a function of the $L/E_\nu$ ratio,
computed for neutrinos propagating in matter
at three different baselines $L=730\rm\ km$, 2900~km and 7400~km, and
also for propagation vacuum (independent of baseline).
Two sets of curves are represented, corresponding to $\delta=+\pi/2$ 
(upper ones) and $\delta=-\pi/2$ (lower ones).
The other oscillation parameters are $\Delta m^2_{32}=3\times 10^{-3}\ \rm
eV^2$, $\Delta m^2_{21}=1\times 10^{-4}\ \rm
eV^2$, $\sin^2 \theta_{23} = 0.5$, $\sin^2 \theta_{12} = 0.5$
and $\sin^2 2\theta_{13} = 0.05$.}
\label{fig:deltle}
\end{figure}

This is clearly visible in Figure~\ref{fig:deltle}, where the rescaled
value of $\Delta_{T}$ is shown for three baselines and for vacuum.
The magnitude of the $T$-violation effect is roughly the same for
vacuum, 730 km and 2900 km, while for very long baselines
(7400 km) the sensitivity is lost. As described in
Section~\ref{le_scaling}, all $\delta$-phase discriminants scale
with $L/E_{\nu}$ until L$\approx$5000 km (see equation~\ref{eq:lmaxmatter})
and because of matter effects, beyond this value the effect decreases
dramatically.

\begin{figure}[p]
\centering
\epsfig{file=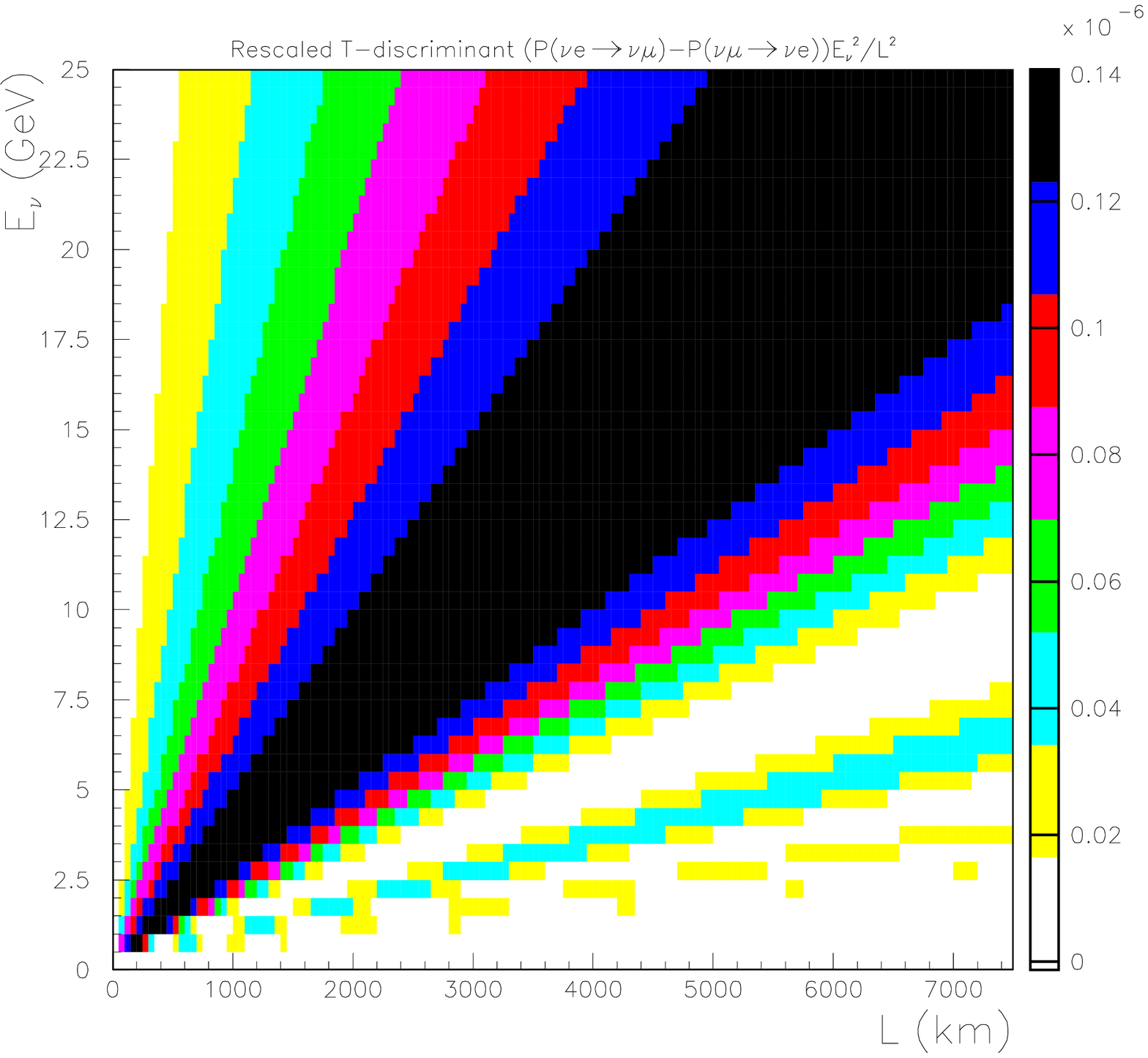,width=10cm}
\vspace*{-0.8cm}
\epsfig{file=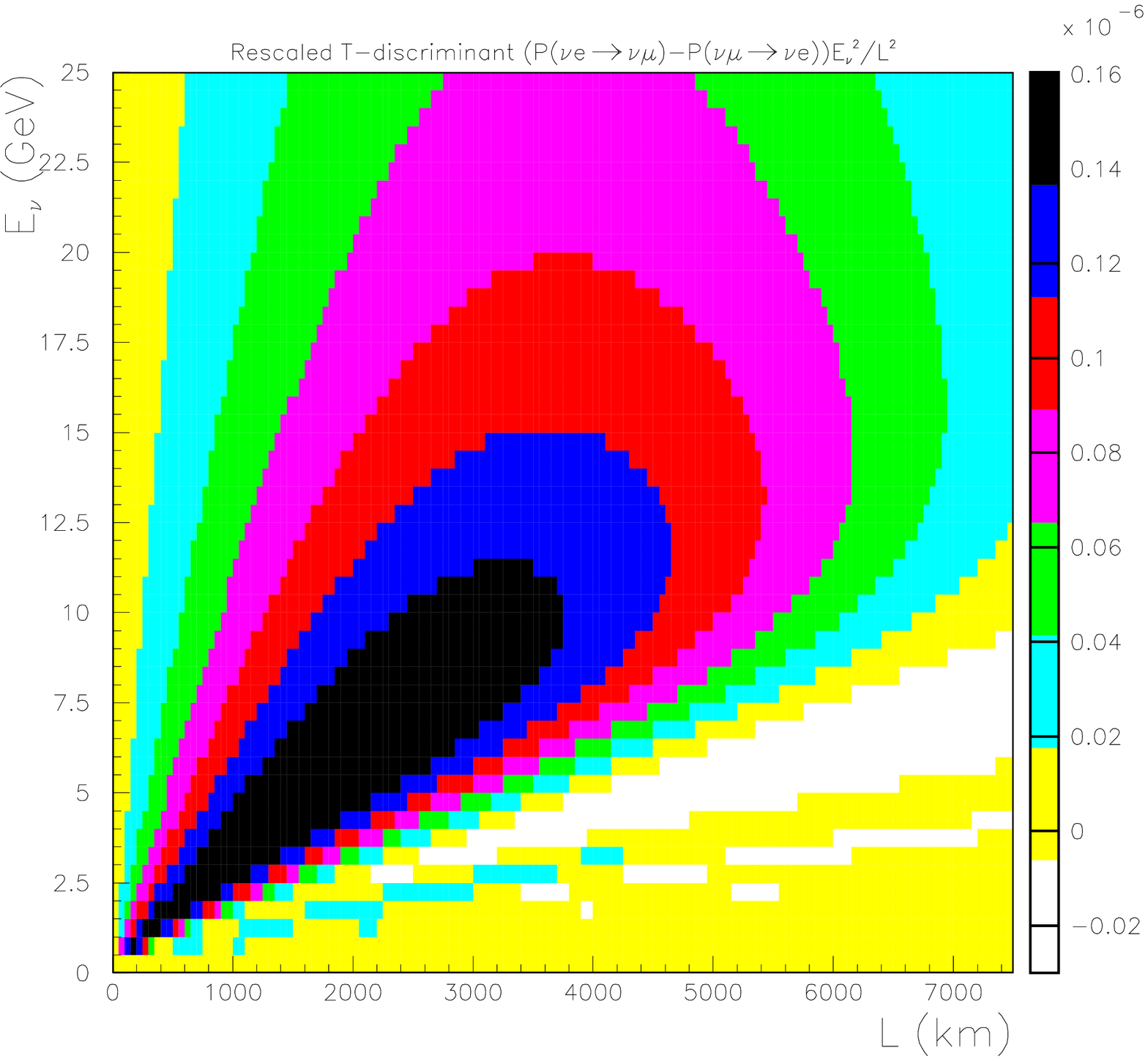,width=10cm}
\vspace*{0.5cm}
\caption{The rescaled $\Delta_{T}$ discriminant (see text for
definition) as a function of the $L$ and $E_\nu$,
computed for neutrinos propagating in vacuum (upper) and
in matter (lower) for $\delta=\pi/2$.
The other oscillation parameters are $\Delta m^2_{32}=3\times 10^{-3}\ \rm
eV^2$, $\Delta m^2_{21}=1\times 10^{-4}\ \rm
eV^2$, $\sin^2 \theta_{23} = 0.5$, $\sin^2 \theta_{12} = 0.5$
and $\sin^2 2\theta_{13} = 0.05$.}
\label{fig:deltcont}
\end{figure}

The drop of the rescaled $\Delta_{T}$ for very long baselines
is also illustrated in Figure~\ref{fig:deltcont}. The rescaled
$\Delta_{T}$ discriminant is shown as a function of the baseline
and of the neutrino energy, for neutrinos in vacuum (top) and
in matter (bottom) for $\delta = +\pi/2$. In vacuum, the same
value of $\Delta_{T} \times E^2/L^2$ is obtained keeping the
$L/E_{\nu}$ ratio constant, while in matter this pattern is destroyed,
loosing in sensitivity for long baselines. For instance, for
$E_\nu \sim 12$ GeV, the discriminant starts dropping
for baselines larger than 4000 km.

As already pointed out, the major difficulty of the $\Delta_{T}$
method is that it requires the measurement of the electron charge
in order to discriminate the large $\nu_e$ background from the beam.
Both, the wrong-sign muon and the wrong-sign electron samples are needed
to build the magnitude.

\section{Direct detection of $CP$-violation at the NF}
We now continue our discussion and consider the $\Delta_{CP}$
discriminant. We recall that the $\Delta_{CP}$ is in principle
quite straight-forward, since it involves
comparing the appearance of $\nu_\mu$ (resp. $\bar\nu_\mu$) in a beam of
stored $\mu^+$ (resp. $\mu^-$) decays as a function of the
neutrino energy $E_\nu$. A non-vanishing $\Delta_{CP}$ obtained from an
observation of the two polarities of the stored muon beam
would be a direct proof for a non-vanishing $\delta$-phase.

However, as already pointed out, this method suffers from the practical
necessity to traverse the Earth matter in order to perform long baseline
experiments.

Since neutrinos and anti-neutrinos are affected by matter with an
``opposite sign'', the matter terms will not cancel in the $\Delta_{CP}$
discriminant. As a result, the $\Delta_{CP}$ will not necessarily vanish
when $\delta=0$ (having the so called ``CP-fake'' violation effects.
Hence, a direct detection of a non-vanishing $\Delta_{CP}$
discriminant does not imply $\delta\neq 0$.

\begin{figure}
\centering
\epsfig{file=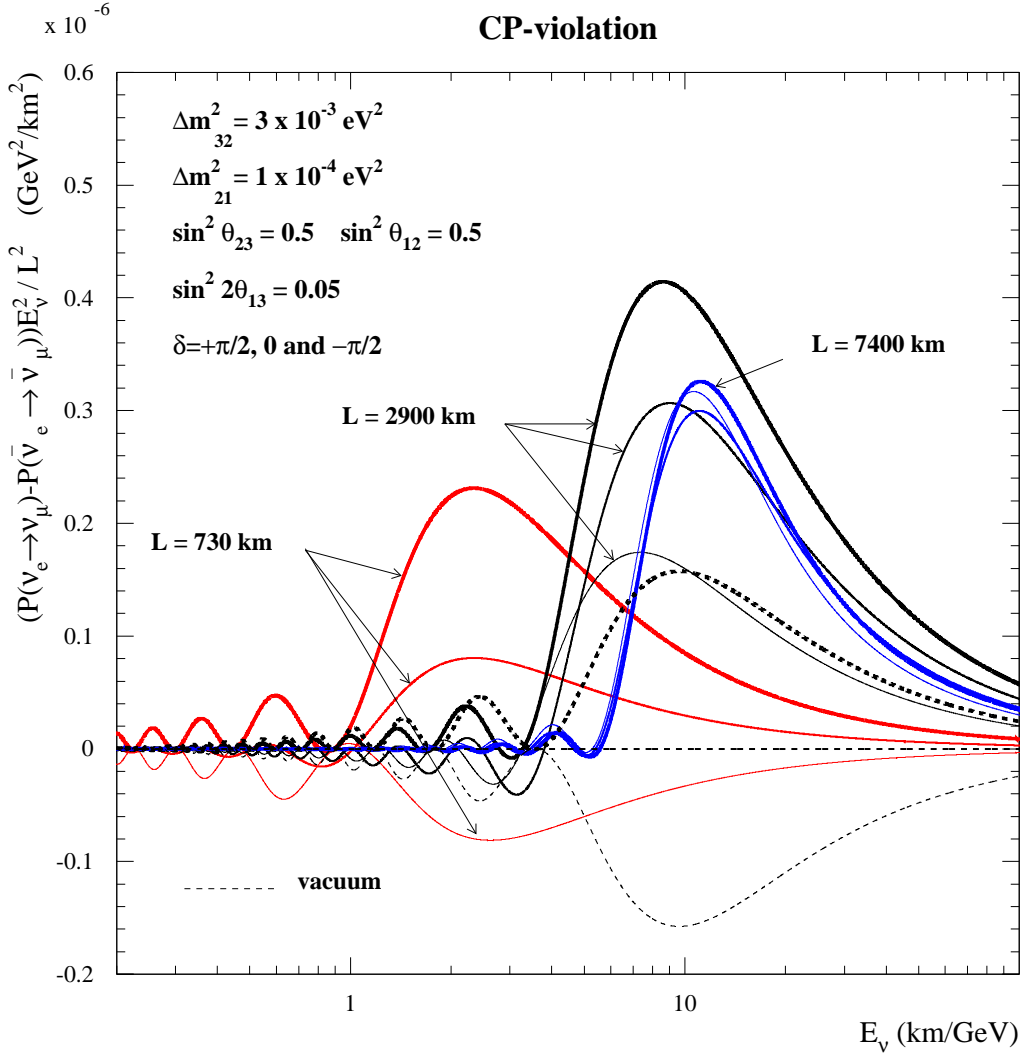,width=15cm}
\caption{The rescaled $\Delta_{CP}$ discriminant (see text for
definition) as a function of the neutrino energy $E_\nu$,
computed for neutrinos propagating in matter
at three different baselines $L=730\rm\ km$, 2900~km and 7400~km.
Three sets of curves are represented, corresponding to $\delta=+\pi/2$ 
(thick lines), $\delta=-\pi/2$ (thin lines) and $\delta=0$.
The other oscillation parameters are $\Delta m^2_{32}=3\times 10^{-3}\ \rm
eV^2$, $\Delta m^2_{21}=1\times 10^{-4}\ \rm
eV^2$, $\sin^2 \theta_{23} = 0.5$, $\sin^2 \theta_{12} = 0.5$
and $\sin^2 2\theta_{13} = 0.05$.}
\label{fig:delcpe}
\end{figure}

We illustrate this situation in Figure~\ref{fig:delcpe}, where
the rescaled $\Delta_{CP}$ discriminant is plotted
as a function of the neutrino energy $E_\nu$,
computed for neutrinos propagating in matter
at three different baselines $L=730\rm\ km$, 2900~km and 7400~km.
Three sets of curves are represented, corresponding to $\delta=+\pi/2$,
$\delta=-\pi/2$ and $\delta=0$.

At the shortest baseline $L=730\rm\ km$, 
we observe that the maximum of the effect is
indeed around the ``first maximum'' at about a neutrino energy of 2~GeV. 
For $\delta=0$ (thin curve), the discriminant does not vanish, due to
matter. This effect is however smaller than the ``genuine'' effect at
$|\delta|=\pi/2$ and we observe that for $\delta=+\pi/2\longrightarrow
-\pi/2$ the discriminant changes sign.

For the medium baseline of $L=2900\rm\ km$, the maximum of the effect
is again near the ``first maximum'' ($E_{\nu} \equiv$ 8~GeV). 
The $\Delta_{CP} \times E^2/L^2$ discriminant is larger (a factor 2)
than for $L=730\rm\ km$ in the case of $\delta=+\pi/2$, but this
includes the ``CP-fake'' contribution coming from matter.
In order to compare at both baselines the ``genuine'' $CP$-violation effect
(the one due to a non zero $\delta$-phase) one has to
compute the difference between $\delta=+\pi/2$ and $\delta=-\pi/2$,
resulting similar at both baselines.
Another interesting feature is that, at $L=2900\rm\ km$ the discriminant
is already positive for any value of $\delta$.

At larger baselines ($L=7400\rm\ km$) $\Delta_{CP}$ is completely
dominated by matter effects. The three values of $\delta$
mix and the sensitivity on $\Delta_{CP}$ to the ``genuine''
$CP$-violation is lost.

\begin{figure}
\centering
\epsfig{file=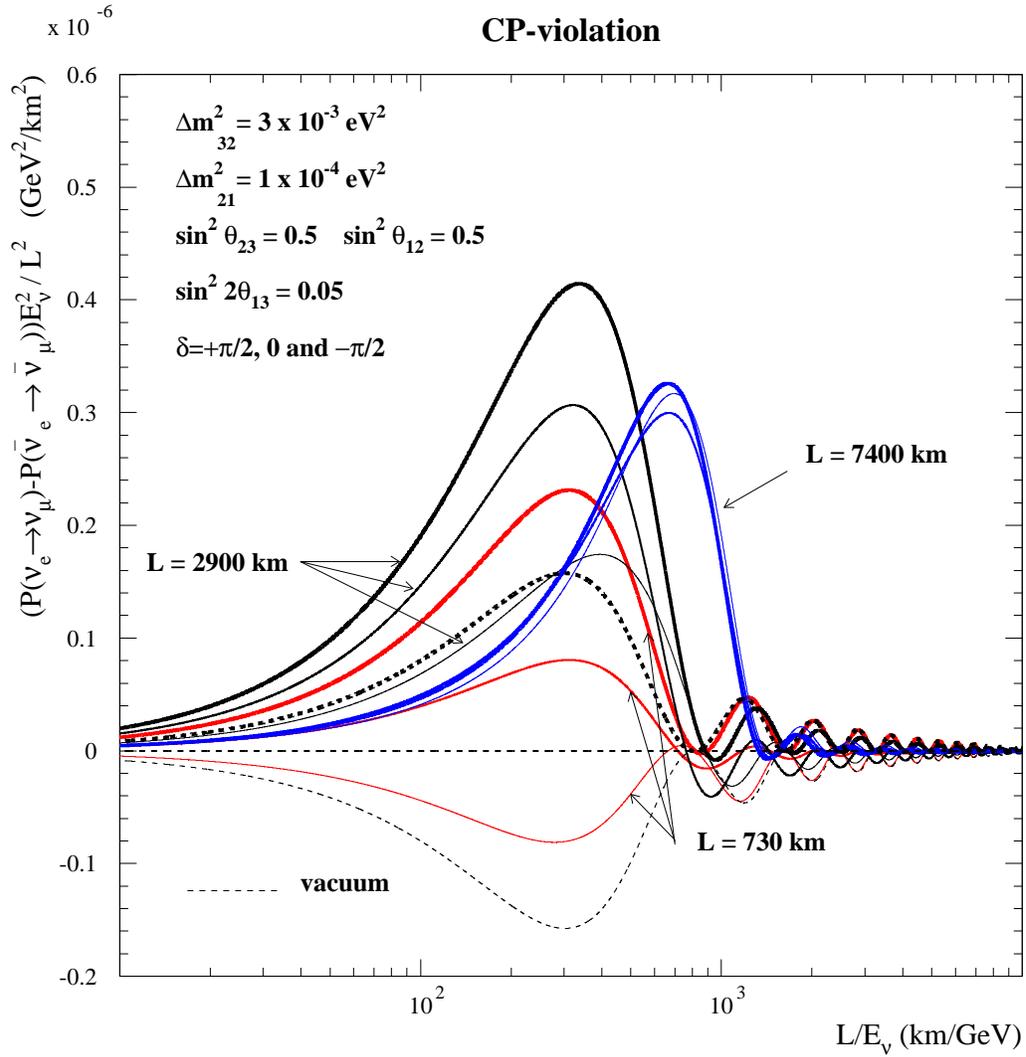,width=15cm}
\caption{Same as Figure~\ref{fig:delcpe} but
as a function of the $L/E_\nu$ ratio. In addition to the three 
different baselines, the case of neutrinos propagating in vacuum 
(independent of baseline) is also shown for comparison.}
\label{fig:delcple}
\end{figure}

Finally, we address the scaling of the effect with
$L/E_\nu$. Figure~\ref{fig:delcple} shows the same set of curve displayed in
Figure~\ref{fig:delcpe}, but as a function of the $L/E_\nu$ ratio. 
In addition to the three 
different baselines, the case of neutrinos propagating in vacuum 
(independent of baseline) is also shown for comparison.
At 730 km and 2900 km the effect is maximum for the same
$L/E_\nu \simeq 400$ km/GeV (for $\Delta m^2_{32}=3\times 10^{-3}\rm\ eV^2$).
In addition, the difference between $\delta=+\pi/2$ and $-\pi/2$ at the
maximum is similar for both baselines.

\section{Behavior for small $\theta_{13}$ values}
In the above sections, the possibility of detecting $CP$-violation, based 
on the comparison of oscillations involving electron and muon 
(anti-)neutrinos has been discussed. This transition is strongly dependent
on the value of the mixing angle $\theta_{13}$, and it is natural to ask
how the sensitivity to $CP$-violation varies with the value of this angle.
It was stated (\cite{romanino}) that, ``as far as the $\Delta m^2_{21}$
effects in the CP-conserving part of the oscillation probability can be
neglected'', the sensitivity to $CP$-violation does not depend on $\theta_{13}$.

We can check this hypothesis directly with the help of our Eq.~\ref{eq:probnenm}.
The probability for oscillations depends on
\begin{eqnarray}
& & 4\underline{c^2_{13}}
 \Bigl[  \sin^2\Delta_{23} s^2_{12} \underline{s^2_{13}}
      s^2_{23} + c^2_{12}
      \left( \sin^2\Delta_{13}\underline{s^2_{13}}s^2_{23} + 
        \sin^2\Delta_{12}s^2_{12}
         \left( 1 - \left( 1 + \underline{s^2_{13}} \right) s^2_{23} \right)  \right)  \Bigr]  
\nonumber \\ 
& \approx & 
4
\Bigl[  \sin^2\Delta_{23} s^2_{12} \underline{s^2_{13}}
      s^2_{23} + c^2_{12}
      \left( \sin^2\Delta_{13}\underline{s^2_{13}}s^2_{23}  
          \right)  \Bigr]  
\end{eqnarray}
where we have neglected $\Delta_{12}$. The parts dependent on $\delta$ go like
\begin{eqnarray}
  -  \frac{1}{2}c^2_{13}\sin (2\theta_{12})\underline{s_{13}}\sin (2\theta_{23})
\cos\delta
     \left[ \cos 2\Delta_{13} - \cos 2\Delta_{23} - 
       2\cos(2\theta_{12})\sin^2\Delta_{12}\right]  \nonumber \\ 
  + \frac{1}{2}c^2_{13}\ \sin\delta
\sin(2\theta_{12})\underline{s_{13}}\sin (2\theta_{23})
\left[ \sin2\Delta_{12} - \sin2\Delta_{13} + 
\sin 2\Delta_{23} \right]
\end{eqnarray}

As can be seen, the $\delta$-dependent part of the probability depends
on $\sin\theta_{13}$, while the leading term (and the total number of
oscillated events) is proportional to $\sin^2\theta_{13}$. 

In the Gaussian approximation, the statistical error is proportional to the square
root of the number of events, so the significance of the $\delta$-effects is
independent of $\theta_{13}$. The validity of this assumption will depend on 
the total number of muon decays and on the total mass of the considered detector.

As far as the $\Delta m^2_{21}$ part is concerned, the approximation is
valid if
\begin{equation}
\sin^2\Delta_{13}\underline{s^2_{13}}s^2_{23} \gtrsim
        \sin^2\Delta_{12}s^2_{12}
         \left( 1 - \left( 1 + \underline{s^2_{13}} \right) s^2_{23} \right) 
\end{equation}
or
\begin{equation}
s_{13} \gtrsim
\sqrt{\frac{\sin^2\Delta_{12}}{2\sin^2\Delta_{13}}}\approx 0.02\frac{\Delta m^2_{12}}{10^{-4}\rm (eV^2)}
\end{equation}
and $\Delta m^2_{23}=3\times 10^{-3} eV^2$.
For smaller values of $\theta_{13}$, the oscillation probability is dominated 
by the constant term, and the cancellation of the $\theta_{13}$ dependence no longer
applies. 

\section{Two concrete cases at 730 and 2900 km}
In order to assess with concrete examples the use of the $\delta$-phase
discriminants, we consider the two baselines, with corresponding muon beam
energy and matter densities:

\begin{center}
\begin{tabular}{|r|c|c|c|} \hline
 Baseline & $E_\mu$ & Muon decays         & Matter density \\ \hline
  732~km  & 7.5~GeV & $10^{21}$           & 2.8~g/cm$^3$ \\
 2900~km  &  30~GeV & $2.5\times 10^{20}$ & 3.2~g/cm$^3$ \\ \hline
\end{tabular}
\end{center}

Both examples were chosen to have the same $L/E_\mu$ and include
matter effects. Because of the linear
rise of the neutrino cross-section with $E_\nu$, the factor 4 in muon
energy between the 732~km and 2900~km case, is ``compensated'' by
an increase of intensity by the same factor in favor of the shorter baseline.

We compute the fluxes assuming unpolarized muons and
disregarding muon beams divergences within the storage
ring. We consider that 50\% of the stored muons decay
on the direction of the detector (``useful'' muons).
 We integrate the expected event rates using
a neutrino-nucleon Monte-Carlo generator\cite{nux}.
The total charged current (CC) cross
section is technically subdivided into
three parts: the exclusive
quasi-elastic scattering channel $\sigma_{QE}$ and
the inelastic cross section $\sigma_{inelasic}$ which
includes all other processes except charm
production which is included separately.
Event rates for the two energy-baseline-flux combinations are shown in 
Tables~\ref{tab:rates1} and \ref{tab:rates2} for a 10~kton fiducial mass
detector.

Our analyses are performed on samples of fully generated Monte-Carlo 
events\cite{nux}, which include proper kinematics of the events, full hadronization
of the recoiling jet and proper exclusive polarized tau decays when 
relevant.
Nuclear effects, which are taken into
account by the FLUKA model, are included
as they are important for a proper estimation of the tau
kinematical identification.

To have a proper account for the detector and background
effects, the specific detector model of a Liquid Argon TPC\cite{icarus}
has been assumed.

The detector response is included in our analyses using a fast simulation which
parameterizes the momentum and angular resolution of the emerging 
particles, using essentially the following values:
electromagnetic shower $3\%/\sqrt{E}\oplus 1\%$, 
hadronic shower $\approx 20\%/\sqrt{E}\oplus 5\%$, and magnetic muon
momentum measurement $20\%$.

\subsection{Lepton charge identification}

Muon identification, charge and momentum
measurement provide discrimination between $\nu_\mu$ 
and $\bar\nu_\mu$ charged current (CC) events. Good
$\nu_e$ CC versus $\nu$ NC
discrimination relies on the fine granularity of the target. Finally, the 
identification of $\nu_\tau$ CC events requires a precise measurement 
of all final state particles.

\newpage
\begin{table}[H]
\begin{center}
\begin{tabular}{|ll|c|c|}\hline
&         & $E_\mu=7.5\rm\ GeV$ & $E_\mu=30\rm\ GeV$          \\
& Process & $L=732\rm\ km$      & $L=2900\rm\ km$             \\ 
&         & $10^{21}$ $\mu^-$   & $2.5\times 10^{20}$ $\mu^-$ \\
\hline
                         & $\nu_\mu$ CC     & 39572 & 35945 \\
Non-oscillated           & $\nu_\mu$ NC     & 10775 & 10688 \\
rates                    & $\bar{\nu}_e$ CC & 14419 & 13911 \\
                         & $\bar{\nu}_e$ NC &  4501 & 4830  \\
\hline
Oscillated               & $\bar{\nu}_e\osc\bar{\nu}_\mu$ CC &  85 &  42 \\
events ($\delta=\pi/2$)  & $\nu_\mu\osc\nu_e$ CC             & 248 & 238 \\
\hline
Oscillated               & $\bar{\nu}_e\osc\bar{\nu}_\mu$ CC & 134 &  72 \\
events ($\delta=0$)      & $\nu_\mu\osc\nu_e$ CC             & 370 & 333 \\
\hline
Oscillated               & $\bar{\nu}_e\osc\bar{\nu}_\mu$ CC & 134 &  69 \\
events ($\delta=-\pi/2$) & $\nu_\mu\osc\nu_e$ CC             & 360 & 323 \\
\hline
\end{tabular}
\end{center}
\caption{Event rates for a 10 kton detector. 
The oscillation parameters are: $\Delta m_{32}^2 = 3 \times 10^{-3}\rm\ eV^2$, 
$\Delta m^2_{12}=1\times 10^{-4}\rm\ eV^2$,
$\sin^2 \theta_{23} = 0.5$, $\sin^2 \theta_{12}=0.5$
and $\sin^2 2\theta_{13} = 0.05$.}
\label{tab:rates1}
\end{table}

\begin{table}[H]
\begin{center}
\begin{tabular}{|ll|c|c|}\hline
&         & $E_\mu=7.5\rm\ GeV$ & $E_\mu=30\rm\ GeV$          \\ 
& Process & $L=732\rm\ km$      & $L=2900\rm\ km$             \\ 
&         & $10^{21}$ $\mu^+$   & $2.5\times 10^{20}$ $\mu^+$ \\
\hline
                         & $\bar{\nu}_\mu$ CC  & 16472 & 16054 \\
Non-oscillated           & $\bar{\nu}_\mu $ NC &  5319 &  5636 \\
rates                    & $\nu_e$ CC          & 35279 & 31399 \\
                         & $\nu_e $ NC         &  9268 &  9245 \\
\hline
Oscillated               & $\nu_e\osc\nu_\mu$ CC              & 408 & 389 \\
events ($\delta=\pi/2$)  & $\bar{\nu}_\mu\osc\bar{\nu}_e $ CC & 123 &  66 \\
\hline
Oscillated               & $\nu_e\osc\nu_\mu$ CC              & 403 & 381 \\
events ($\delta=0$)      & $\bar{\nu}_\mu\osc\bar{\nu}_e $ CC & 128 &  71 \\
\hline
Oscillated               & $\nu_e\osc\nu_\mu$ CC              & 266 & 273 \\
events ($\delta=-\pi/2$) & $\bar{\nu}_\mu\osc\bar{\nu}_e $ CC &  84 &  43 \\
\hline
\end{tabular}
\end{center}
\caption{Same as Table 1, but $\mu^+$ decays.}
\label{tab:rates2}
\end{table}
\newpage

\begin{table}[H]
\begin{center}
\begin{tabular}{|llll|}
\multicolumn{4}{c}{$\mu^- \; \rightarrow \; e^- \; \nu_{\mu} \;\; \bar{\nu}_e$}\\
\hline
Right sign muons ($\mu^-$): & & & \\
   & $\nu_\mu$               & CC &  \\
   & $\nu_\mu \osc \nu_\tau$ & CC & ($\tau^-  \rightarrow \mu^-$) \\
   & $\bar{\nu}_e$           & NC & ($\pi^- \rightarrow \mu^-$) \\
   & $\nu_\mu$               & NC & ($\pi^- \rightarrow \mu^-$) \\
\hline
Wrong sign muons ($\mu^+$): & & & \\
   & $\bar{\nu}_e \osc \bar{\nu}_\mu$   & CC &  \\
   & $\bar{\nu}_e \osc \bar{\nu}_\tau$  & CC & ($\tau^+  \rightarrow \mu^+$) \\
   & $\bar{\nu}_e$  & NC & ($\pi^+ \rightarrow \mu^+$) \\
   & $\nu_\mu$      & NC & ($\pi^+ \rightarrow \mu^+$) \\
\hline
Electrons: & & & \\
   & $\bar{\nu}_e$                    & CC & ($\bar{\nu}_e \rightarrow e^+$) \\
   & $\nu_\mu \osc \nu_e$             & CC & ($\nu_e  \rightarrow e^-$) \\
   & $\nu_\mu \osc \nu_\tau$          & CC & ($\tau^- \rightarrow e^-$) \\
   & $\bar{\nu}_e \osc \bar{\nu}_\tau$& CC & ($\tau^+ \rightarrow e^+$) \\
\hline
Wrong sign electrons ($e^-$): & & & \\
   & $\bar{\nu}_e$           & CC & $(\bar{\nu}_e \rightarrow e^+) \times p_{conf}$ \\
   & $\nu_\mu \osc \nu_e$    & CC & $(\nu_e  \rightarrow e^-) \times (1 - p_{conf})$ \\
   & $\nu_\mu \osc \nu_\tau$ & CC & ($\tau^- \rightarrow e^-) \times (1 - p_{conf})$ \\
   & $\bar{\nu}_e \osc \bar{\nu}_\tau$ & CC & ($\tau^+ \rightarrow e^+) \times p_{conf}$ \\
\hline
Neutral Currents (no leptons): & & & \\
   & $\nu_\mu$       & NC & (hadrons) \\
   & $\nu_\mu \osc \nu_\tau$            & CC & ($\tau^-  \rightarrow$ hadrons) \\
   & $\bar{\nu}_e$   & NC & (hadrons) \\
   & $\bar{\nu}_e \osc \bar{\nu}_\tau$  & CC & ($\tau^+  \rightarrow$ hadrons) \\
\hline
\end{tabular}
\end{center}
\caption{Event classes in case of $\mu^-$ beam.}
\label{tab:eventclass}
\end{table}

It is natural to classify the events in various classes depending
on their final state configuration\cite{us1}. We
illustrate them for the case of $\mu^-$ stored in the ring. 
\begin{enumerate}
\item {\bf Right sign muons ($rs\mu$):} the leading muon has {\it the
same charge} as those circulating inside the ring. We include
the non-oscillated $\numu$~CC events, the $\numu\ra\nutau$ oscillations
charged current events followed by $\tau\ra\mu$ and the background from
hadron decays in neutral currents induced by all neutrino flavors.

\item {\bf Wrong sign muons ($ws\mu$):} the leading muon has
{\it opposite charge} to those circulating inside the ring.
Opposite-sign leading muons can only be produced by neutrino oscillations, since
there is no component in the beam that could account for them.
This includes $\bar{\nu}_e\to\bar{\nu}_\mu$ oscillations and
$\bar{\nu}_e\to\bar{\nu}_\tau$ oscillations with $\tau^+\ra\mu^+$ decays.
We also include hadron decays in neutral currents induced by all neutrino flavors.

\item {\bf Electrons ($e$):} events with a prompt electron ({\it both charges})
and no primary muon identified.
Events with leading electron or positron are produced by the 
charged-current interactions of the
following neutrinos: non-oscillated $\bar{\nue}$ neutrinos,
$\numu\ra\nue$ oscillations
and $\bar{\nu}_e\to\bar{\nu}_\tau$ or $\numu\ra\nutau$ oscillations
with $\tau\ra e$ decays

\item {\bf Wrong sign electrons ($wse$):} the leading electron has an
identified charge which has {\it the same sign} as that of the muons circulating
inside the ring. We define 
\begin{itemize}
\item the efficiency $\epsilon_e$ to identify the charge;
\item the probability $p_{conf}$ for charge confusion.
\end{itemize}
Hence, the wrong sign electron sample is built from unoscillated
$\bar\nue$'s from the beam with a weight $P(\bar\nue\ra\bar\nue)\times
\epsilon_e\times p_{conf}$ and from muon neutrinos oscillations with the
weight $P(\numu\ra\nue)\times \epsilon_e\times (1-p_{conf})$.

\item {\bf No Lepton ($0\ell$):} events corresponding to NC interactions or
$\nu_\tau$ CC events followed by a hadronic decay of the tau lepton.
Events with {\it no leading electrons or muons} will be used to 
study the $\numu\ra\nutau$ oscillations. These events 
can be produced in neutral current processes or in
$\bar{\nu}_e\to\bar{\nu}_\tau$ or $\numu\ra\nutau$ oscillations
with $\tau\ra hadrons$ decays.
\end{enumerate}

Table~\ref{tab:eventclass} summarizes the different processes that contribute
to the five event classes. $p_{conf}$ is the electron charge confusion
probability.

For electron or muon charged current events, the visible event
energy reconstructs the incoming neutrino energy and is therefore a very
important variable to study the energy dependence of the oscillations.

The last three classes can only be cleanly studied in
a fine granularity detector.

The energy loss of a minimum ionizing particle in
liquid Argon is about 200 MeV/m, while the nuclear interaction length is
84 cm. That means that a muon with a momentum of 0.5 GeV will travel about
2.5 meters before being absorbed, corresponding to about 3 nuclear interaction
lengths. The probability that a pion will travel so long without interacting
is small, and also detailed simulations show that at this level of
background rejection pions do not represent a problem. 

Based on preliminary estimates\cite{echarge}, we assume that for electrons
with energy less than 7.5~GeV, the
charge identification efficiency is 20\% with a charge confusion
probability of $0.1\%$, unless otherwise noted.

\subsection{Direct extraction of the oscillation probabilities}
\label{extroscprob}
From the visible energy distributions of the events, one can extract the
oscillation probabilities. The visible energy of the events are plotted
into histograms with 10 bins in energy. The $\nue\ra\numu$ oscillation 
probability in each energy bin $i$ can be computed as
\begin{eqnarray}
{\cal P}_i(\nue\ra\numu)=\frac{N_i(ws\mu)-N^0_i(ws\mu)}
{\epsilon_i(p_\mu>p_\mu^{cut})N_i^0(e)}
\,\,\,\,\,\,\,\,\,\, [\mu^+ \,\,\, decays]
\end{eqnarray}
where $N_i(ws\mu)$ is the number of wrong-sign muon events in the i-th bin
of energy, $N^0_i(ws\mu)$ are the background events in the i-th bin
of energy, $\epsilon_i(p_\mu>p_\mu^{cut})$ is the efficiency of the muon
threshold cut in that bin, and $N_i^0(e)$ is the number of electron events
in the i-th bin of energy in absence of oscillations. The number of events
corresponds to the statistics obtained from $\mu^+$ decays. A similar
quantity for antineutrinos ${\cal P}_i(\bar\nue\ra\bar\numu)$ will be computed with
events coming from  $\mu^-$ decays.

Similarly, the $\numu\ra\nue$ oscillation 
probability in each energy bin $i$ can be computed as
\begin{eqnarray}
\label{pnumunue}
{\cal P}_i(\numu\ra\nue)=\frac{N_i(wse)-N^0_i(wse)}
{\epsilon_e (1-p_{conf})N_i^0(rs\mu)}
\,\,\,\,\,\,\,\,\,\, [\mu^- \,\,\, decays]
\end{eqnarray}
where $N_i(wse)$ is the number of wrong-sign electron events in the i-th bin
of energy, $\epsilon_e$ is the efficiency for charge discrimination,
$p_{conf}$ the charge confusion,
and $N_i^0(rs\mu)$ is the number of right sign muon events
in the i-th bin of energy in absence of oscillations. The number of events
corresponds to the statistics obtained from $\mu^-$ decays. A similar
quantity for antineutrinos ${\cal P}_i(\bar\numu\ra\bar\nue)$ will be computed with
events coming from  $\mu^+$ decays.

These binned probabilities could be combined in an actual experiment in
order to perform direct searches of the effects induced by the
$\delta$-phase.

\subsection{Direct search for T-asymmetry}

For measurements involving the discrimination of the electron charge, we
limit ourselves to the lowest energy and baseline configuration ($E_\mu =
7.5\rm\ GeV$ and $L=732\rm\ km$), since we expect the discrimination of the
electron charge to be practically possible only at these lowest energies.

The binned $\Delta_T(i)$ discriminant for neutrinos is defined as
\begin{eqnarray}
\Delta_T(i)={\cal P}_i(\numu\ra\nue)-{\cal P}_i(\nue\ra\numu)
\end{eqnarray}
and a similar discrimant $\bar\Delta_T(i)$ can be computed for antineutrinos.

These quantities are plotted for neutrinos
and antineutrinos for three values of the
$\delta$-phase ($\delta=+\pi/2$, $\delta=0$ and $\delta=-\pi/2$)
in Figure~\ref{fig:directt}.
The errors are statistical and correspond to a normalization of $10^{21}$
muon decays and a baseline of $L=732\rm\ km$. 
A 20\% electron efficiency with a charge confusion
probability of 0.1\% has been assumed.
The full curve corresponds
to the theoretical probability difference.

A nice feature of these measurements is the change of sign of the effect
with respect of the change $\delta\ra -\delta$ and also with respect
to the substitution of neutrinos by antineutrinos. These changes of sign
are clearly visible and would provide a direct, model-independent,
proof for $T$-violation in neutrino oscillations.

In order to cross-check the matter behavior, one can also contemplate the
$CPT$-discriminants defined as
\begin{eqnarray}
\Delta_{CPT}(i)={\cal P}_i(\numu\ra\nue)-{\cal P}_i(\bar\nue\ra\bar\numu)
\nonumber\\
\bar\Delta_{CPT}(i)={\cal P}_i(\nue\ra\numu)-{\cal P}_i(\bar\numu\ra\bar\nue)
\end{eqnarray}
These quantities are plotted in Figure~\ref{fig:directcpt}, with the same
assumptions as in Figure~\ref{fig:directt}.
The full curve corresponds to the theoretical probability difference.
As expected, these quantities are
{\it independent from the $\delta$-phase and probe only the matter effects}.
The change of sign of the effect
with respect to the substitution of
neutrinos by antineutrinos is clearly visible.

It should be however noted that in the case of the $CPT$ discriminant, the
statistical power is rather low, since this measurement combines the
appearance of electrons (driven by the efficiency for detecting the
electron charge) and involves antineutrinos, which are suppressed by matter
effects. Hence, the statistical power is reduced compared to the
$T$-discriminant.

\subsection{Direct search for CP-asymmetry}
\label{sec:directsearchcp}

In the direct search for the CP-asymmetry, we rely only on the appearance
of wrong-sign muons. We compare in this case the two energy and baselines
options. 

The binned $\Delta_{CP}(i)$ 
discriminant for the shortest baseline $L=732\rm\ km$, $E_\mu=7.5\rm\ GeV$
and longest baseline $L=2900\rm\ km$, $E_\mu=30\rm\ GeV$
(lower plots) for three values of the
$\delta$-phase ($\delta=+\pi/2$, $\delta=0$ and $\delta=-\pi/2$)
are shown in Figure~\ref{fig:directcp}.
The errors are statistical and correspond to a normalization of
$10^{21}$($2.5\times 10^{20}$)
for $L=732(2900)\rm\ km$. The full curve corresponds
to the theoretical probability difference.
The dotted curve is the
theoretical curve for $\delta=0$ and represents the effect of 
propagation in matter.

As was already pointed out, the $\Delta_{CP}$ does not vanish even in
the case $\delta=0$, since matter introduces an asymmetry. 
At the shortest baseline ($L=732\rm\ km$), these effects are rather
small, as illustrated in Figure~\ref{fig:respcp1}. This has the advantage
that the observed asymmetry would be positive for $\delta>0$, but would
still change sign in the case $\delta\approx -\pi/2$. In the fortunate case
in which Nature has chosen such a value for the $\delta$-phase, the
observation of the negative asymmetry would be a striking sign for
$CP$-violation, since matter could never produce such an effect.

For other values of the $\delta$-phase, the effect is positive. It is also
always positive at the largest baseline $L=2900\rm\ km$, since at those
distances the effect induced by the $\delta$-phase is smaller than the
asymmetry introduced by the matter.

\subsection{Comparison of two methods}
\label{compare}
The binned $\Delta_T(i)$ and $\Delta_{CP}(i)$ discriminant can be used to
calculate the $\chi^2$ significance of the effect, given the statistical
error on each bin. We compute the following $\chi^2$'s:
\begin{eqnarray}
\chi^2_T = \sum_{i}
\frac{\left(\Delta_T(i,\delta)-\Delta_T(i,\delta=0)\right)^2}
{\sigma(\Delta_T(i,\delta))^2}+
\frac{\left(\bar\Delta_T(i,\delta)-\bar\Delta_T(i,\delta=0)\right)^2}
{\sigma(\bar\Delta_T(i,\delta))^2}
\end{eqnarray}
where $\sigma(\Delta_T(i,\delta))$ is the statistical error in the bin.
Since $\Delta_{T}(\delta)$ and $\bar\Delta_{T}(\delta)$ use independent
sets of data, the global $\chi^2_T$ can be computed in this way,
as linear sum of both contributions.

Similarly, the $\chi^2$ of the CP-asymmetry is
\begin{eqnarray}
\chi^2_{CP} = \sum_{i} \frac{\left(\Delta_{CP}(i,\delta)-\Delta_{CP}(i,\delta=0)\right)^2}
{\sigma(\Delta_{CP}(i,\delta))^2}
\end{eqnarray}

We study the significance of the effect as a function of the solar mass
difference $\Delta m^2_{21}$, since the effect associated to the
$\delta$-phase will decrease with decreasing $\Delta m^2_{21}$ values. We
consider the range compatible with solar neutrino experiments,
$10^{-5}\lesssim \Delta m^2_{21}\lesssim 10^{-4}\rm\ eV^2$.

The exclusion regions obtained 
at the 90\%C.L. (defined as $\Delta \chi^2 = +1.96$)
in the $\delta$-phase vs $\Delta
m^2_{21}$ plane are shown in Figure~\ref{fig:excltcp}.
The rest of parameters are fixed to the reference values
($\Delta m^2_{32}=3\times 10^{-3}\ \rm eV^2$,
 $\Delta m^2_{21}=1\times 10^{-4}\ \rm eV^2$,
 $\sin^2 \theta_{23} = 0.5$, $\sin^2 \theta_{12} = 0.5$, and
 $\sin^2 2\theta_{13} = 0.05$).
An electron charge confusion probability of 0.1\% and an electron
detection efficiency of 20\% has been assumed.
The normalizations assumed are
$10^{21}$ and $5\times 10^{21}$ muon decays with energy 
$E_\mu=7.5\rm\ GeV$ and a baseline of $L=732\rm\ km$.

The results are very encouraging. With $10^{21}$ muon decays, the region
$\Delta m^2_{21}\gtrsim 4\times 10^{-5}\rm\ eV^2$ is covered. For $5\times
10^{21}$ muons, this region extends down to $2\times 10^{-5}\rm\ eV^2$.
If we consider that the value of $\Delta m^2_{21}$ is known and that
it has a value
of $\Delta m^2_{21}=10^{-4}\rm\ eV^2$, one can constrain the values
of the $\delta$-phase within the range $|\delta|\lesssim 0.57$ or
$|\delta| \gtrsim 2.6$ for $10^{21}$ muons and
$|\delta|\lesssim 0.12$ and $|\delta| \gtrsim 3.0$ for $5\times 10^{21}$ 
muon decays at the 90\%C.L.

 Figure~\ref{fig:exclcp2base} shows the same exclusion plot
but comparing the two considered $L/E_{\nu}$ values.
For L = 730~km, the normalizations are $10^{21}$
and $5 \times 10^{21}$ muons, while for L = 2900~km the considered
fluxes are four times smaller.
Contrary to what happens at short baselines, where a nice symmetry
between $+ \delta$ and $- \delta$ is observed, at L = 2900~km 
matter effects introduce a clear asymmetry between the two excluded regions.

 We conclude that an exhaustive direct,
model-independent exploration of the $\delta$-phase, within the full range 
$10^{-5}\lesssim \Delta m^2_{21}\lesssim 10^{-4}\rm\ eV^2$ requires an
intensity of $5\times 10^{21}$ muon decays of each sign.

As explained in Section~\ref{extroscprob},
the value of the $\nu_{\mu} \rightarrow \nu_e$ oscillation probability
(and hence the $\chi^2_T$ discriminant) depends on both 
the electron detection efficiency, $\epsilon_e$, and
the electron charge confusion, $p_{conf}$ (see equation~\ref{pnumunue}).
In Figure~\ref{fig:chiconfeff} we show the iso-curves at
1$\sigma$, 2$\sigma$ and 3$\sigma$ levels of
significance\footnote{The $n\sigma$ region is reached when
$\sqrt{\chi^2_{\Delta T} + \chi^2_{\Delta \bar{T}}} \geq n$.}
that can be obtained for $\delta = +\pi/2$ and $-\pi/2$.
For instance, in the case of
($\epsilon_e$, $p_{conf}$)$=$(20\%,1\%) one could exclude both
$\delta = \pm \pi/2$ at 2$\sigma$ level.
Looking at the shape of the iso curves we can conclude that,
even if the result depends on both parameters, 
it is much more sensitive to the value of the charge confusion,
which should be smaller than $\sim$0.1\% for efficiencies of $\sim$20\%
to have an exclusion power above 3$\sigma$.

\subsection{Comparison with the fit of the visible energy distributions}

The most effective way to fit the oscillation parameters is
to study the visible energy distribution of
the four classes of events\footnote{In this case, we do not use
the information coming from the electron charge.}, since assuming
the unoscillated spectra are known, they contain
direct information on the oscillation probabilities.

Of course, for electron or muon charged current events, the visible
energy reconstructs the incoming neutrino energy. In the
case of neutral currents or the charged current of tau
neutrinos, the visible energy is less than the visible
energy because of undetected neutrinos in the final state.
The information is in this case degraded but can
still be used.

In case no $CP$-violation is observed, the result of the fit
in terms of 2-dimensional 90\% C.L. contours
in the $\Delta m^2_{12}-\delta$ plane is shown in 
Figure~\ref{fig:evisfit}. The {\tt Y} axis spans the $\Delta m^2_{12}$
range allowed by LMA solar neutrinos, and the whole range $-\pi<\delta<\pi$
has been consider.
For each pair of values ($\Delta m^2_{12}$,$\delta$), the fit to the
reference distributions\footnote{The reference values are
 $\Delta m^2_{32}=3\times 10^{-3}\ \rm eV^2$,
 $\sin^2 \theta_{23} = 0.5$,
 $\sin^2 \theta_{12} = 0.5$,
 $\sin^2 2\theta_{13} = 0.05$ and
 $\delta = 0$.}
was performed fixing all parameters (lower curves) and
leaving $\theta_{13}$ free (upper curves).
The result for the two considered baselines and energies are shown.

As expected, this method is clearly more powerful than the direct search 
for $CP$-asymmetry discussed in Section~\ref{sec:directsearchcp}.
The comparison between both results reveals that the non-excluded zones
near the extremes ($|\delta| \sim \pi$) in Figure~\ref{fig:excltcp}
are now covered and can be explored with this second method
(Figure~\ref{fig:evisfit}).
However, even if higher exclusive, the fit of the visible energy distributions
requires a good knowledge of the oscillation parameters, while the
direct search for $CP$-asymmetry is essentially model independent.
Nevertheless, even assuming that the value of $\theta_{13}$ is not known
precisely (upper curves in Figure~\ref{fig:evisfit})
the excluded regions extend beyond the limits set by the first method
for $|\delta| \sim \pi$.

The result obtained with the full simulation fully supports
all the above phenomenological considerations: for baselines such that
the maximum of the interesting effect lies well below the MSW resonance,
the destructive effect of matter almost does not play any role, and
for a given machine power, if $L/E_\mu$ is kept constant, the same
sensitivity is reached.

\section{Summary and Conclusions}

 In this document, we have discussed general strategies to detect
$CP$-violation effects related to the complex phase $\delta$ of the neutrino
mixing matrix, in the framework of a neutrino factory.
Here is the summary and conclusions:

$\bullet$
 In order to directly compare effects at different energies and baselines,
 we have defined the ``rescaled probability'':
 $P(\nu_\alpha\osc\nu_\beta;E_\nu,L)\times \frac{E^2_\nu}{L^2}$. It
 approximately weighs the probability by the neutrino energy spectrum of the
 neutrino factory and by the attenuation of the neutrino flux with the
 distance (Section~\ref{rescaledprob}).

$\bullet$
 In vacuum, the region of the ``first maximum'' on the
 $\nu_{\mu} \rightarrow \nu_e$ oscillation probability yields
 $E^{max}_{\nu}\simeq 2\rm\ GeV$ at 730~km, $\simeq 8\rm\ GeV$
 at 2900~km and $\simeq 20\rm\ GeV$ at 7400~km
 for $\Delta m^2_{32}=3\times 10^{-3}\rm\ eV^2$.
 For neutrinos propagating through matter, the oscillation probabilities
 are distorted and the resonant energy is
 $E^{res}_\nu \simeq 14.1$, $12.3$ and $10.7\rm\ GeV$ for $\Delta
 m^2_{32}=3\times 10^{-3}\rm\ eV^2$ and $\rho$
 equal to $2.7$, $3.2$ and $3.7\rm\ g/cm^3$ respectively.

$\bullet$
 {\it The most favorable choice of neutrino energy $E_\nu$ and
 baseline $L$ is in the region of the ``first maximum''} given by 
 $(L/E_\nu)^{max}\simeq 400$ km/GeV for
 $|\Delta m^2_{32}|=3\times 10^{-3}\rm\ eV^2$.
 As discussed in Section~\ref{le_scaling}, if one wants to
 study oscillations in this region
 one has to require that the energy of the ``first-maximum'' be smaller than
 the MSW resonance energy:
 $2\sqrt{2}G_Fn_eE^{max}_\nu\lesssim\Delta m^2_{32}\cos 2\theta_{13}$.
 This fixes a limit on the baseline $L_{max} \approx$5000 km
 beyond which matter effects spoil the sensitivity.

 {\it This implies that we concentrate on medium $L/E_{\nu}$}.

$\bullet$
 We have considered three quantities which are good discriminators
 for a non-vanishing $\delta$-phase (Section~\ref{discriminators}):
 $\Delta_{\delta}$, $\Delta_{CP}$ and $\Delta_{T}$.

 $\Delta_{\delta}$ can provide excellent determination of the phase, limited
 only by statistics, but needs a precise knowledge of the other oscillation
 parameters and possible correlations with the $\theta_{13}$ at high energy
 can appear.

 $\Delta_{CP}$ has the advantage that involves the appearance of wrong-sign
 muons only, experimentally easy to detect, but the disadvantage of involving 
 both, neutrinos and antineutrinos, which requires a good understanding of
 the effects related to matter.
 Matter effects will largely spoil $\Delta_{CP}$ at long distances.

 $\Delta_{T}$ is not affected by matter effects because it involves only
 neutrinos, but it requires to discriminate the electron charge, which is
 experimentally very challenging.
 A non-vanishing $\Delta_{T}$ would be a direct proof for a non-vanishing
 $\delta$-phase.

 {\it We have demonstrated that the discriminants of the $\delta$-phase
 in vacuum scale with $L/E_{\nu}$} (in matter, this scaling remains valid
 for $L$ smaller than $\sim$5000 km).

$\bullet$
 In one hand, because of the linear rise of the neutrino cross-section with
 neutrino energy at high energy, the statistical significance of the effect
 scales with $E_{\nu}$, so it grows linearly with $L$
 (for $L/E_{\nu}$ constant).

 On the other, the study of $\Delta_{T}$ requires the electron charge
 identification.
 {\it For a constant ($L/E_\nu$) ratio, the need of low energy electrons
 points towards lower-energy beams and shorter distances.}

$\bullet$
 As an example, we have considered two concrete cases keeping the same 
 $L/E_\mu$ ratio:
 ($L = $ 730 km, $E_{\mu} =$ 7.5 GeV) and
 ($L = $ 2900 km, $E_{\mu} =$ 30 GeV).
 We have classified the events in five classes (right and wrong sign muons,
 right and wrong sign electrons and no leptons) assuming a 10 kT Liquid Argon
 detector with an electron charge confusion of 0.1\% .
 We have computed the exclusion regions in the
 $\Delta m^2_{12} - \delta$ plane using the $\Delta_{CP}$ and $\Delta_{T}$
 discriminants, obtaining a similar excluded region provided that the
 electron detection efficiency is $\sim$20\%.
 The $\Delta m^2_{12}$ 
 compatible with the LMA solar data can be tested with a flux of
 5$\times 10^{21}$ muons.

$\bullet$
 Finally, we have computed the exclusion regions in the
 $\Delta m^2_{12} - \delta$ plane fitting the visible energy distributions.
 This method, more powerful than the previous one but not model independent,
 extends the excluded regions up to values of $|\delta|$ close to $\pi$,
 even when $\theta_{13}$ is left free.

%
%

~\newpage
\begin{figure}[p]
\centering
\epsfig{file=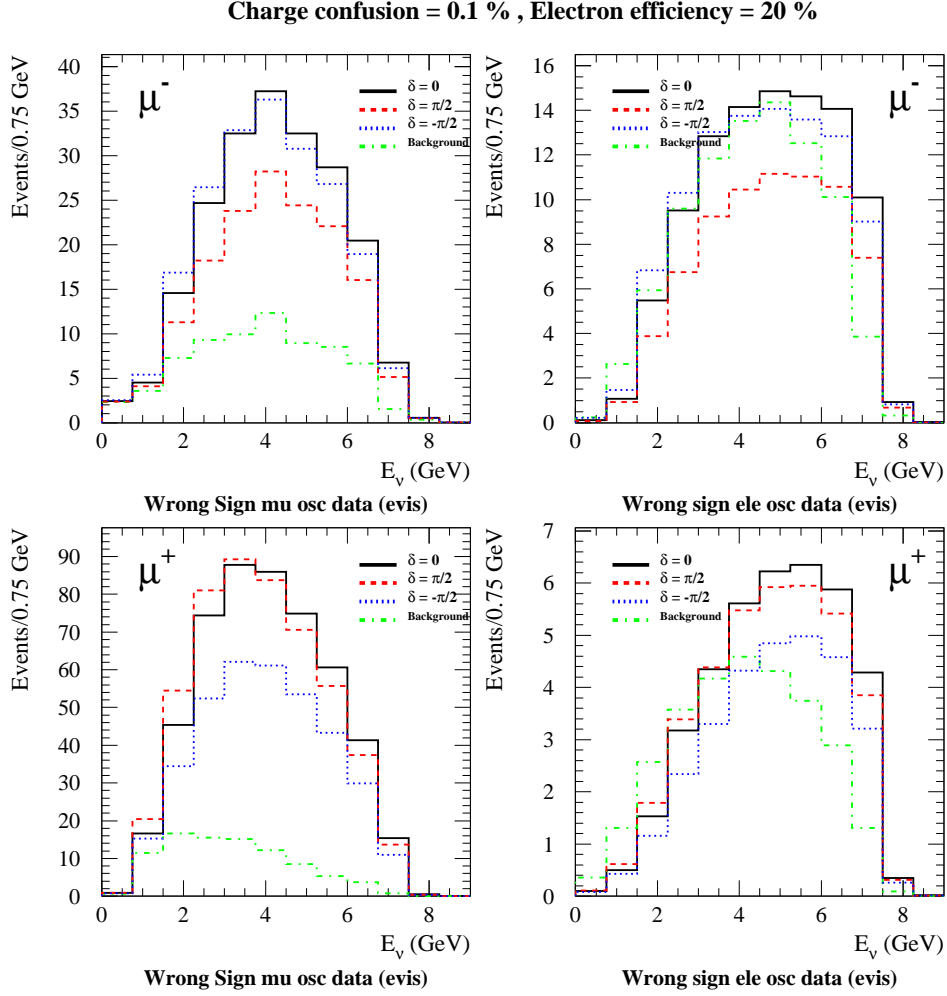,width=14cm}
\caption{Visible energy distribution for wrong-sign muons (left) and
wrong-sign electrons (right) normalized to $10^{21}$ muon decays.
The electron efficiency $\epsilon_e$ is assumed to be 20\% and charge
confusion probability is set to 0.1\%.
Three sets of curves are represented, corresponding to $\delta=+\pi/2$
(dashed line), $\delta=0$ (full line) and $\delta=-\pi/2$ (dotted line).
The background contribution from tau decays is also shown.
The other oscillation parameters are $\Delta m^2_{32}=3\times 10^{-3}\ \rm
eV^2$, $\Delta m^2_{21}=1\times 10^{-4}\ \rm
eV^2$, $\sin^2 \theta_{23} = 0.5$, $\sin^2 \theta_{12} = 0.5$
and $\sin^2 2\theta_{13} = 0.05$.}
\end{figure}


\begin{figure}[p]
\centering
\epsfig{file=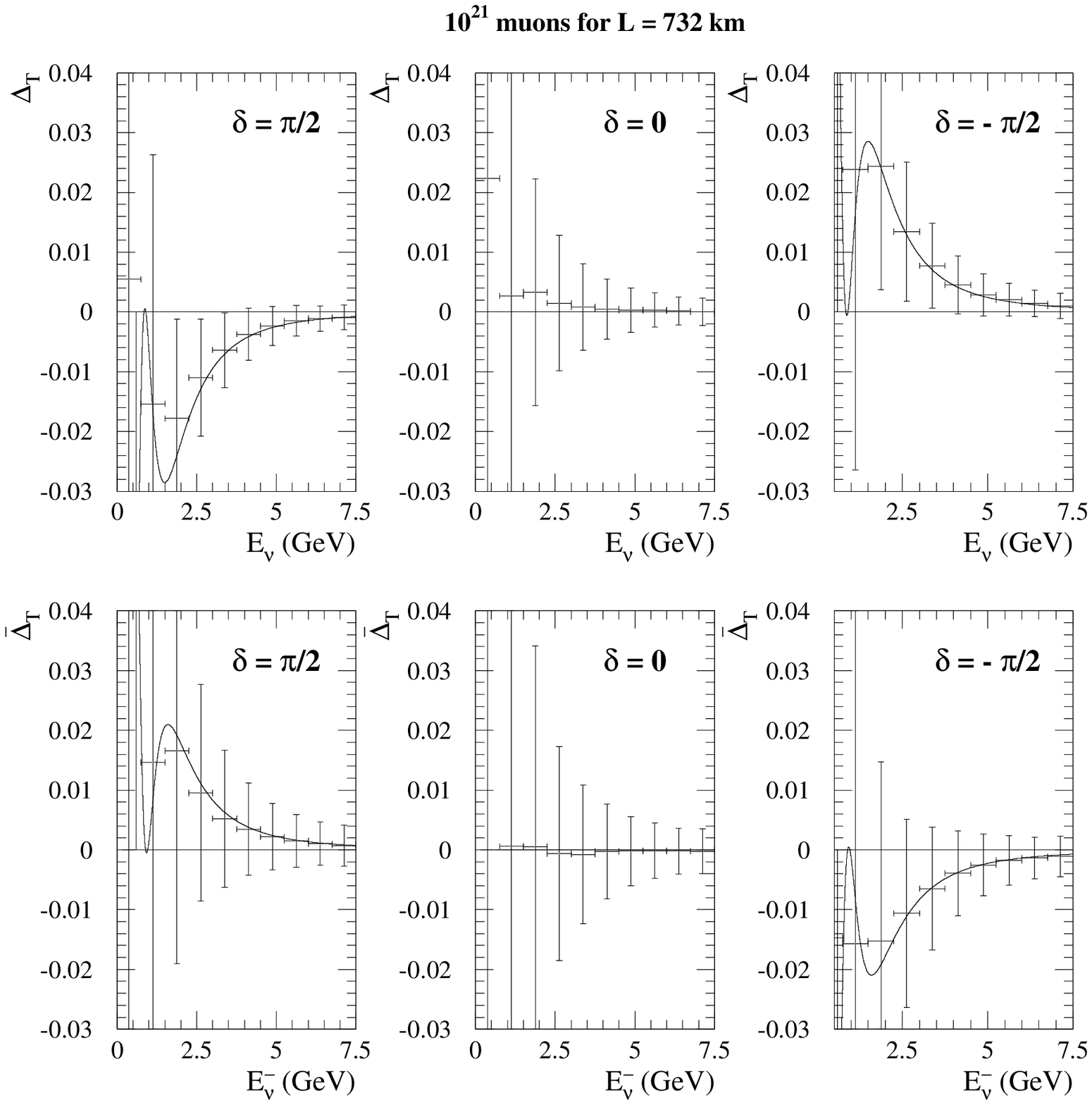,width=15cm}
\caption{Direct $T$-violation: Binned $\Delta_T(i)$ discriminant for neutrinos (upper
plots) and antineutrinos (lower plots) for three values of the
$\delta$-phase: $\delta=+\pi/2$, $\delta=0$ and $\delta=-\pi/2$.
The errors are statistical and correspond to a normalization of $10^{21}$
muon decays and a baseline of $L=732\rm\ km$. 
A 20\% electron efficiency with a charge confusion
probability of 0.1\% has been assumed. The change of sign of the effect
with respect of the change $\delta\ra -\delta$ and the substitution of
neutrinos by antineutrinos is clearly visible. The full curve corresponds
to the theoretical probability difference.}
\label{fig:directt}
\end{figure}

\begin{figure}[p]
\centering
\epsfig{file=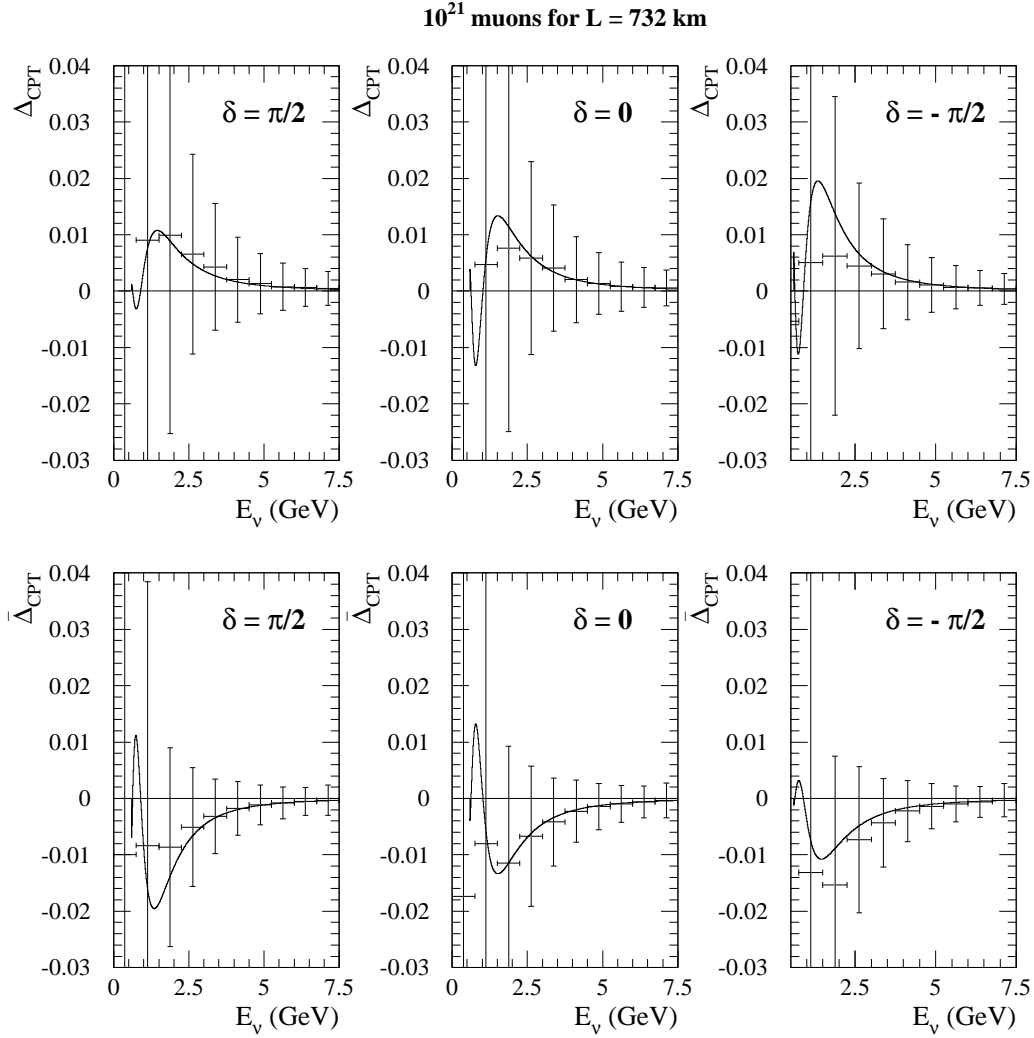,width=15cm}
\caption{Direct $CPT$-violation: Binned $\Delta_{CPT}(i)$ (upper
plots) and $\bar\Delta_{CPT}(i)$  (lower plots) for three values of the
$\delta$-phase: $\delta=+\pi/2$, $\delta=0$ and $\delta=-\pi/2$.
The errors are statistical and correspond to a normalization of $10^{21}$
muon decays and a baseline of $L=732\rm\ km$. 
A 20\% electron efficiency with a charge confusion
probability of 0.1\% has been assumed. The change of sign of the effect
with respect to the substitution of
neutrinos by antineutrinos is clearly visible. The full curve corresponds
to the theoretical probability difference.}
\label{fig:directcpt}
\end{figure}

\begin{figure}[p]
\centering
\epsfig{file=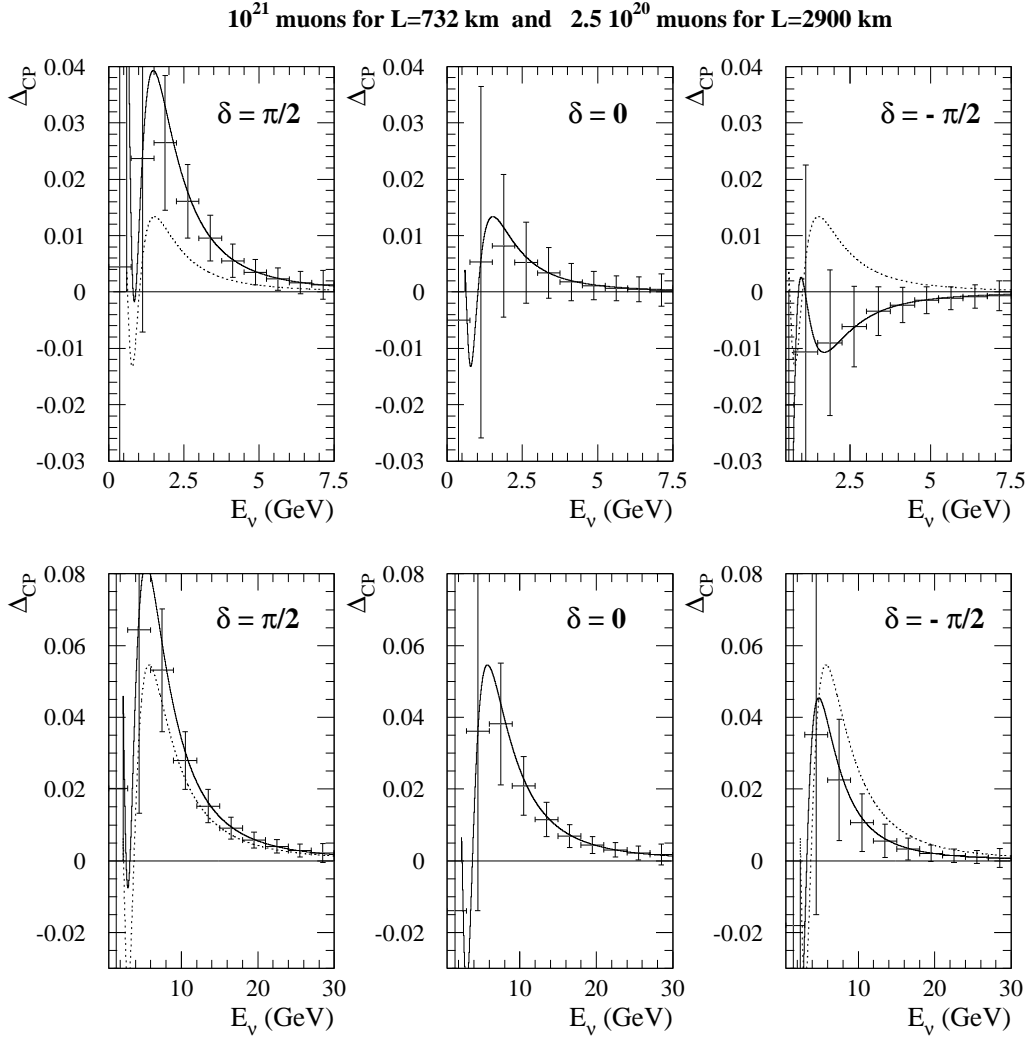,width=15cm}
\caption{Direct $CP$-violation: Binned $\Delta_{CP}(i)$ 
discriminant for the shortest baseline $L=732\rm\ km$, $E_\mu=7.5\rm\ GeV$ (upper
plots) and longest baseline $L=2900\rm\ km$, $E_\mu=30\rm\ GeV$
(lower plots) for three values of the
$\delta$-phase: $\delta=+\pi/2$, $\delta=0$ and $\delta=-\pi/2$.
The errors are statistical and correspond to a normalization of
$10^{21}$($2.5\times 10^{20}$)
for $L=732(2900)\rm\ km$. The full curve corresponds
to the theoretical probability difference. The dotted curve is the
theoretical curve for $\delta=0$ and represents the effect of propagation
in matter.}
\label{fig:directcp}
\end{figure}

\begin{figure}[p]
\centering
\epsfig{file=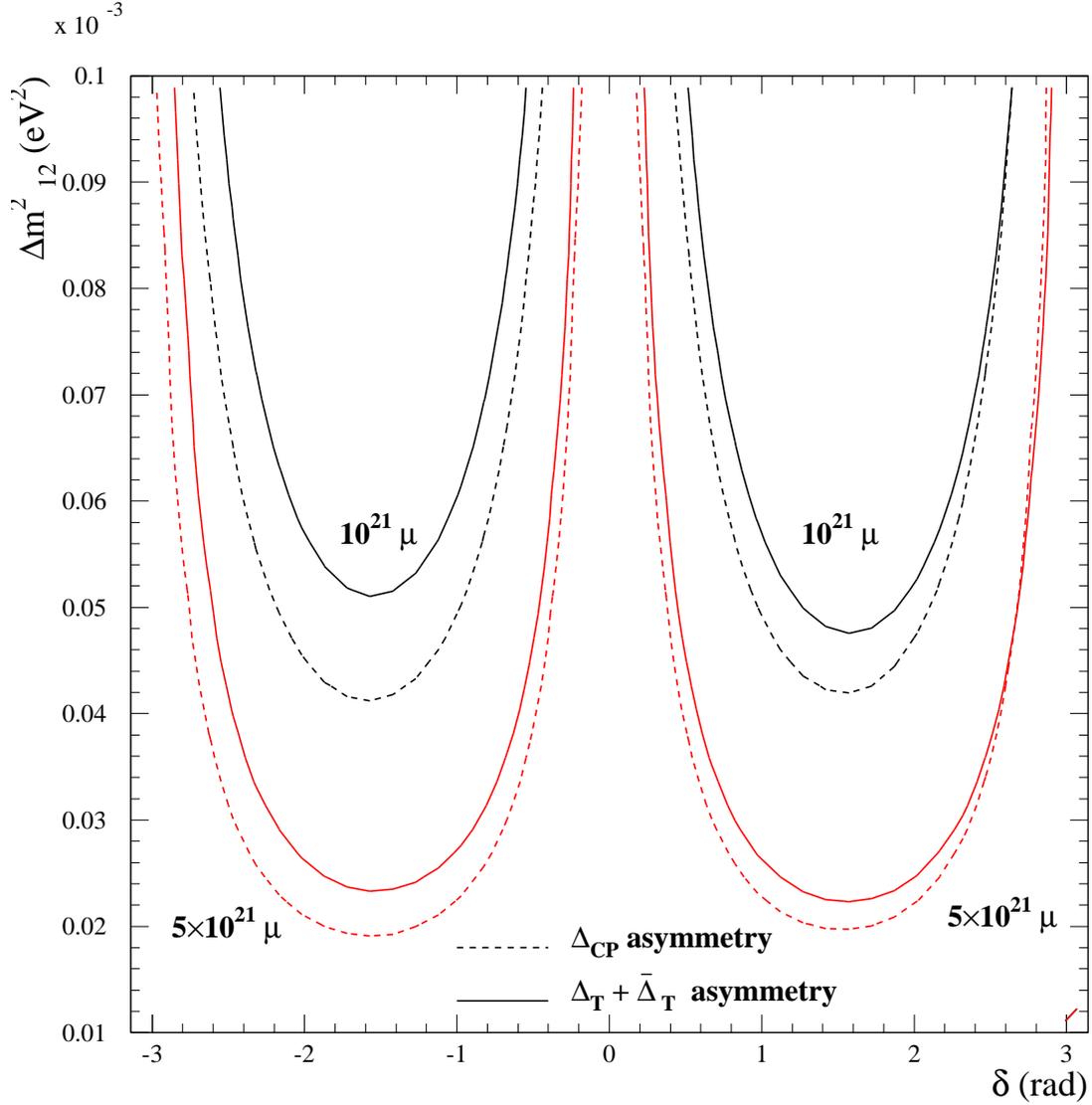,width=16cm}
\caption{Exclusion region at the 90\%C.L. ($\chi^2 > \chi^{2}_{min} + 1.96$)
in the $\delta$-phase vs $\Delta m^2_{21}$ plane.
Two regions obtained with the $\Delta_{CP}$ and 
the sum of $\Delta_{T}$ and $\bar{\Delta}_{T}$
discriminants are shown. A charge confusion probability of 0.1\%
and an electron efficiency of 20\% has been assumed.
The result is shown for two normalizations,
$10^{21}$ and $5 \times 10^{21}$ muon decays of each type
with energy $E_\mu=7.5\rm\ GeV$ and a baseline of $L=732\rm\ km$.
The reference oscillation parameters are
$\Delta m^2_{32}=3\times 10^{-3}\ \rm eV^2$,
$\sin^2 \theta_{23} = 0.5$,
$\sin^2 \theta_{12} = 0.5$ and
$\sin^2 2\theta_{13} = 0.05$.}
\label{fig:excltcp}
\end{figure}

\begin{figure}[p]
\centering
\epsfig{file=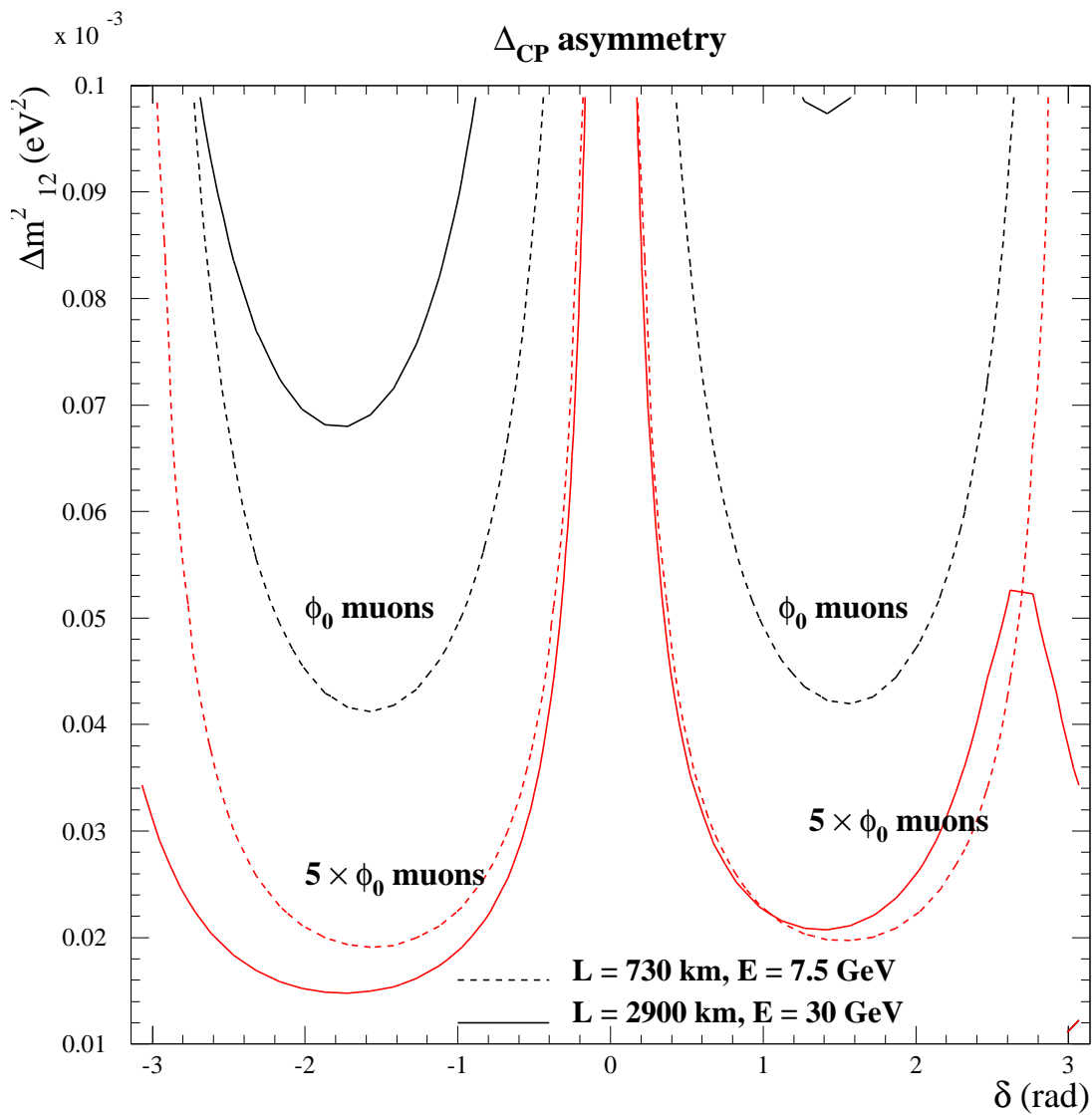,width=16cm}
\caption{Exclusion region at the 90\%C.L. ($\chi^2 > \chi^{2}_{min} + 1.96$)
in the $\delta$-phase vs $\Delta m^2_{21}$ from the $\Delta_{CP}$ discriminant.
The result is shown for two baselines and two normalizations:
 ($L=732\rm\ km$, $E_\mu=7.5\rm\ GeV$, $\phi_0 = 10^{21} \mu$ flux) and 
 ($L=2900\rm\ km$, $E_\mu=30\rm\ GeV$, $\phi_0 = 2.5 \times 10^{20} \mu$ flux).
The reference oscillation parameters are
$\Delta m^2_{32}=3\times 10^{-3}\ \rm eV^2$,
$\sin^2 \theta_{23} = 0.5$,
$\sin^2 \theta_{12} = 0.5$ and
$\sin^2 2\theta_{13} = 0.05$.}
\label{fig:exclcp2base}
\end{figure}

\begin{figure}[p]
\centering
\epsfig{file=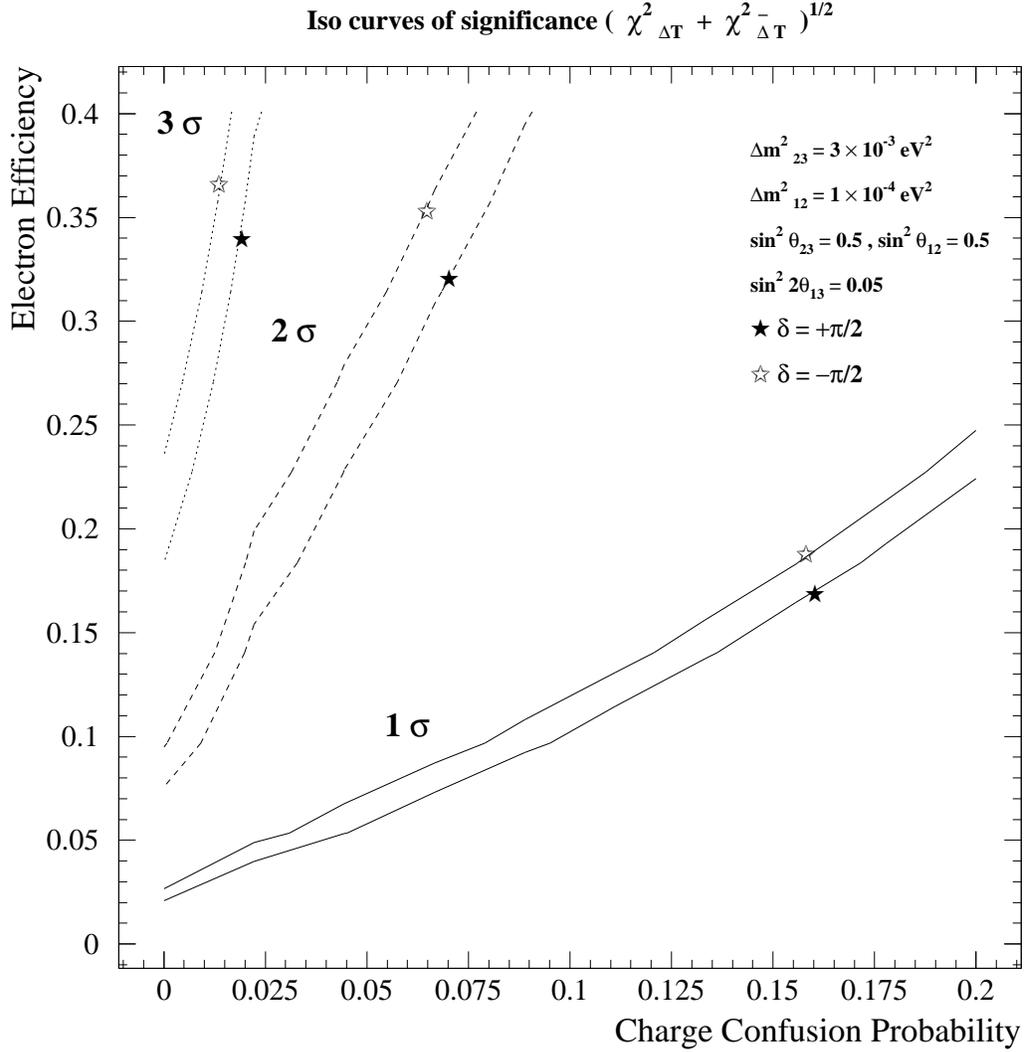,width=15cm}
\caption{The $1\sigma$, $2\sigma$ and $3\sigma$ $\chi^2$ contours
of the combined $\Delta_T$ and $\bar{\Delta}_T$ discriminants as a function
of the electron efficiency and the probability for charge confusion.
The result is shown for $\delta=+\pi/2$ and $\delta=-\pi/2$.
The other oscillation parameters are $\Delta m^2_{32}=3\times 10^{-3}\ \rm
eV^2$, $\Delta m^2_{21}=1\times 10^{-4}\ \rm
eV^2$, $\sin^2 \theta_{23} = 0.5$, $\sin^2 \theta_{12} = 0.5$
and $\sin^2 2\theta_{13} = 0.05$.}
\label{fig:chiconfeff}
\end{figure}

\begin{figure}[p]
\centering
\epsfig{file=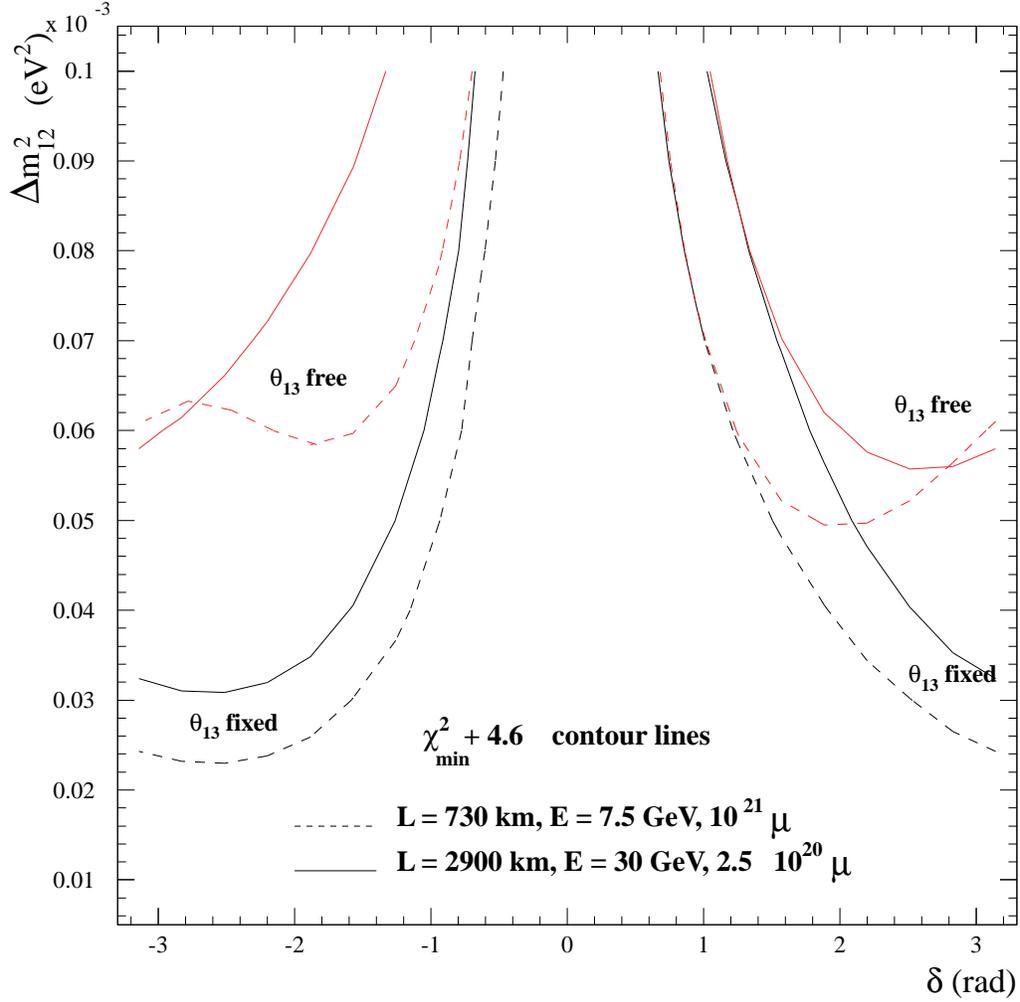,width=15cm}
\caption{90\% C.L. sensitivity on the $CP$-phase $\delta$ as a function of
$\Delta m^2_{21}$ for the two considered baselines.
The reference oscillation parameters are
$\Delta m^2_{32}=3\times 10^{-3}\ \rm eV^2$,
$\sin^2 \theta_{23} = 0.5$,
$\sin^2 \theta_{12} = 0.5$,
$\sin^2 2\theta_{13} = 0.05$ and
$\delta = 0$.
The lower curves are made fixing all parameters to the reference values
while for the upper curves $\theta_{13}$ is free.}
\label{fig:evisfit}
\end{figure}

\end{document}